\documentclass[]{PBCreport}

\usepackage{placeins}

\usepackage{booktabs}
\usepackage{afterpage}
\usepackage[T1]{fontenc}
\usepackage{textcomp}
\usepackage{color, colortbl}
\usepackage{booktabs}
\usepackage{afterpage}
\usepackage{tabularx}
\usepackage{mathtools}
\usepackage{siunitx}
\usepackage{upgreek}
\usepackage{xcolor}
\usepackage{hyperref}
\usepackage{graphicx,mathptmx,amsmath,amsfonts,amscd,amsthm,amssymb,eucal,psfrag,color,subfig,url,multirow,gensymb}
\usepackage{rotating}
\usepackage{epsfig}
\usepackage{latexsym}
\usepackage{xspace}
\usepackage{comment}
\usepackage{indentfirst}
\usepackage{enumerate}
\usepackage{url}

\DeclareSIUnit\c{\text{c}}
\DeclareSIUnit\eVperc{\eV\per\c}
\DeclarePairedDelimiterXPP\BigOSI[2]%
  {\mathcal{O}}{(}{)}{}%
  {\SI{#1}{#2}}

\usepackage[acronym,toc]{glossaries}

\makeglossaries

\hypersetup{
     colorlinks = true,
     linkcolor = black,
     anchorcolor = blue,
     citecolor = blue,
     filecolor = blue,
     urlcolor = blue
     }

\definecolor{LightCyan}{rgb}{0.88,1,1}
\definecolor{LightYellow}{rgb}{1,0.97,0.9}

\documentlabel{CERN-PBC Report-2025-003}

\begin{document}
\newcommand{\leadFS}{\ensuremath{^{208}\mathrm{Pb}^{82+}}\xspace}
\newcommand{\leadPS}{\ensuremath{^{208}\mathrm{Pb}^{54+}}\xspace}
\newcommand{\magnesiumFS}{\ensuremath{^{24}\mathrm{Mg}^{12+}}\xspace}
\newcommand{\magnesiumPS}{\ensuremath{^{24}\mathrm{Mg}^{7+}}\xspace}
\newcommand{\oxygenFS}{\ensuremath{^{16}\mathrm{O}^{8+}}\xspace}
\newcommand{\oxygenPS}{\ensuremath{^{16}\mathrm{O}^{4+}}\xspace}
\newcommand{\boronelePS}{\ensuremath{^{11}\mathrm{B}^{3+}}\xspace}
\newcommand{\boroneleFS}{\ensuremath{^{11}\mathrm{B}^{5+}}\xspace}
\newcommand{\borontenFS}{\ensuremath{^{10}\mathrm{B}^{5+}}\xspace}
\newcommand{\borontenPS}{\ensuremath{^{10}\mathrm{B}^{3+}}\xspace}
\newcommand{\lumprot}[1]{$#1\times10^{34}$~cm$^{-2}$s$^{-1}$\xspace}
\newcommand{\bstar}{\ensuremath{\beta^{*}}\xspace}
\makeatletter
\def\bstctlcite{\@ifnextchar[{\@bstctlcite}{\@bstctlcite[@auxout]}}
\def\@bstctlcite[#1]#2{\@bsphack
  \@for\@citeb:=#2\do{%
    \edef\@citeb{\expandafter\@firstofone\@citeb}%
    \if@filesw\immediate\write\csname #1\endcsname{\string\citation{\@citeb}}\fi}%
  \@esphack}
\makeatother

\title{Summary Report of the Physics Beyond Colliders Study at CERN}

\author{\parbox{\textwidth}{\small\it
R.~Alemany~Fernández$^{1}$, M.~Au$^{1}$, G.~Arduini$^{*,1}$, L.~Bandiera$^{2}$, D.~Banerjee$^{1}$, H.~Bartosik$^{1}$, J.~Bernhard$^{1}$, D.~Boer$^{\&,3}$, J.~Boyd$^{1}$, O.~Brandt$^{4}$, M.~Brugger$^{1}$, O.~Buchm\"uller$^{5,6}$, F.~Butin$^{1}$, S.~Calatroni$^{1}$, C.~Carli$^{1}$, N.~Charitonidis$^{1}$, P. Crivelli$^{7}$, D.~Curtin$^{8}$, R.~T.~D'Agnolo$^{9,10}$, G.~De~Lellis$^{11,12}$, O. Denisov$^{1,13}$, P. Di Nezza$^{14}$, B.~D\"obrich$^{15}$, Y.~Dutheil$^{1}$,  J.~R.~Ellis$^{1,16}$, S.~A.~R.~Ellis$^{17}$, T.~Ferber$^{\dagger,18}$, M.~Ferro-Luzzi$^{1}$, M.~Fraser$^{1}$, T.~Galatyuk$^{\&,19}$, D.~Gamba$^{1}$, C.~Gatti$^{14}$, S.~Gilardoni$^{1}$, A.~Glazov$^{\&,20}$, M.~Gorshtein$^{21}$, E.~Granados$^{1}$, E.~Gschwendtner$^{1}$, P.~Hermes$^{1}$, J.~Jaeckel$^{*,22}$,  M.~A.~Jebramcik$^{1}$,  Y.~Kadi$^{1}$, F.~Kahlhoefer$^{+,23}$, F.~Kling$^{20}$, M.~Kowalska$^{1}$, M.~W.~Krasny$^{1,24}$,  B.~Ketzer$^{25}$, M.~Lamont$^{1}$, P.~Lenisa$^{2,26}$, A.~Lindner$^{20}$, M.~Maćkowiak-Pawłowska$^{27}$, A.~Martens$^{28}$, F.~Martinez Vidal$^{29}$, C. Matteuzzi$^{30}$, N.~Neri$^{31}$, L.~J.~Nevay$^{1}$, M.~Ovchynnikov$^{1}$, Y.~Papaphilippou$^{1}$, L.~L.~Pappalardo$^{2,26}$, J.~Pawlowski$^{\&,22}$, M.~Perrin-Terrin$^{32}$, J.~Pinfold$^{33}$, L.~Ponce$^{1}$, M.~Pospelov$^{+,34,35}$, T.~Prebibaj$^{1}$, J.~Pretz$^{36,37}$, S.~Redaelli$^{1}$, J.~Rojo$^{38}$, G.~Rumolo$^{1}$, A.~Rybicki$^{39}$, G.~Schnell$^{*,40,41}$, S.~Schuh$^{1}$, I.~Schulthess$^{20}$, E.~Scomparin$^{13}$, S.~Stapnes$^{1}$, P.~N.~Swallow$^{4}$, F.~Terranova$^{30}$, S.~Ulmer$^{42,43}$, G.~Usai$^{44}$, C.~Vall\'ee$^{32}$, M.~van Dijk$^{1}$, G. Venanzoni$^{45,46}$, M.~Wing$^{47}$, E.~D.~Zimmerman$^{48}$
}}

\abstract{\noindent The Physics Beyond Collider (PBC) Study Group was initially mandated by the CERN Management to prepare the previous European Particle Physics Strategy Update for CERN projects other than the high-energy frontier colliders. The main findings were summarized in an PBC Summary Report submitted to the Strategy Update.
Following the Update process, the PBC Study Group was confirmed on a permanent basis with an updated mandate
taking into account the strategy recommendations. The Study Group is now in charge of supporting the proponents of new ideas to address the technical issues and physics motivation of the projects ahead of their review by the CERN Scientific Committees and decision by the Management. The present document updates the previous PBC summary report
to inform the new ongoing European Particle Physics Strategy Update process, taking into account the evolution of the CERN and worldwide landscapes and the new projects under consideration within the Study Group. 
}

\maketitle

{\fontsize{9.4}{11.2}\selectfont 
\noindent
$^{1}$CERN, Geneva, CH\\
$^{2}${INFN Ferrara, Ferrara, IT} \\
$^{3}$Van Swinderen Institute for Particle Physics and Gravity, University of Groningen, Groningen, NL\\
$^{4}$Department of Physics, University of Cambridge, Cambridge, UK\\
$^{5}$Blackett Laboratory, Imperial College London, London, UK\\
$^{6}${University of Oxford, South Parks Road, Oxford, UK} \\
$^{7}${ETH Zurich, Institute for Particle Physics and Astrophysics, Zurich, CH}\\
$^{8}$Department of Physics, University of Toronto, Toronto, CAN\\
$^{9}${Institut de Physique Théorique, Université Paris Saclay, CNRS, CEA, Gif-sur-Yvette, FR} \\
$^{10}${Laboratoire de Physique de l’École Normale Supérieure, ENS, Université PSL, CNRS, Sorbonne Université, Université Paris Cité, Paris, FR}\\
$^{11}$INFN-Sezione di Napoli, Napoli, IT\\
$^{12}$Università degli Studi di Napoli Federico II, Napoli, IT\\
$^{13}${INFN, Sezione di Torino, Torino, IT}\\
$^{14}${Laboratori Nazionali di Frascati, INFN,  Frascati, IT} \\
$^{15}$Max-Planck-Institut  f\"ur Physik (Werner-Heisenberg-Institut),  Garching, DE\\
$^{16}${King’s College London, Strand, London, UK} \\
$^{17}${D\'epartement de Physique Th\'eorique, Universit\'e de Gen\`eve, Gen\`eve, CH}\\
$^{18}$Institute of Experimental Particle Physics, Karlsruhe Institute of Technology, Karlsruhe, DE\\
$^{19}${Department of Physics, Technische Universit\"{a}t Darmstadt, Darmstadt, DE}\\
$^{20}${Deutsches Elektronen-Synchrotron DESY, 22603 Hamburg, DE}\\
$^{21}${Johannes Gutenberg-Universität Mainz, 55128 Mainz, DE}\\
$^{22}${Institut f\"{u}r Theoretische Physik, Universit\"{a}t Heidelberg, Heidelberg, DE}\\
$^{23}$Institute for Astroparticle Physics, Karlsruhe Institute of Technology, Karlsruhe, DE\\
$^{24}${LPNHE, Sorbonne Université, Université de Paris, CNRS/IN2P3, Paris, FR} \\
$^{25}${Universität Bonn, Helmholtz-Institut für Strahlen- und Kernphysik, 53115 Bonn, DE} \\
$^{26}${University of Ferrara, Ferrara, IT} \\
$^{27}$Warsaw University of Technology, Warsaw, PL\\
$^{28}$Université Paris-Saclay, CNRS/IN2P3, IJCLab, Orsay, FR\\
$^{29}${Instituto de Física Corpuscular and Universitat de València, Catedratico Jose Beltrán, 2, València, SP}\\
$^{30}${INFN and Università di Milano Bicocca, Milano, IT}\\
$^{31}${INFN and Università di Milano, Milano, IT}\\
$^{32}$Aix Marseille Université, CNRS/IN2P3, CPPM, Marseille, FR\\
$^{33}$Centre for Particle Physics Research,
Physics Department,
University of Alberta,
Edmonton,
CAN\\
$^{34}$William I. Fine Theoretical Physics Institute, School of Physics and Astronomy, University of Minnesota, Minneapolis, US\\
$^{35}$School of Physics and Astronomy, University of Minnesota, Minneapolis, US\\
$^{36}$RWTH Aachen University, Aachen, DE \\
$^{37}$Forschungszentrum Jülich, Jülich, DE\\
$^{38}$Nikhef National Institute for Subatomic Physics, Science Park 105, 1098 XG Amsterdam, NL\\
$^{39}$Institute of Nuclear Physics, Polish Academy of Sciences, Radzikowskiego 152, 31-342 Krakow, PL\\
$^{40}$Department of Physics \& EHU Quantum Center, University of the Basque Country UPV/EHU, Bilbao, SP\\
$^{41}$IKERBASQUE, Basque Foundation for Science, Bilbao, SP\\
$^{42}${Heinrich-Heine Universit\"at, D\"usseldorf, DE}\\
$^{43}${RIKEN, Wako-Saitama, Tokyo, JP}\\
$^{44}${Universit\`a di Cagliari and INFN, Sezione di Cagliari, Monserrato, IT}\\
$^{45}$University of Liverpool Liverpool, UK\\
$^{46}$INFN-Sezione di Pisa, Pisa, IT\\
$^{47}${Department of Physics and Astronomy, UCL, London, UK} \\
$^{48}${University of Colorado, Boulder, Colorado, US}\\[+.15cm]

\noindent 
$^{*}$Physics Beyond Collider (PBC) coordinator\\
$^{+}$PBC Feebly Interacting Particle Physics Centre (FPC) convener\\
$^{\&}$PBC QCD Working Group convener\\
$^{\dagger}$PBC Beyond Standard Model Working Group convener
}
\newpage

\tableofcontents

\newpage

\section{Executive summary}

Physics Beyond Colliders (PBC) is a multidisciplinary study group to promote and facilitate a diverse physics programme at CERN. It was established by the CERN Directorate in 2016 to explore physics opportunities beyond high-energy frontier colliders and provide input into the 2020  European Particle Physics Strategy Update (EPPSU). Following the EPPSU process, the~PBC Study Group was confirmed on a permanent basis with an updated mandate~\cite{PBC_mandate}. Taking into account the strategy recommendations, it has flourished since then and now serves as a gateway for new experimental ideas building on CERN infrastructure, technology and expertise. It helps developing experiments from their inception to a level ready for discussion in the relevant scientific committees. 
Notable recent examples in this direction are FASER, SMOG2, SHiP/NA67,  and MuonE, the first two taking data since the beginning of LHC Run~3, the third in full swing in the TDR phase, and the fourth doing test measurements. 
This report provides an update to previous findings~\cite{PBC_summary_2020}, outlines ongoing projects, the planned and potential evolution of CERN’s accelerator complex, and reviews existing proposals with their potential physics reach considering the international physics landscape.

PBC aims at addressing open questions in fundamental physics. Two main physics areas discussed within PBC are quantum chromodynamics (QCD)  and beyond the Standard Model (BSM) physics. Both of them feature their own working groups to develop the experimental realization as well as the physics case. The BSM area is supported by a theory-experiment exchange hub, the Feebly Interacting Particle Physics Center (FPC), which also interacts with the worldwide community and scrutinizes and ensures the leading role of PBC experiments in this wider context.
Dedicated accelerator and technology working groups assist preparing the proposals with integration into the CERN complex as well as with technological, siting, and engineering aspects relevant for implementation at CERN or outside. 

\subsection*{Worldwide and CERN Physics Beyond Colliders landscape }

In the BSM area, the main targets are fundamental open questions such as: Are there new particles? What is Dark Matter? How to probe Dark Sectors? Can we shed light on the mysterious neutrino sector?
Benefiting from both high-energy and high-intensity beams, 
PBC experiments are using a whole range of strategies to push this frontier forward. FASER, NA62, and NA64 have already taken data providing world-leading constraints on a wide range of new feebly interacting particles (FIPs). FASER and SND@LHC have been using the main LHC collisions but exploited them in a novel far-forward position. This enabled the detection of the first collider produced TeV neutrinos, opening new avenues for studying neutrino interactions at the TeV scale. NA62, in addition to observing the decay of a charged kaon into a charged pion and a neutrino-antineutrino pair, one of the rarest processes ever observed,  took data with an SPS beam in a beam dump mode where the full primary proton beam is
dumped on a collimator. In this mode, NA62 has provided world-leading constraints on hypothetical particles such as dark photons and axion-like particles. NA64 used high-energy electron and muon beams with an active beam dump able to measure missing energy. NA64 is a key player in the search for sub-GeV dark sector particles and probed a sizable range of the parameter space that would have been able to explain the 
deviation of the muon anomalous magnetic moment $a_{\mu}$ from the Standard Model prediction. Recent pilot runs with muon beams have yielded critical constraints on lepton-flavor violating interactions.

QCD poses a number of critical challenges to our understanding. Being a strongly coupled system it requires developing and testing new computational methods beyond perturbation theory. Open questions include the phase diagram of QCD as well as predictions on the detailed structure of hadrons. The former is already a focus of NA61/SHINE. The AMBER/NA66 experiment focuses on the latter and the fundamental origins of mass within QCD. In its presently approved programme, AMBER/NA66 investigates the proton charge radius with high-energy muon-proton scattering, contributing to the resolution of long-standing discrepancies in proton radius measurements, as well as measures antiproton production cross sections. 
Likewise, SMOG2, a fixed-target program of LHCb at the LHC, utilizing a gas target immediately upstream of the experiment, offers a unique opportunity to study proton-nucleus collisions under controlled conditions. This experiment provides key insights into cosmic ray fluxes and cosmic ray showers, as well as on the QCD phase diagram and nuclear structure in general.

\subsection*{Developments in the CERN accelerator complex}

An overview of the CERN accelerator complex is shown Figure \ref{fig:AccComplex}.
Since the past EPPSU the accelerator complex was augmented with the LHC Injector Upgrade (LIU) aimed at providing high-brightness beams for the LHC in view of its high-luminosity upgrade (HL-LHC). The quality and intensity of the beams delivered to the fixed-target experimental areas of the complex from ISOLDE to the North Area~(NA) have also improved as a result of these modifications. This benefits PBC experiments working with ions and protons to understand QCD as well as those looking for new physics using protons but also secondary beams of electrons, muons and other particles. Further specific improvements are discussed within PBC to address specific needs, e.g., the ability to quickly switch between ion species to allow experiments such as NA61/SHINE to explore quark gluon plasma (QGP) properties as a function of the system size while enabling DICE/NA60+ to further explore the chiral and QGP phase transitions. 
The ongoing NA consolidation programme together with the high intensity upgrade of ECN3 will extend the physics opportunities in this area and improve the beam availability. Importantly, the recently approved SHiP/NA67 experiment will use the underground cavern ECN3 to search for FIPs with a state-of-the-art beam dump absorber capable of absorbing high intensity and to operate reliably over several years.
The HL-LHC upgrade taking place during LS3 will create unprecedented opportunities for hidden-sector searches, high-energy neutrino studies, and proton/nucleus structure measurements. These can be enhanced and fully realized by a range of suitable supplementary detectors, insertion of fixed targets and extraction of parts of the beam halo by means of crystals.

\subsection*{Future facilities}
A number of proposed and foreseen new facilities are considered within PBC. 
The envisioned Forward Physics Facility (FPF), discussed in more detail further below,
would allow for a variety of experiments located in a new cavern some 600~m down the LHC collision
axis from the ATLAS IP. These experiments, shielded by rock could make use of a massive amount of
TeV-level neutrinos and possibly of dark matter particles coming ``for free'' from HL-LHC
operations. In order to fully exploit the potential during the HL-LHC era,
infrastructure work would have to start during LS3.

The need for a high-precision neutrino beam for cross-section measurements to reduce systematic uncertainties of future neutrino oscillation experiments like DUNE and Hyper-Kamiokande has been recognized since the 2020 EPPSU. 
Recent studies within PBC of a short-baseline neutrino facility (SBN) have shown the possibility of providing a 
high-precision monitored and tagged medium-intensity neutrino beam, where the properties of the individual neutrinos are known. This would enable percent-level neutrino cross-section measurements at the GeV scale with a moderate amount of protons on target.
A new slow-extraction system towards a new experimental area close to the HiRadMat facility would be required, allowing the use of existing transfer tunnels. 
A conceptual design of the new slow-extraction scheme from the SPS needs to be further developed, demonstrating its feasibility and the compatibility with LHC operation.
The study, including the civil engineering and technical infrastructure design of the corresponding experimental area, should be completed by the end of LS3 in order to conduct the necessary modifications during LS4, if such proposal is approved.

First proposed within the PBC context, the Gamma Factory aims at reaching a new level of intensity and beam quality for MeV scale gamma rays. Partially stripped ions are accelerated to LHC energies and collided with a resonant optical laser. The de-excitation photons then are focused with an energy enhanced by the square of the ion relativistic boost. After first tests at the SPS and LHC, a proof-of-principle experiment is in preparation at the SPS and a wide range of physics applications from material science to fundamental physics are being developed. 

PBC also closely follows developments for the FCC-ee injectors, CLIC, and AWAKE, exploring possibilities to use the respective beams directly for physics applications.

\subsection*{Status of experiments within PBC and opportunities}
NA64 plans to increase the number of particles on targets and using different species (electrons, positrons, muons, hadrons) thereby widening the range of accessible physics. In particular, the positron programme will enable the first probe of the so-called ``thermal target lines'', which represent the dark sector coupling values predicted by astrophysical and cosmological arguments for selected model parameter values in the mass range \SIrange[range-units=single]{135}{250}{\MeV}. MUonE, aiming to measure the hadronic contributions to $a_{\mu}$ in a novel way, has progressed to supervision by SPSC and is performing pilot runs in view of a possible approval for operation post-LS3. SHiP/NA67 is expected to start data taking from 2031 and it is designed to detect neutrino interactions and scattering events of
hypothetical light dark matter particles produced in proton-target collisions as well as for searching for decays of FIPs.

A whole range of proposal aims at improving the exploitation of LHC in its high-luminosity phase with auxiliary detectors in the forward (SND@HL-LHC, FPF), intermediate \linebreak (MoEDAL-MAPP2), and transverse (ANUBIS, CODEX-b, MATHUSLA40) directions with respect to the collision axis of the main LHC interaction points. Test setups are either being built or already taking data. Optimization of the physics reach and cost inside but also between experiments is driving the development. In particular, the FPF would provide space
for a suite of far-forward small/medium-size detectors during the HL-LHC era. Experiments at the FPF can discover a wide variety of new particles that cannot be discovered at fixed-target facilities or other LHC experiments. The FPF is the only facility that will be able to detect millions of neutrinos with TeV energies, enabling precision probes of neutrino properties for all three flavors. These neutrinos will improve our understanding of proton and nuclear structure, enhancing the power of new particle searches at ATLAS and CMS, and providing valuable input for astro-particle experiments like IceCube, Auger and others. The studies conducted so far indicate that the civil engineering for the construction of the experimental cavern could take place during HL-LHC operation though construction should start during LS3 in order to allow commissioning with beam of the FPF detectors before the end of LHC Run~4.

Exploitation and extension of the capabilities to perform experiments with ions in the North Area are pursued by DICE/NA60+ and NA61/SHINE with the aim to perform studies of the phase structure of QCD including the confinement and chiral phase transitions. The future programme at AMBER/NA66 focuses on studying hadron structure and the fundamental origins of mass within QCD. Through precision measurements of Drell-Yan and J/$\psi$ production, it aims to extract parton distribution functions of pions and kaons, shedding light on the internal dynamics of strongly bound systems. 

SMOG2 has opened the way to fixed-target physics at the LHC. The LHCspin proposal envisages to conduct unique spin-dependent measurements of Drell-Yan and heavy-flavour production processes, by studying collisions of both proton and lead beams on polarized hydrogen or deuterium gas targets. This would
provide insight into details of the multi-dimensional structure of nucleons in terms of transverse momentum-dependent parton distributions.

Machine studies conducted at the SPS and LHC have demonstrated the suitability of crystals for the control of the beam halo and these devices are now used regularly in operation to enhance the collimation efficiency for LHC lead ion beams. The Double Crystal setup Proof of Principle Experiment (TWOCRYST) has the goal to demonstrate the feasibility of channeling the proton beam halo with a crystal onto an in-vacuum target and to bend TeV charmed baryons generated in the interaction by means of a second large bending-angle crystal that acts like a very powerful precession magnet.  The TWOCRYST experiment, scheduled for 2025, will provide vital information for the technical design of ALADDIN, a proposed fixed-target experiment at the Insertion Region (IR) 3 of the LHC
aiming to measure the magnetic and electric dipole moments of baryons, such as $\Lambda_c^+$ and $\Xi_c^+$. These measurements are highly sensitive probes for BSM physics but have remained unexplored so far due to the extremely short lifetimes of these particles. The ALADDIN collaboration is working on a technical proposal of the detector aiming to start the installation of the required accelerator hardware and detector during LS3, if approved, to conduct beam commissioning and start data taking before the end of LHC Run~4.

A broad range of further experimental opportunities at CERN and elsewhere are being developed within PBC. Notably a number of small experiments to detect axion(-like particle) dark matter benefit from magnet and RF cavity expertise (RADES, FLASH and an experiment aiming for a heterodyne detection approach). Other proposals aim at the detection of CP-violating electric dipole moments of the proton (cpEDM) and nuclei, with the latter exploiting the existing ISOLDE facility. Stronger connections with PBC and further exploitation of the Antiproton Decelerator are also being investigated. Infrastructure such as large and deep access shafts (e.g., those of the LHC) may open possibilities for atomic interferometry (AION) to reach a new level in the exploration of gravitational waves.

\subsection*{Physics targets and opportunities}
While by their very nature of being a broad and diverse effort, the PBC experiments aim at and have powerful sensitivity to a number of important physics targets. 

An important one is a broad and deep exploration of FIPs, i.e., dark sector particles. PBC experiments extend the reach of collider experiments to significantly smaller couplings (e.g., NA64, SHiP/NA67), complementary flavour and generally different coupling structures (e.g., NA64), longer lifetimes (e.g., ANUBIS, CODEX-b, MATHUSLA40), as well as different mass regimes. They can thereby target candidates for dark matter of thermal (e.g., NA64, SHiP/NA67) and non-thermal origin (e.g., AION, FLASH, RADES). 

Gravitational waves by now are proven to be an important driver in exploring astrophysics and the cosmos, but also fundamental processes such as phase transitions in the early Universe and the interactions giving rise to them. Using techniques such as atomic interferometry (AION) to enter new frequency ranges, allows to discover new sources as well as probing new underlying physics scales.

CP violation is a necessary ingredient for baryogenesis, but its absence is also a puzzle of the strong interactions. Suitable observables such as electric dipole moments (e.g., ALADDIN, cpEDM, ISOLDE EDM) can provide sensitivity to new physics at mass scales often many orders of magnitude in excess of the TeV scale.

Neutrino masses and mixings, possible non-standard interactions as well as the possible existence of sterile states are promising avenues to find physics beyond the Standard Model. At the same time, deep-inelastic neutrino scattering can serve as a unique probe of nucleons and nuclei. Both forward (e.g., FPF) and fixed-target experiments (e.g., NA64, SHiP/NA67) within PBC can contribute novel kinematic regions and unique sensitivities to these aspects. Novel precision neutrino beams (e.g., SBN) may provide crucial cross-section measurements for other endeavors  such as DUNE and Hyper-Kamiokande. 

Understanding the low-energy behavior of QCD and notably its phase structure remains an important challenge. Using the versatility of the SPS ion beams allows for the exploration of this question in promising kinematic regimes (e.g., DICE/NA60+) and with different ion species (e.g., NA61/SHINE), complementing other facilities worldwide. Spectroscopic information on QCD bound states (e.g., AMBER/NA66) can give complementary information on the low-energy QCD regime. Fixed-target experiments at the LHC supply valuable parton distribution functions for Standard Model measurements as well as BSM searches.

Highly precise measurements of electron muon scattering (MUonE) can shed light on the hadronic contributions to $a_{\mu}$, thereby providing new insights into this currently open puzzle, helping to decide whether it is a Standard Model effect or a sign of new physics.

\subsection*{Concluding strategic remarks}
Recent years have demonstrated that covering the full range of possibilities for fundamental particle physics requires a broad range of experimental strategies, complementing those at colliders. This notably includes dark matter and, connected with this, the search for FIPs, but also a better understanding of QCD. PBC realizes the idea of a diverse experimental programme (at CERN or elsewhere), making full use of the facilities, infrastructure and expertise at CERN. It thereby serves as a demonstrator for the possible role that large laboratories can play in this context. In addition, PBC also shows that large labs can serve as an incubator and facilitator bringing together theory, experiment as well as engineering. Together with the physics benefits, a diverse experimental landscape, including numerous smaller and medium scale experiments,  also serves to improve the health and long-term viability of the community, allowing young researchers to take leading roles within an experiment, garnering experience in all aspects, from planning, construction to data taking and overseeing the whole experiment. Some of those experiments can be performed making use of the existing infrastructures with minimal adaptions, others require further investments in facilities like the FPF or the SBN with timelines that require decisions as early as during the upcoming LS3.

Altogether, supporting a powerful and diverse programme at CERN and similar programmes at other labs is a crucial decision for a successful long-term strategy.

\clearpage

\section{Physics landscape and current open questions targeted by PBC}
\noindent

\subsection{CERN PBC Experimental landscape}

A number of experimental proposals have seen the light and have been supported within the framework of the~\acrshort{pbc} initiative since its launch in 2016. The status of these proposals and their impact on the physics landscape are briefly summarized in this Section.

\subsubsection{NA61/SHINE}
The~\acrfull{na61/shine} is the only~\acrfull{ft} hadron spectrometer operating
at the~\acrshort{cern}~\acrshort{sps}, served by the H2 beam line in the~\acrshort{sps}~\acrfull{na}. It has the unique capability for large-acceptance measurements over a versatile set of beams and targets in the specific regime of collision energy, $\SI{5}{~\GeV}<\sqrt{{s}_\mathrm{NN}}<\SI{17}{\GeV}$.
Its accumulated data set includes $p$+$p$, $p$+C, $\pi$+C, $K$+C, $p$+Pb, Be+Be, C+C, Ar+Sc, Xe+La, and Pb+Pb collisions, each recorded at up to six beam momenta.

The most recent achievements of the experiment include (see~Section~\ref{sec:PSPSPS_Opportunities}):
  the first-ever direct measurement of open charm production in nucleus-nucleus collisions at~\acrshort{sps} energies~\cite{Merzlaya:2024cbt}, 
the observation of a large excess of charged over neutral $K$ meson production in Ar+Sc collisions at $\sqrt{{s}_\mathrm{NN}}=\SI{11.9}{\GeV}$, interpreted as evidence for an unexpectedly large violation of isospin (flavour) symmetry in nucleus-nucleus reactions~\cite{NA61SHINE:2023azp}, 
and the completion of a 2D scan of hadron production in $p$+$p$ and nucleus-nucleus collisions, each at six beam momenta,
 which brings 
 insight into the changeover from confined to deconfined matter as a function of collision energy and colliding system size~\cite{NA61SHINE:2017fne,NA61SHINE:2020czq, NA61SHINE:2021nye}. 
 
Presently, the upgraded~\acrshort{na61/shine} detector
operates a high-statistics Pb+Pb data-taking, bringing the first-ever differential measurement of open charm production in heavy-ion collisions close to the threshold (to be completed in 2026)~\cite{Gazdzicki:2799311}.

At the same time,~\acrshort{na61/shine}
provides a substantial set of reference data to improve the precision of neutrino experiments and new measurements for cosmic-ray physics programs~\cite{NA61SHINE:2022uxp,NA61SHINE:2023bqo,NA61SHINE:2022tiz,NA61SHINE:2024rzv}.

\subsubsection{NA62}
The~\acrlong{na62} was built to measure precisely the branching ratio ${\cal B}(K^+\to\pi^+\nu\bar\nu$), and has recently measured this decay with a 5 sigma significance \cite{press_NA62,NA62:2024pjp}. 
Thanks to its high intensity beam and detector performance (redundant particle-identification capability,  extremely  efficient  veto  system  and  high  resolution  measurements  of  momentum, time, and energy), 
~\acrshort{na62} has also achieved sensitivities to long-lived light mediators in a variety of new-physics scenarios 
by operating in the so-called~\acrfull{bd} mode, where the full primary proton beam is dumped on a collimator downstream of the production target, and a dedicated K12 beamline setup~\cite{Rosenthal:2019qua,CBWG-report-2025} allows to reduce the muon background in the detector.
Two analyses involving \acrfull{hs} particle decays to di-electrons \cite{NA62:2023nhs} and di-muons \cite{NA62:2023qyn} have been published and a search for hadronic decays of~\acrfull{ds} particles has been presented in 2024 \cite{hadrons_NA62}.

\subsubsection{{NA64} ($-e$, $-\mu$, $-h$) -- Phase 1}   

The \acrfull{na64} was designed to conduct a sensitive search for light, sub-GeV~\acrshort{ds} particles that could interact with ordinary matter via light mediators, potentially explaining the origin of the \acrfull{dm} in the universe. These theoretically well-motivated and cosmologically viable scenarios are difficult to probe using traditional~\acrshort{dm} detection methods. \acrshort{na64} effectively combines active target and missing-energy techniques to search for \acrfull{ldm} particles. Using electron and positron beams with energies of~\SIrange[range-units=single,range-phrase=--]{50}{100}{\GeV}, \acrshort{na64} has achieved the best sensitivity to an important class of models based on the hypercharge kinetic mixing of a dark photon mediator. World-leading constraints have been set \cite{NA64:2023wbi}, allowing the experiment to explore the most compelling parameter space for \acrshort{ldm} models, consistent with the observed~\acrshort{dm} relic abundance.
The \acrshort{na64} searches for \acrshort{ldm} provide important inputs for other new physics scenarios, including $^8$Be anomaly \cite{NA64:2018lsq}, \acrshort{alps} \cite{NA64:2020qwq}, inelastic \acrshort{dm} \cite{NA64:2021acr, Mongillo:2023hbs}, $B-L$ \cite{NA64:2022yly}  and $L_\mu-L_\tau\,\,\, Z'$ models \cite{Andreev:2024lps}, and \acrfull{lfv} processes \cite{Ponten:2024grp}. Additionally, the experiment is highly complementary to other approaches, such as underground direct detection, neutrino,~\acrshort{bd}, and high-energy collider experiments. These results offer a crucial physics input and motivation for the current and future experimental research programme at~\acrshort{cern}.
In 2018, \acrshort{na64} proposed to the~\acrshort{cern} \acrfull{spsc} the possibility to conduct a sensitive search for \acrshort{ldm} in a higher mass range using a muon beam \cite{NA64:2024nwj}. A key intermediate goal was to probe the parameter space of the $L_\mu-L_\tau\,\,\, Z'$  model, which addresses the muon $g-2$ anomaly while explaining the~\acrshort{dm} relic density. \acrshort{na64} completed its first muon pilot runs in 2022, nearly ruling out the $g-2$ explanation and setting first constraints on \acrshort{ldm} for dark sectors primarily coupled to the second lepton generation \cite{NA64:2024nwj}. These results demonstrate \acrshort{na64}'s capabilities to run with the M2 \acrshort{na} muon beamline \cite{CBWG-report-2025} and provide a strong motivation to continue this unique research programme in the coming years.
Furthermore, a short 2023 test run with a pion beam \cite{NA64:2024azv}, searching for invisible decays of neutral $\eta$ and $\eta'$ mesons, revealed that \acrshort{na64}’s sensitivity to leptophobic~\acrshort{dm} (predominantly coupled to quarks) could be significantly enhanced with future active-target runs. This opens a new research direction for \acrshort{na64} searches with hadron beams.
Since its approval in 2016, \acrshort{na64} has been pioneering searches for~\acrshort{ldm} across various modes: electron \cite{NA64:2023wbi}, positron \cite{NA64:2023ehh}, muon \cite{NA64:2024nwj}, and hadron \cite{NA64:2024azv}. The experiment has successfully met its primary objectives, as outlined in the input to the 2020~\acrfull{eppsu}, producing results which demonstrate its ability to operate in a near-background-free environment. These accomplishments have been recognized by the \acrshort{pbc} and \acrfull{fips} communities as original, complementary to other ongoing and planned projects and worthy of continued exploration.

\subsubsection{{NA66/AMBER} -- Phase 1}
Selected topics of the physics programme of the \acrfull{amber}  were included in~\cite{PBC_summary_2020} and~\cite{ESPPU2020}.
The~\acrshort{amber}~Phase-1 proposal was approved by the~\acrfull{cernrb} in December 2020.
This \acrfull{ft} experimental setup is located at the~\acrshort{na} M2 beam line \cite{CBWG-report-2025}, which is capable of delivering beams of muons, pions, kaons, and protons as well as their anti-particles with beam momenta up to~\SI{250}{\GeV}.
In spite of the extremely important discovery of Higgs boson, the origin of~99~\% of the mass of all visible universe is still unclear. The Higgs mechanism does not explain why the proton is so heavy and the pion is so light. The underlying physics process of the emergence of the hadron mass is likely tightly related to the unique dynamical mass generation mechanism in continuum~\acrfull{qcd}.
 A determination of the valence-quark~\acrfull{pdf} of the pion and kaon---through~\acrfull{dy} and J/$\psi$ di-muon production, as well as direct-photon production---and high precision hadron spectroscopy measurements would provide   sensitivity to the  mechanisms responsible for the emergence of hadron mass in~\acrshort{qcd}. The present pre-\acrshort{ls3} measurements include that of the proton's charge radius (from its form factor in $\mu p$ elastic scattering) and of antiproton production cross sections in proton-\(^4\)He/H\(_{2}\)/D\(_{2}\) collisions. A determination of the valence-quark~\acrshort{pdf} of the pion and kaon, through~\acrshort{dy} and J/$\psi$ di-muon production will follow as one of three Phase-1 approved measurements.

\subsubsection{{Fixed-target} experiments at the LHC}\label{sec:SMOG2}

The ``forward'' detector configuration of the~\acrfull{lhcb} at the~\acrshort{lhc}~\acrfull{ip}8, originally designed to perform detailed studies of beauty hadron decays (\acrfull{cpv} and rare final states)~\cite{doi:10.1142/S0217751X15300227}, was exploited to extend the physics programme to a broad range of~\acrshort{ft} measurements. The programme started by using the~\acrfull{smog}, a device developed to determine the~\acrshort{lhc} colliding beams luminosity by beam-gas imaging~\cite{LHCbPrecisionLumi_2014}. A gas target was obtained by injecting small amounts of light noble gases just upstream of the~\acrfull{velo} vacuum chamber. Pioneering~\acrshort{lhc}~\acrshort{ft} measurements were performed with He, Ne and Ar targets, obtaining unique results such as production cross sections of antiprotons in $p$-He collisions~\cite{PhysRevLett.121.222001,EPJC-83-543-2023} and of charm hadrons in $p$-He/Ne collisions~\cite{PhysRevLett.122.132002,EPJC-83-541-2023,EPJC-83-625-2023}.
The performance of the gaseous target was enhanced by installing a~\SI{1}{\cm} diameter and \SI{20}{\cm} long open-ended storage cell around the beam path (\acrshort{smog} upgrade, called~\acrshort{smog2}), just upstream of the~\acrshort{velo}, and by injecting the gas directly into the middle of the cell via a capillary. The \acrshort{smog} upgrade, in addition to increasing the target areal density by up to two orders of magnitude (depending on gas type), also included a more sophisticated gas-feed system which allowed to rapidly switch between injected gas types and which brought the possibility to deliver more gas target species (in particular  H$_2$, D$_2$, O$_2$)~\cite{garcia2024}.  
These improvements were facilitated by the studies performed in the~\acrshort{lhcft}~\acrfull{wg}, where various challenges, such as aperture limitations, wake-field effects, dynamic vacuum phenomena and beam losses from beam-gas interactions, were addressed and adequate solutions were devised (such as an acceptable tube radius, appropriate wake-field suppression contacts, amorphous carbon coating, etc.)~\cite{BoscoloMeneguolo:2651289,Parragh:2836903}.
Most importantly, thanks to the clear separation between the~\acrshort{ip}8 luminous region and the storage cell, and thanks to the~\acrshort{lhcb} upgraded detector capabilities~\cite{Aaij_2024}, \acrshort{ft} data and collider data are now acquired {\em simultaneously} at~\acrshort{lhcb}.
A unique and diverse~\acrshort{ft} programme is now under way at~\acrshort{lhcb} which covers measurements of particles production spectra and cross sections with a variety of targets (mentioned above) and different beams ($p$, Pb, O) at several~\acrfull{com} energies (from~\SIrange[range-units=single]{72}{115}{\GeV}) with a detector that covers a pseudorapidity range from about 1.8 to 5.5. These measurements are particularly important for the understanding of cosmic ray fluxes, cosmic ray showers, the~\acrfull{qgp} and nuclear structure in general. 

\subsubsection{FASER}
\label{sec:faser}
The~\acrfull{faser}~\cite{Feng:2017uoz} was designed to search for light, weakly-coupled new particles, produced in the~\acrshort{lhc} collisions, and to study high-energy collider neutrinos of all flavours. The experiment is located in the junction between the~\acrfull{ti}12 service tunnel and the~\acrshort{lhc}, aligned with the~\acrfull{atlas} collision-axis line of sight, which maximises the neutrino flux and energy, and the potential flux of light new particles. 

After initial studies within~\acrshort{pbc}, the collaboration submitted a~\acrfull{loi}~\cite{FASER:2018ceo} and then a technical proposal~\cite{FASER:2018bac} to the~\acrfull{lhcc} in September/November 2018, and was approved by the \acrshort{cern} research board in March 2019. The detector systems were designed, constructed and underwent standalone commissioning on the surface in 2019 and 2020, before being installed into the~\acrshort{ti}12 tunnel in the~\acrshort{lhc} complex in March 2021. During 2020 the experimental area was prepared by~\acrshort{cern} technical teams, with a small trench excavated to allow the detector to be positioned on the collision-axis line of sight. The needed electrical, transport, cooling and communication infrastructure was installed. After the installation the experiment underwent a sustained period of in situ commissioning with cosmic-ray and single-beam data. Physics operation commenced in July 2022, at the start of~\acrshort{lhc}~Run~3, with very smooth detector operations and overall above 97\% of the delivered luminosity has been recorded (with a total dataset size of nearly~\SI{200}{\femto\barn^{-1}} recorded by the end of 2024).

The~\acrshort{faser} detector~\cite{FASER:2022hcn} has two main parts: an electronic detector composed of a veto scintillator system, a tracking spectrometer, and an electromagnetic calorimeter, designed for long-lived particle searches; and a dedicated~\SI{1}{\tonne} tungsten/emulsion detector called~\acrfull{fasernu} designed for neutrino measurements. The detector performance is very good, with all of the main design requirements achieved.

Based on the excellent detector operations and performance the experiment has been able to deliver impactful physics results in a timely fashion, with first search and neutrino results using the 2022 data released in time for the 2023 winter conferences. These were the search for dark photons, with unique sensitivity in an unexplored region of parameter space motivated by~\acrshort{dm}~\cite{FASER:2023tle} and the first observation of muon neutrinos at a collider~\cite{FASER:2023zcr}, with 153 neutrino candidate events observed. Further results have been released in 2024, with a search for~\acrfull{alps}~\cite{FASER:2024bbl}, the first measurements of the interaction cross section at~\SI{}{\TeV} energies for electron and muon neutrinos~\cite{FASER:2024hoe} using the \acrshort{fasernu} emulsion detector, and the first differential cross section measurement of neutrinos and antineutrinos~\cite{FASER:2024ref} at~\SI{}{\TeV} energies. 

The \acrshort{faser} collaboration is considering an upgrade of its neutrino detector covering the rapidity range $\eta>7$, hence allowing to probe both the on-axis region, currently probed by \acrshort{faser}, and the off-axis region, currently probed by \acrfull{sndlhc}, in one detector~\cite{FASER-Run4-ESPP}.

\subsubsection{SND@LHC}

\acrshort{sndlhc}~\cite{Ahdida:2750060,SNDLHC:2022ihg} is a compact experiment exploiting the high flux of energetic neutrinos of all flavours from the~\acrshort{lhc}. It covers the pseudo-rapidity range of $7.2<\eta<8.4$, where a large fraction of neutrinos originates from charmed-hadrons decays. 
The experiment is located 480\,m downstream of~\acrshort{ip}1 in the unused~\acrshort{ti}18 tunnel. The detector is composed of a hybrid system based on a~\SI{800}{\kilo\gram} target mass of tungsten plates, interleaved with emulsions and electronic trackers, followed downstream by an hadronic calorimeter and a muon identification system. The configuration allows efficiently distinguishing between all three neutrino flavours, thus probing physics of heavy-flavour production at the~\acrshort{lhc} in a region inaccessible to other experiments and enabling \acrfull{lfu} tests in the neutrino sector. The detector concept is also well suited to search for~\acrshort{fips} via signatures of scattering in the detector target. 

The data-taking, which started at the beginning of the~\acrshort{lhc} Run 3, was immediately successful with the collection of an integrated luminosity of~\SI{36.8}{\femto\barn^{-1}} in 2022. 
The data sample from 2022 had a total of $8.4 \times  10^9$ incoming muon tracks \cite{SNDLHC:2023mib}. A rejection power at the level of $10^{-12}$ was achieved by combining the veto system and the~\acrfull{scifi} stations of the two most upstream walls of the target. The only background left was due to neutral hadrons produced by muon interactions in the upstream rock which would in turn generate hadronic jets with a reconstructed muon track in the final state: this background was estimated to 0.086 expected events. 
Given the 8 candidate events found in the data, an observation of collider neutrinos was established with a significance of about 7$\sigma$ \cite{SNDLHC:2023pun}. By adding the data taken in 2023 \cite{SNDLHC:2024bzp}, an integrated luminosity of~\SI{68.6}{\femto\barn^{-1}} was achieved and the search for neutrino interactions without a muon in the final state was carried out. This sample consists essentially of \acrfull{nc} neutrino interactions and \acrfull{cc} electron neutrinos. Electron neutrinos produce typically an electron in the final state with more than~\SI{300}{\GeV} energy on average, with a large energy density  deposited in the neutrino target. The selection could therefore be tuned to enrich the sample of electron neutrinos by using the energy density. The data accumulated in a test beam carried out at the~\acrshort{sps} in 2023 for the calorimeter calibration were used to validate the Monte Carlo simulation. A control region outside of the signal selection cuts was defined which showed a good agreement between data and simulation. As a result of the selection, 9 candidate events were found with an expected background of 0.3 events, dominated by muon neutrinos where the muon is not identified. This has led to the observation of neutrino interactions without a muon in the final state with a significance above 6$\sigma$ and to the evidence for electron neutrinos with 3.7$\sigma$ \cite{SNDLHC:2024qqb}.

During 2022/2023 the Veto system did not completely cover the target area. In order to solve this problem and further reduce the veto inefficiency, a small trench was dug, the veto system was lowered and a third Veto plane was
installed during the 2023-2024 \acrfull{yets} \cite{SNDLHC:2025ujo}. The inefficiency was reduced to $(8.7 \pm 3.5) \times 10^{-9}$ on the full target area in 2024 when the experiment recorded a luminosity of about 120 fb$^{-1}$. The analysis of 2024 data is ongoing.

\subsection{Theory and phenomenology support and interface}

\acrshort{pbc} provides both motivation and a forum for direct interactions between theory, phenomenology and experiment. This exchange enhances the physics case of the individual experiments, improving both the breadth and depth of the physics reach by working with the community to better understand and quantify existing signatures and to explore interesting new signatures suggested by theoretical considerations. At the same time these discussions also facilitate a comparison of different experimental capabilities and of the resulting sensitivities.

A crucial result of this effort during the first phase of~\acrshort{pbc} was the development of a set of benchmark cases~\cite{Beacham:2019nyx} for new physics based on theoretically well-motivated models that can be viewed as simplifications of several broad classes of fundamental theories of ~\acrfull{bsm} physics. By providing a detailed phenomenological description and a solid calculational foundation for these models, it became possible to perform a consistent comparison of sensitivities across different experiments within~\acrshort{pbc} and across the world-wide landscape. Indeed, the benchmark cases and resulting discussions proved to have significant impact beyond~\acrshort{pbc}.

The original~\acrshort{pbc} benchmark models centered around the idea of~\acrfull{fips} with masses at or below the GeV scale, which can be targeted by high-intensity and high-precision experiments. This focus led to the establishment of a dedicated~\acrshort{pbc} working group, the~\acrfull{fpc}, which serves as an interface between theory/phenomenology community and the experimental efforts within and beyond~\acrshort{pbc}. A series of dedicated workshops~\cite{Agrawal:2021dbo,Antel:2023hkf} as well a school at Les Houches~\cite{FIPschool} strengthened this collaboration.

The~\acrshort{fpc} continually develops and refines the benchmark models and maintains an up-to-date collection of sensitivity projections, directly interacting with the experiments and the~\acrshort{bsm} working group. Crucially, this activity provided input for the discussion on the future use of the~\acrfull{ecn3}~\cite{Ahdida:2867743} by strengthening and sharpening the physics cases of the proposals for the~\acrfull{hike}~\cite{HIKE_PROPOSAL_2023},~\acrfull{shadows}~\cite{SHADOWS_PROPOSAL_2023} and~\acrfull{ship}~\cite{Albanese:2878604} experiments.

Interfacing with theory is also crucial for the~\acrshort{qcd}-oriented experiments. This is done directly in the~\acrshort{qcd} working group. In the first~\acrshort{pbc} phase~\cite{QCDWorkingGroup:2019dyv} this helped sharpen the targets of experiments in the area of~\acrshort{qcd} phase transition physics, but also discussed the sensitivity of experiments such as~\acrfull{muone} to the hadronic vacuum polarization contribution to $(g-2)_{\mu}$ with its potential impact on searching new physics via this observable. 

Currently the~\acrshort{qcd} working group, and therefore also the phenomenological discussion within it, is focused on the opportunities presented by ion experiments in the~\acrshort{na}. But efforts to support other experiments (e.g., \acrshort{lhcft},~\acrshort{amber}) continue to play an important role. Notably this also includes the exploitation of neutrinos that can be detected in~experiments such as~\acrshort{faser}, ~\acrshort{sndlhc}, and~\acrshort{ship}.

\subsection{Worldwide landscape}

The particle physics initiatives outlined by the \acrshort{pbc} contribute to a broader international research landscape. This Section briefly introduces a number of relevant global efforts, highlighting those with closely related scientific objectives and significant evolution since the 2020 \acrshort{eppsu} 
(see~\cite{PBC_summary_2020,Beacham:2019nyx,QCDWorkingGroup:2019dyv} for the relevant \acrshort{pbc} reports).

Among the various projects, the \acrshort{qcd} experimental landscape is dominated by large experiments that evolve on a relatively large time scale and most of the world-wide experiments presented in Refs.~\cite{PBC_summary_2020,QCDWorkingGroup:2019dyv} are still active or foreseen. The time line of the future heavy-ion experiments \acrshort{cbm} (\acrshort{fair}) and CEE+ (HIAF) has been kept and they start taking data in the next 2 years. 

The next-generation lepton-hadron machine, the \acrfull{eic} \cite{AbdulKhalek:2021gbh}, is to be built at \acrshort{bnl} in the US, to start physics collisions in the first half of the 2030s. Key performance aspects include \cite{EIC-ORG-PLN-010}: 
    \begin{itemize} 
        \item center-of-mass energy range: 29 to 140~GeV;
        \item  ion beams from deuterons to the heaviest stable nuclei (e.g., Au, Pb, U);
        \item  high collision luminosity from $10^{33}$ to $10^{34} \text{cm}^{-2}~\text{s}^{-1}$ for electron-proton collisions;
        \item highly polarized (70\%) electron, proton, and light-ion beams;
        \item an interaction region and integrated detector capable of nearly 100\% kinematic coverage;
        \item the capability of incorporating a second such interaction region as needed;
    \end{itemize}
    allowing for a wide \acrshort{qcd} physics programme, ranging from precision multi-dimensional imaging of nucleons with leptonic probes, the spin structure of nucleons and novel spin effects in \acrshort{qcd}, exploration of gluon saturation at low parton momentum fractions, to modification of parton structure in the nuclear environment.

\acrfull{jlab} has completed its luminosity-frontier 12 GeV energy update, enabling simultaneous delivery of highly (up to 90\%) polarized electrons of up to 11 GeV to experiments conducted in Hall A, B, and C, and up to 12 GeV into Hall D, at luminosities of up to $10^{35} \text{cm}^{-2}~\text{s}^{-1}$ and potentially up to $10^{39} \text{cm}^{-2}~\text{s}^{-1}$ with the future SoLID fixed-target experiment.
The experimental focus is on precision \acrshort{qcd} studies---complementary to those at the future \acrshort{eic} in a different kinematic regime (e.g., valence-quark region)---and low-energy tests of the \acrshort{sm} and \acrshort{bsm} physics. Especially relevant in the context of this document is the approved \acrfull{bdx}~\cite{BDX:2016akw} searching for \acrshort{ldm}. It is scheduled to run from 2026-2029 in a newly constructed facility (to be built) in front of the Hall-A beam dump, which will be served by a 65~$\mu$A electron beam of 11~GeV.

Several upgrades of the complex are envisioned: A positron source could feed an unpolarized positron beam of order 1~$\mu$A, less so for polarized positrons (of order 60\% polarization), earliest in the first half of the 2030s \cite{Afanasev:2019xmr}. An energy upgrade to 22 GeV beam energy could be realized with most of the existing accelerator installation for running as early as the second half of the 2030s, with hardly any impact on polarization and beam current as compared to the present configuration and with a broad physics program~\cite{Accardi:2023chb}.

An exciting new direction are $\eta$/$\eta'$ factories, which aim to exploit the special role of these particles in the \acrshort{sm} to search for new physics in their rare decays. A first such experiment, JEF at \acrshort{jlab}~\cite{Somov:2024iB}, is currently running and aims to extend the physics potential of the GlueX detector~\cite{GlueX:2020idb}.
The proposed REDTOP experiment~\cite{Gatto:2024Ti} at the European Spallation Source plans on collecting more than $10^{13} \eta$/yr ($10^{11} \eta'$/year). The primary target is to search for new particles with mass below 950 MeV, such as dark photons or \acrshort{alps}.

The \acrfull{ldmx} is a proposed beam-dump experiment at \acrshort{slac} using the 8 GeV $e^-$ beam from the \acrshort{linac} diverted to end station A~\cite{Akesson:2022vza}. The experiment aims to fully reconstruct the energy deposition of each incident electron using a high-granularity electromagnetic calorimeter and a hermetic hadronic calorimeter. \acrshort{ldmx} then reconstructs the missing energy and transverse momentum to achieve (nearly) zero background with $10^{14}$ electrons on target. The primary target of \acrshort{ldmx} are invisibly decaying dark photons, but the experiment will also be sensitive to visible (displaced) decays. A first beam test with a prototype detector is planned for late 2025 and a detailed design report is currently in preparation. Another beam-dump experiment proposed in the US is {DarkQuest}~\cite{Apyan:2022tsd}, which plans to upgrade the SpinQuest experiment, a fixed-target experiment using a high-intensity 120 GeV proton beam at the \acrshort{fnal} Main Injector. Using an electromagnetic calorimeter, {DarkQuest} aims to search for long-lived particles decaying to electron-positron pairs. A suitable detector has been identified and studied with a test beam in 2024~\cite{Apyan:2025ldk}.

 The \acrfull{alpsii} experiment~\cite{Bahre:2013ywa} also searches for axions and axion-like particles. It is based on the “light-shining-through-a-wall” principle and has been set up over the past years in a currently unused part of the tunnel of \acrshort{desy}’s former \acrfull{hera}  around the \acrshort{hera} North Hall. The 250 m long experiment combines a string of 24 modified superconducting \acrshort{hera} dipole magnets with two 124 m long high-precision optical cavities~\cite{Kozlowski:2024jzm,Ortiz:2020tgs} and extremely low-power light detection~\cite{Hallal:2020ibe,RubieraGimeno:2022pjx}.

\acrshort{alpsii} has the goal to improve the sensitivity on the axion-photon coupling by a factor of 1,000 compared to its predecessors. For the first time, \acrshort{alpsii} will reach a parameter region in a model-independent fashion where the axion could also explain the dark matter in our universe and astrophysical anomalies. With its first run – concluded in May 2024 – \acrshort{alpsii} has already reached an axion-photon coupling sensitivity roughly 30 times larger than earlier experiments (publication in preparation). The target sensitivity shall be reached in 2027.

The {\acrfull{iaxo}} is a next-generation axion helioscope designed to search for solar axions with unprecedented sensitivity. In particular, it will explore \acrshort{qcd} axion models in the mass range from meV to eV, covering scenarios motivated by astrophysical observations without relying on the dark matter paradigm. \acrshort{iaxo} will build upon the two-decade experience gained with CAST, the detailed studies for \acrshort{babyiaxo}, which is expected to start construction at \acrshort{desy} soon, as well as new technologies. Several studies in recent years have demonstrated that \acrshort{iaxo} has the potential to probe a wide range of new physics beyond solar axions, including dark photons, chameleons, gravitational waves, and axions from nearby supernovae.

The {\acrfull{luxe}}~\cite{Abramowicz:2019gvx,Abramowicz:2021zja,Bai:2021gbm} at the European XFEL (EuXFEL) facility aims to explore collisions between ultra-relativistic electrons (or photons) and intense laser pulses to study Strong-Field Quantum Electrodynamics (SFQED) and searches for new particles that couple primarily to photons.
The installation timeline depends on EuXFEL's operational schedule, particularly the T20 electron beam extraction line, requiring a facility shutdown of approximately 12 weeks. Following approval, \acrshort{luxe} could start operations  around 2030.

There is currently no proposal for a non-collider experiment dedicated to measuring charged or neutral kaon decays within the \acrshort{pbc} program. The \acrshort{cern} fixed-target kaon physics program is presently centered on the \acrlong{na62}, which recently reported the first observation of the rare $K^+ \to \pi^+ \nu \bar{\nu}$ decay~\cite{NA62:2024pjp}. The measured branching ratio is consistent with the \acrshort{sm} prediction within 1.7$\sigma$, though its central value is approximately 50\% higher than the \acrshort{sm} expectation. This result is particularly intriguing in light of recent measurements by the \acrshort{belleii} collaboration, which reported evidence for the $B^+ \to K^+ \nu\bar{\nu}$ decay with a central value $2.7\sigma$ above the \acrshort{sm} expectation~\cite{Belle-II:2023esi}. These flavor-changing neutral current transitions, with a neutrino-antineutrino pair in the final state, are theoretically well understood; thus, significant deviations from the \acrshort{sm} predictions could indicate new physics.

While the \acrshort{hike} experiment~\cite{HIKE_PROPOSAL_2023}, proposed as a follow-up to \acrshort{na62}, was not approved by \acrshort{cern}, kaon physics continues with the study of neutral $K_L$ decays at the KOTO experiment at \acrshort{jparc}. Although KOTO continues to improve its sensitivity, it will not reach the level required to observe the \acrshort{sm}-predicted $K_L \to \pi^0 \nu \bar{\nu}$ decay. A letter of intent has recently been submitted for a successor experiment, KOTO II, which aims to achieve the necessary sensitivity to observe this decay~\cite{KOTO:2025gvq}.

\acrshort{belleii}, a next-generation flavor physics experiment at KEK's SuperKEKB collider in Tsu\-ku\-ba, Japan, has collected electron--positron ($e^+e^-$) collision data since 2018. With peak luminosities exceeding $5.0 \times 10^{34}\,\mathrm{cm}^{-2}\mathrm{s}^{-1}$ and strong European participation, it focuses on high-precision measurements in the fields of \acrshort{qcd} and flavor physics, as well as in searches for New Physics. Data-taking resumed in 2024, aiming for an integrated luminosity of $10\,\mathrm{ab}^{-1}$ by 2032, followed by a detector upgrade to reach a total of $50\,\mathrm{ab}^{-1}$~\cite{Aihara:2024zds}.

Since the last European Strategy Update, neutrino oscillations have been confirmed as a central particle physics topic with a vivid competition. The main players in the field are \acrshort{fnal} in the \acrshort{us}, with the current NOvA experiment and future~\acrfull{dune}, \acrshort{jparc} in Japan with the current T2K and future~\acrfull{hk} experiments, and China with the JUNO reactor project. The ongoing NOvA and T2K programs are currently favoring the neutrino normal mass ordering and a non-0 \acrshort{cpv}. In the coming years, their antineutrino (NOvA) and total (T2K) data samples are expected to be doubled, with the goal to approach the   $3\sigma$ level for the determination of the mass ordering and \acrshort{cpv} detection.~\acrshort{dune} and~\acrshort{hk} are now both officially approved and in full construction swing, with start of data-taking expected before 2030.~\acrshort{dune} should quickly establish the mass ordering at $5\sigma$ level within 1 to 3 years of data-taking. On the longer term, \acrshort{cpv} is expected to be established at more than $3\sigma$ within 75\% of the possible ranges, with a measurement better than $20\degree$ in 10 years of data-taking for favorable values. The construction of the JUNO experiment in China is now close to completion and start of data-taking imminent. The mass hierarchy is expected to be established within 6 years at the $3\sigma$ level.

\section{Evolution of the CERN accelerator complex and post-LS3 outlook}\label{sec:acceleratorevolution}
\subsection{CERN Accelerator complex after LS2, planned upgrades during LS3 and future perspectives}

\subsubsection{LHC Injectors}

\begin{figure}[h!]
    \centering
    \includegraphics[width=1\linewidth]{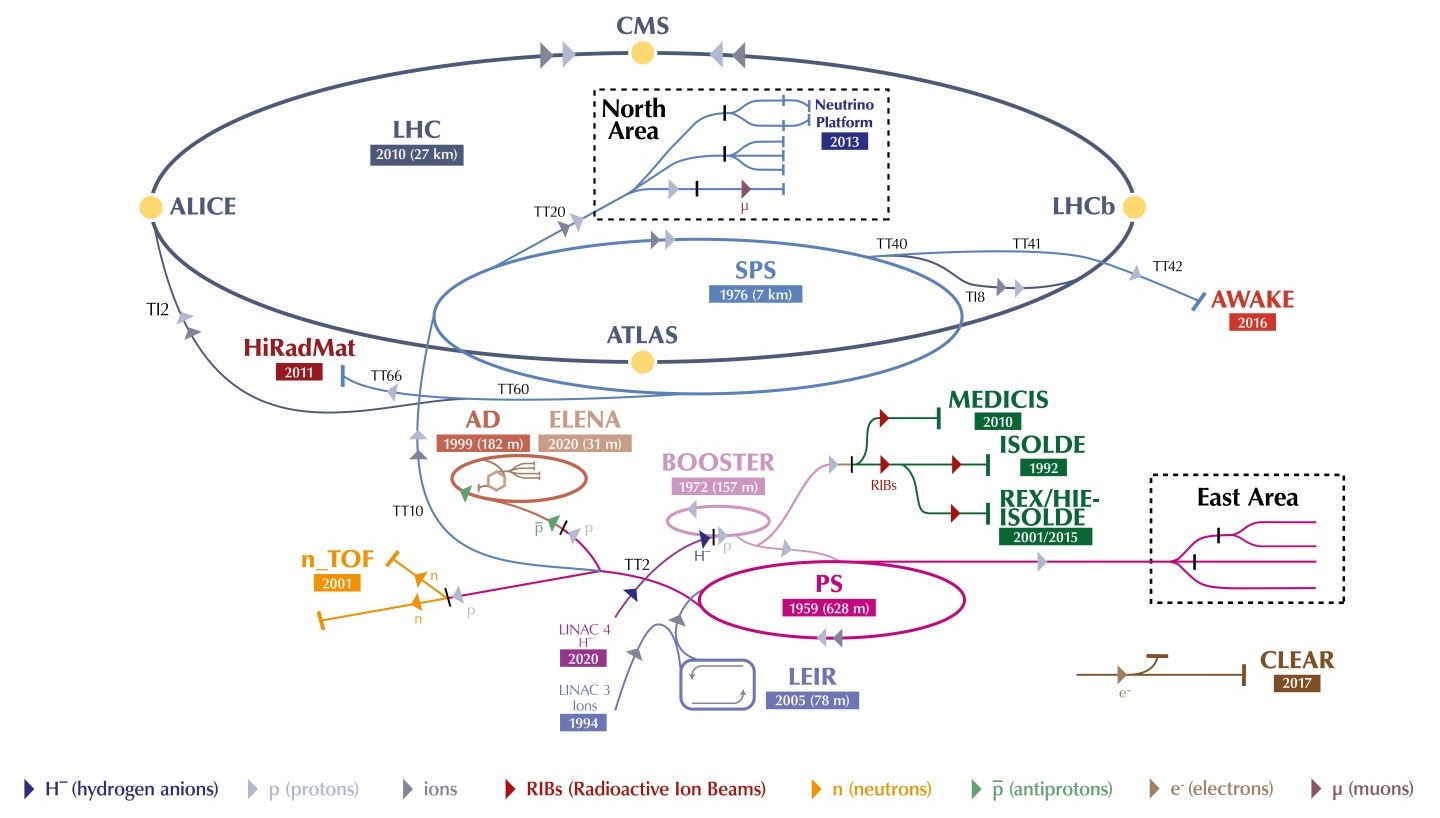}
    \caption{Overview of the LHC injectors chains, both for protons and heavy ions after~\acrshort{ls2}. Courtesy: \acrshort{cern}.}
    \label{fig:AccComplex}
\end{figure}

The~\acrshort{cern} injector complex underwent an extensive upgrade during the~\acrfull{ls2}, which lasted about two years (2019-20) and was driven by the~\acrfull{liu} Project~\cite{LIU1,LIU2} (Figure~\ref{fig:AccComplex}). The project included several baseline items to enable the~\acrshort{lhc} injector chain to deliver the high-brightness proton beams required by the~\acrfull{hllhc} to reach a leveled luminosity of 5$\times10^{34}$ cm$^{-2}$s$^{-1}$. Here is a list of the main ones:
\begin{itemize} 
\item The connection of the~\acrfull{psb} to the new 160~MeV H\(^-\) linear accelerator,~\acrfull{linac4}, and the installation of the corresponding charge-exchange injection scheme.
\item The increase of the extraction energy of the~\acrshort{psb} to~\SI{2}{\GeV}.
\item The upgrade of the injection region and~\acrfull{rf} systems of the~\acrfull{ps} to allow~\SI{2}{\GeV} injection and mitigate longitudinal instabilities.
\item The power and~\acrfull{llrf} systems upgrade of the main~\acrshort{sps}~\acrshort{rf} system (\SI{200}{\MHz}), a new~\acrshort{sps} beam dump and new protection devices in the transfer lines to~\acrshort{lhc}.
\end{itemize}

\noindent In the ion chain, the~\acrfull{leir} was the object of an intense performance improvement campaign, which enabled it to double the amount of ions produced per cycle. In the~\acrshort{sps}, the upgrade of the~\SI{200}{\MHz}~\acrshort{rf} system allowed performing momentum slip-stacking, thus reducing the bunch spacing from 100 to~\SI{50}{\ns} and doubling the number of bunches that can be packed in the~\acrshort{lhc}, another~\acrshort{hllhc} goal. During the present run (Run~3), the injector chain has proven to be able to achieve the~\acrshort{liu} targets in terms of maximum bunch population~$N$ and minimum normalized emittance~$\epsilon^n_{x,y}$, and sometimes even surpass them as shown in Table~\ref{tab:achieve_target}. 
\begin{table}[h!]
\centering
\begin{tabular}{lcccc}
\hline\hline
& $N$ [10$^{11}$ p/b] & $\epsilon^n_{x,y}$ [$\mu$m] & Bunch spacing [ns] & Train structure\\
\hline
\acrshort{liu} target & 2.3 & 2.1 & 25 & 4 $\times$ 72 bunches\\ 
Achieved & 2.3 & 1.7 & 25 & 4 $\times$ 48 bunches\\ 
\hline
& $N$ [10$^{8}$ ions/b] & $\epsilon^n_{x,y}$ [$\mu$m] & Bunch spacing [ns] & Train structure\\
\hline
\acrshort{liu} target & 2.0 & 1.5 & 50 & 7 $\times$ 8 bunches\\
Achieved & 2.8 & 1.8 & 50 & 7 $\times$ 8 bunches\\  
\hline\hline
\end{tabular}
\caption{Beam parameters at~\acrshort{sps} extraction for protons and Pb ions,~\acrshort{liu} target and best performance achieved so far in Run~3.}
\label{tab:achieve_target}
\end{table}

Beams used in the injectors for~\acrshort{ft}~physics have also largely benefited from the~\acrshort{liu} upgrades. Some of the new benefits have been put to fruition by improving the quality or increasing the intensity of the beams delivered to the experiments. For example, the production scheme of the beams for the~\acrshort{na} has been improved in the~\acrshort{ps} thanks to the commissioning of the barrier bucket at extraction (to be used together with the~\acrfull{mte}, already operational for many years), made possible by using the same broad band cavity
installed in the~\acrshort{ps} ring as a part of the feedback system against
longitudinal instabilities. The losses at~\acrshort{ps} extraction
of this beam have been thus significantly reduced, such that it will be possible to remove the element protecting the extraction septum (so-called \textit{dummy septum}), improving the aperture for all beams~\cite{vadai2022implementationsynchronisedpsspstransfer}. Another example is the production of 5 bunches instead of 4 in the~\acrshort{psb} (with one ring producing two bunches instead of one) for the~\acrfull{ad}~beam, made possible by the lower transverse emittances achieved with the~\SI{160}{\MeV} injection. This has led to 20~\% larger intensity per pulse on the~\acrshort{ad} target and consequently a higher number of antiprotons into~\acrshort{ad} and the~\acrfull{elena}~\cite{ADbeams}.

Other benefits have been studied during~\acrfull{md} sessions and are kept in store when potential requests for higher intensities might come after~\acrshort{ls3}. For instance, before~\acrshort{ls2} the~\acrshort{psb} accelerated a maximum of
$10^{13}$~p/ring, but with large losses during injection and~\acrshort{rf}
capture (up to 40\%). After the connection to~\acrshort{linac4}, the~\acrshort{psb} has already unveiled its potential to produce 60\% higher
intensities (i.e., up to $1.6\times 10^{13}$~\acrfull{ppp}/ring, a value so far successfully demonstrated in two of the four rings) with few percent losses only occurring in the very first phase of the acceleration~\cite{Asvesta:2900904}. Its routine operation has however stayed at $0.8\times 10^{13}$~p/ring, as this value currently fulfills the needs of the~\acrfull{isolde} users with 30-40\% cycles in the supercycle.
Table~\ref{tab:achieve_FT} summarizes the currently achieved performances for the main~\acrshort{ft} beams.

\begin{table}[h!]
\centering
\begin{tabular}{lcccc}
\hline\hline
& Intensity [10$^{10}$~\acrshort{ppp}] & $\epsilon^n_{x},\epsilon^n_{y}$ [\SI{}{\micro\m}] & Bunch length [\SI{}{\ns}] & \# bunches\\
\hline
\acrshort{isolde}  & 5000 & 8, 8 & 240 & 4\\ 
\acrshort{ntof} parasitic  & 800 & 9, 7 & 44 & 1\\ 
\acrshort{ntof} dedicated & 1020 & 14, 11 & 28/44 & 1\\
\acrshort{ad} & 2000 & 9, 6 & 35 & 5 \\
\acrshort{ea} & 80 & 1.5, 1.3& - & unbunched\\
\acrshort{na} & 4200 & 7, 4 & - & unbunched\\
\acrshort{awake} & 30 & 2, 2 & 0.5 - 0.7& 1\\
\hline\hline
\end{tabular}
\caption{Beam parameters at extraction for protons, best performance achieved so far in Run~3. Here, \acrshort{ppp} =~\acrlong{ppp}; ~\acrshort{ntof}~=~\acrlong{ntof};~\acrshort{awake}~=~\acrlong{awake}.}
\label{tab:achieve_FT}
\end{table}

From Table \ref{tab:achieve_FT} we can see that dedicated \acrshort{ntof} beams have been produced with intensities slightly above 1000~$\times$10$^{10}$~\acrshort{ppp}, however around this value transverse stability becomes marginal upon transition crossing. This is the reason why the highest operational value is assumed to be 920~$\times$10$^{10}$~\acrshort{ppp}. The vertical instability limiting the intensity on the \acrshort{ntof} beam has been widely studied in the past and found to be a~\acrfull{tmci} mainly driven by the impedance of the \acrshort{ps} kickers (e.g.,~\cite{PhysRevAccelBeams.21.120101}). To efficiently mitigate it, an impedance reduction campaign of these kickers would be needed. 

The beam to the~\acrshort{na} was operated routinely with an intensity of 4200~$\times$~10$^{10}$~\acrshort{ppp} extracted from the~\acrshort{sps} over an extended period of time in 2022, but this led to important losses in the region of the extraction electrostatic septum and at the splitter magnets in the transfer lines to the experiments, which resulted in high levels of radiation in these areas. Crystal shadowing and channeling measures have been meanwhile widely tested and optimized, and are planned to be put in place for the future operation at this intensity level~\cite{Velotti:2886012,ArrutiaSota:2724487}. Higher intensities on this beam are being studied, with a high brightness version developed at the~\acrshort{psb} and up to 3300~$\times$~10$^{10}$~\acrshort{ppp} stably accelerated in the~\acrshort{ps} and extractable with \acrshort{mte} and barrier bucket~\cite{Cuvelier:2876316}. In the~\acrshort{sps} detailed studies of working point optimization along the cycle are ongoing before pushing the intensity to higher than operational values in order to check possible stability limitations or aperture bottlenecks. 

Finally, it should be stressed that, although it was mentioned above that 1600~$\times$10$^{10}$~\acrshort{ppp} were successfully accelerated in two of the four \acrshort{psb} rings and can thus be seen as a peak performance to be aimed at for future operation,  the intensity value quoted for the \acrshort{isolde} beams in Table \ref{tab:achieve_FT} considers 1250~$\times$10$^{10}$~\acrshort{ppp}/ring, a value stably and reproducibly achieved in autumn 2024 and used for yield measurements as a function of intensity on the \acrfull{gps} target.

No other major upgrades are foreseen in the injectors during the~\acrshort{ls3} and the~\acrshort{lhc} ion experiments will operate with lead ions in Run~4. However, a number of proposals contemplating beam with ions lighter than lead, in addition to lead ones, have been made~\cite{PBC_ionreport}. The~\acrshort{lhc} experiments, and in particular~\acrshort{alice} with an upgraded detector (\acrshort{alice3})~\cite{aliceLOI:2022} propose to maximize the nucleon-nucleon luminosity with ions of intermediate mass. The discovery at the~\acrshort{lhc} of the~\acrshort{qgp} properties in small systems, like proton-proton and proton-ion collisions, is motivating the physics community to study in more detail light ion collisions at the~\acrshort{lhc}~\cite{indico:1436085, Wiedemann:2024}. The~\acrshort{na61/shine}~\cite{NA61/SHINEaddendum} collaboration has proposed the continuation of their ion physics programme with beams of different species (lead and light ions) and energies and the~\acrfull{na60+} has proposed experiments with lead beams at different energies~\cite{NA60+LoI,PBC_ionreport} after~\acrshort{ls3}. ~\acrfull{hearts}, a facility to study the effects of radiation to electronics, is being proposed at~\acrshort{cern}~\cite{hearts}. It aims at delivering different ion species during the same day with switching times of maximum 15 minutes, which is not possible with the current ion complex. One of the reasons being the availability of a single ion source. 
Extensive studies have been conducted~\cite{light_ions_feasibility_ats_note,oxy4lhc_ats_note,Slupecki:2929494} showing the feasibility of producing and accelerating different ion species as requested by the above proposals. The intensities for boron, magnesium, oxygen and lead requested by the above-mentioned~\acrshort{na} experiment proposals are expected to exceed the limits imposed by~\acrfull{rp} considerations in the~\acrshort{na}. 
A similar study is being conducted to address the performance reach of the ion complex for additional ion species requested by~\acrshort{alice3} and~\acrshort{hearts} (argon, calcium, krypton, indium and xenon). 

A pre-conceptual analysis has been performed to determine the upgrades of the ion complex~\cite{indico:1481171} that would be required to deliver different ion species with switching times of around 15 minutes, mitigate space-charge effects in the injectors to increase the beam brightness and to halve the bunch spacing to almost double the number of bunches injected into the~\acrshort{lhc}.

\subsubsection{Experimental Areas}

\paragraph{\acrshort{ps}~\acrfull{ea}}
The~\acrshort{ea} at~\acrshort{cern} was designed to use the protons from the~\acrshort{ps} and has served the physics community for over 50 years. With its perfect complementarity with the beam energies available at the~\acrshort{cern}~\acrshort{na}, which uses the protons from the~\acrshort{sps}, it remains extremely popular and necessary for many different experiments and test beams. Given the upgrades needed for the~\acrshort{hllhc}, there is a growing request for test beams to calibrate the detectors over a wide range of energies. In addition, physics experiments like~\acrfull{cloud} and irradiation facilities, like IRRAD and \acrfull{charm}, depend on the reliability and stable operation of the~\acrshort{ea}. An extensive renovation of the~\acrshort{ea} took place during the~\acrshort{ls2} following a study phase between 2009 and 2015. The renovated~\acrshort{ea} started its operation in 2021 and has been serving the diverse user community since. The main goal of the project was to ensure the long-term operation of the~\acrshort{ea} experimental facilities in line with the operation needs and the physics requirements. It included a new beamline layout, a new cycled powering scheme and refurbished beam elements and associated infrastructure~\cite{EastArea}. In an effort to reduce the annual energy consumption of the facility, the~\acrshort{ea} building was upgraded with improved wall and roof cladding to better insulate it. The massive magnet yokes have been replaced with laminated ones to cycle all the magnets, and the power converters have been upgraded with energy recovery capacitor banks. The energy consumption has thus been reduced from~\SI{11}{\giga\watt\hour} to~\SI{0.6}{\giga\watt\hour}. To improve reliability, and maintainability the number of different families of the magnets has also been reduced from 22 to only 12. The upgrades and the new layout have also resulted in improved beam purities especially for the electron beam in the secondary beamlines at low energies which complements the energy ranges available in the~\acrshort{na}.

\paragraph{\acrshort{isolde} and~\acrshort{ntof}}

The~\acrshort{ntof} facility underwent major renovation works during~\acrshort{ls2}, with the second-generation spallation lead target replaced by a new  assembly. With an innovative design, the target consists of lead plates cooled by nitrogen gas, and it can receive a higher number of~\acrshort{ppp}, up to~$10^{13}$, as an increased average proton flux. A new target replacement is planned for~\acrshort{ls4}, presently planned in 2034, most probably with minor design improvements with respect the current one. For the moment there are no other major modifications planned for the facility, which proved to be  adequate to deliver the required flux of neutrons for the next decade while maintaining its unique features, mainly the longest flight path with short proton pulses (\SI{7}{\ns}), the highest instantaneous proton (neutron) intensity and widest range of neutron energies spanning from ~\SI{}{\meV} to \SI{}{\GeV}~\cite{nTOFLOI}.

\acrshort{isolde} is the only radioactive ion beam facility using~\SI{}{\GeV} protons, giving access to several processes (fission, spallation, fragmentation) that contribute to radioisotope production. The \acrshort{isolde} primary area will undergo a major renovation in the coming years. The replacement of the two front-end beam dumps will take place during~\acrshort{ls3}. The two devices are approaching the end of their operational lifetime, and they will be replaced with two beam dumps capable of receiving higher power beams.  This, together with the proposed upgrade of the primary proton beam line from 1.4 GeV to~\SI{2}{\GeV},  will increase both the capability and the capacity of the facility. The higher power on target will open up the possibility of exploring new isotopes thanks to the increase of the production cross section with the higher energy. In addition, the increased yields will facilitate more detailed and higher precision studies, as well as increasing the throughput of experiments. Such advantages will be amplified if the potential of the~\acrshort{psb} after the~\acrshort{liu} upgrades can be used to deliver higher proton intensities~\cite{ISOLDELOI}.  The planned introduction of nanomaterial-based target also promises the possibility to produce shorter-lived isotopes. Other interventions to the facility will be focused on maintaining and improving its availability and performance, with particular attention to the superconducting linac of the~\acrfull{hieisolde} facility. Studies will be launched to make beams of radioactive molecules available in the framework of~\acrshort{pbc} (see Section~\ref{sec:radiomolecules}).

\paragraph{\acrshort{sps}~\acrfull{na}}

The~\acrshort{na} beamlines \cite{CBWG-report-2025} are a cornerstone of the~\acrshort{pbc} programme, providing a versatile~\acrshort{ft} facility for exploring fundamental interactions. These beamlines deliver a broad mix of particle beams, ranging from protons and hadrons to electrons, muons, and ions, which are tailored to support both precision measurements and innovative new physics searches. Their design emphasizes flexibility and high performance, ensuring that experiments can be optimized for a wide range of research objectives. Table~\ref{tab:NA-beam-characteristics} summarises the beam characteristics of the~\acrshort{na} beamlines. 

\begin{table}[tbp]
\footnotesize
    \centering
       \begin{tabularx}{\linewidth}{lXXXXXXX}
       \hline \hline
       
       \textbf{Parameter} & \textbf{H2} & \textbf{H4} & \textbf{H6} &
\textbf{H8} & \textbf{M2}  & \textbf{K12} & \textbf{P42}\\
       \midrule
       Max. momentum [$\mathrm{GeV}/c$] & $400/360$ & $400/360$ & $205$ &
$400/360$ & $280$ & $75$ & $400$ \\
      Max. acceptance [$\upmu\mathrm{Sr}$]& $1.5$ & $1.5$ & $2$ & $2.5$ &  $5$(h) & $12.7$ & $1.4$(p) \newline $0.3$(e,h)\\
      Maximum $\mathrm{\Delta p/p}$ [\%]  & $\pm2.0$ & $\pm1.4$ & $\pm1.5$ &
$\pm1.5$ & $\pm4.0$ & $\pm2.0$ & $\pm0.5$(p)\\
Typical $\mathrm{\Delta p/p}$ [\%]  & & & & & & $\pm1.0$ & \\
      Maximum intensity/spill & $10^7/10^5$ & $10^7/10^5$ & $10^7/10^5$ &
$10^7/10^5$ &  $5\cdot10^8$ & $2\cdot10^9$ & $6\cdot10^{12}$\\
      Particle types (\textcolor{teal}{\textbf{typical}}) & $p$, $h$, $\mu$, \newline $e$, ions & $p$, $h$, $\mu$,\newline $e$, ions &
$h$, $\mu$, $e$ & $p$, $h$, $\mu$,\newline  $e$, ions & \textcolor{teal}{$\mathbf{\mu}$}, \textcolor{teal}{$h$}, $e$ & \textcolor{teal}{$h$}, $\mu$  & \textcolor{teal}{$p$}, $h$, $\mu$, \newline $e$, ions\\

       \hline \hline
    \end{tabularx}
    \caption{Overview of the characteristics of the~\acrshort{na} beamlines. If multiple entries are provided for a specific value, the first one corresponds to primary beam operation, while the second one refers to secondary beams. The following abbreviations are used for particle types: $p$ (protons), $h$ (hadrons), $e$ (electrons), $\mu$ (muons). Intensities are given for a typical spill of \SI{4.8}{\second}.}
    \label{tab:NA-beam-characteristics}
 \end{table}

\begin{figure}[h!]
  \centering
  \includegraphics[width=\linewidth]{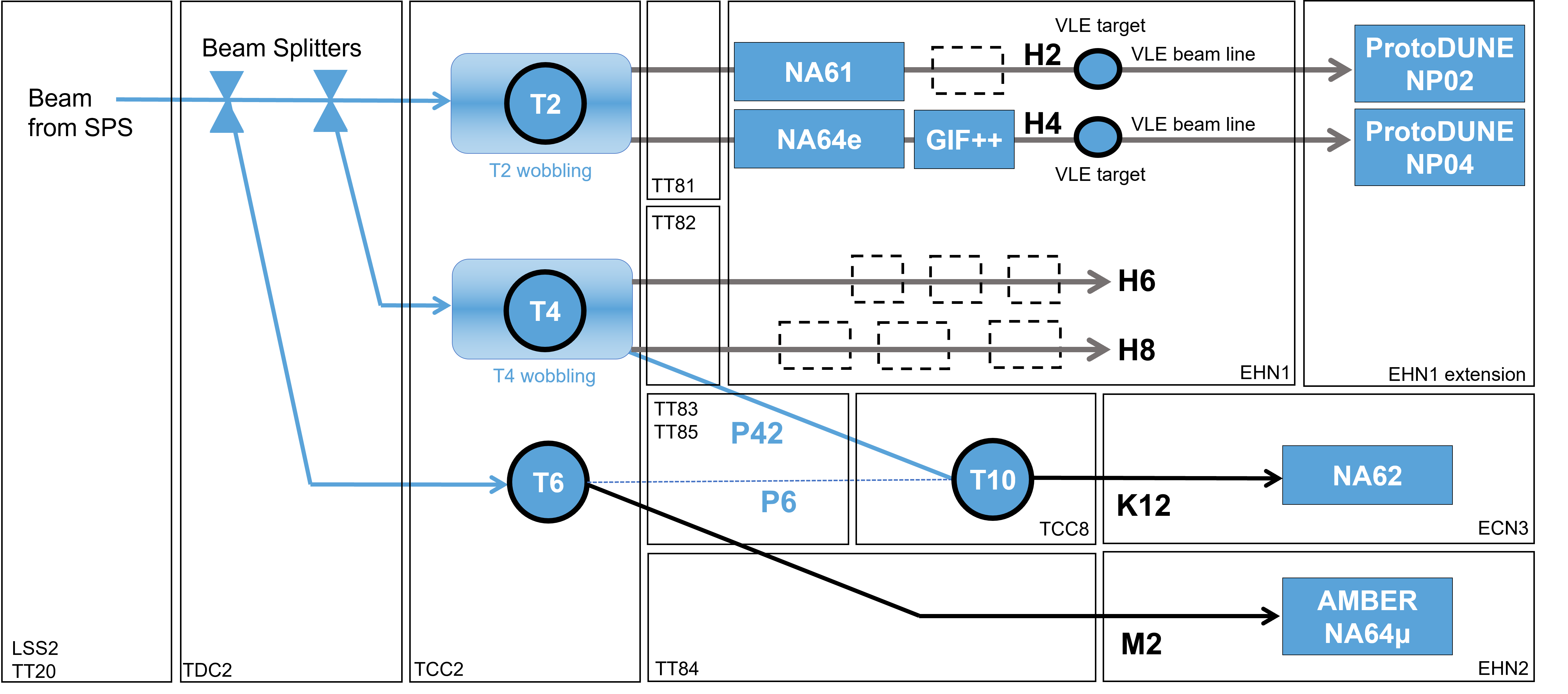}
  \caption{A schematic layout of the NA beamline and experiment complex as of 2025~\cite{Ahdida:2867743}.}
  \label{fig:NA-layout}
\end{figure}

The~\acrshort{na} is fed by a slow‐extracted beam from the~\acrshort{sps}, typically delivered in~\SI{4.8}{\s} spills at~\SI{400}{\GeV\per~c}. The beam impinges on the primary production targets T2, T4, and T6, located in the~\acrfull{tcc2}. From these targets, secondary beams are produced and distributed via dedicated beamlines. The H2, H4, H6, and H8 lines, each extending over ~\SIrange[range-units=single,range-phrase=--]{600}{700}{\m}, deliver secondary and tertiary beams to~\acrfull{ehn1} for test‐beam applications and fixed‐target experiments [e.g., \acrshort{na61/shine}, \acrshort{na64}--e, and the~\acrfull{gif++}]. In contrast, the M2 beamline, which spans nearly~\SI{1.2}{\km}, is designed for high‐intensity muon production and also provides hadron and low‐energy electron calibration beams to~\acrfull{ehn2}, serving currently the~\acrshort{amber} and~\acrshort{na64}--$\upmu$ experiments. Furthermore, the K12 beamline delivers beams to~\acrshort{ecn3}, serving the~\acrshort{na62} experiment. Unlike the other beamlines, the K12 beamline starts in~\acrfull{tcc8}, which receives the non-interacting protons surviving the T4 target that are transported via the P42 beamline to the T10 production target. Figure~\ref{fig:NA-layout} shows a schematic overview of the different beamlines, experiments, facilities, and experimental areas as of 2025.
\\

\noindent\textit{~\acrfull{nacons} }\vspace{0.1cm}

\begin{figure}[tb]
  \centering
  \includegraphics[width=\linewidth]{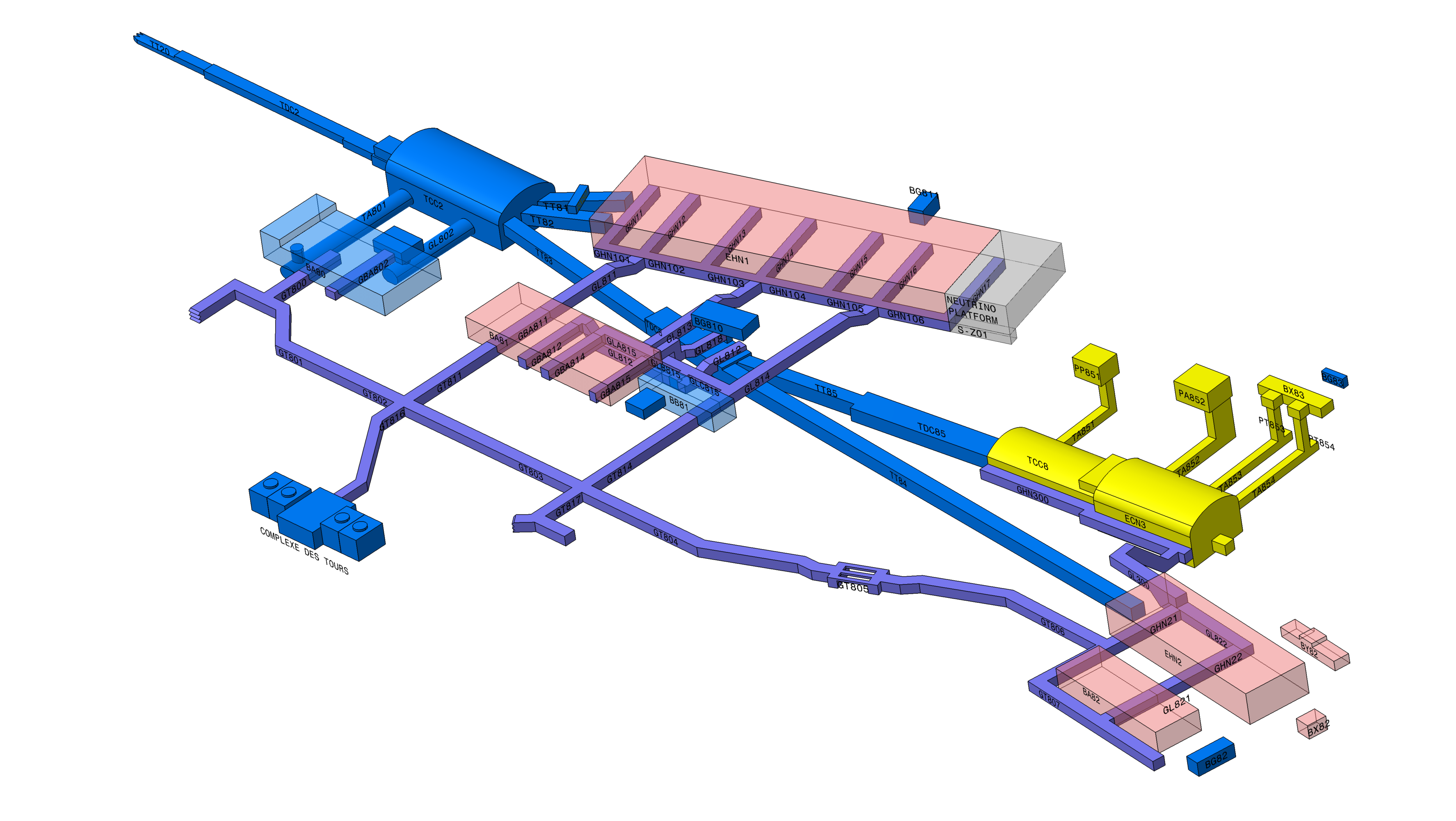}

  \caption[NA-CONS overview]{Overview of the experimental areas, beam tunnels, target caverns, and technical galleries in the~\acrshort{cern}~\acrshort{na}. The locations affected by Phase~1 of~\acrshort{nacons} are depicted in blue (caverns, tunnels, service buildings) and in violet (technical galleries), areas targeted by Phase 2 are shown in orange, and locations under \acrshort{hiecn3} are in light green.  Courtesy: CERN.}
  \label{fig:NA-CONS-overview}
\end{figure}

A~\acrfull{nacons} has been approved in 2021 with two phases covering the renovation of infrastructures and services to recover reliability and bring the facility into compliance with modern safety requirements. The first phase of the consolidation programme is targeting the single-point-of-failure with particular focus on the primary beam delivery systems. The second phase of the consolidation will guarantee the compatibility with future upgrades and new experiments. Figure~\ref{fig:NA-CONS-overview} gives an overview on the locations that are targeted by both consolidation phases: Phase 1 of~\acrshort{nacons} (2019 -- 2029), in blue, aims at the consolidation of the primary beam areas 
and the high-intensity beamline towards~\acrshort{tcc8} and~\acrshort{ecn3}. Phase~2 (2030 -- 2035), in orange, will address the consolidation of~\acrshort{ehn1} and~\acrshort{ehn2} together with their associated beamlines and technical buildings. 

The~\acrshort{nacons} baseline does not cover a full renovation of~\acrshort{tcc2} but discussions are ongoing to extend its scope to address a series of faults in~\acrshort{tcc2} hinting to a potential end-of-life scenario of various components.
It is proposed to carry out a full renovation of~\acrshort{tcc2} including the entire re-alignment of the beamlines, the replacement of the old 
magnets and production of additional spares, the replacement of plug-in supports and damaged magnet supports, extensive cabling/de-cabling, shielding removal and re-routing of all the services to provide better accessibility. Such an extension of scope will ensure a safe and reliable operation. 
\\

\noindent\textit{\acrfull{hiecn3}}\vspace{0.1cm}

\begin{figure}[th]
\includegraphics[width = \textwidth]{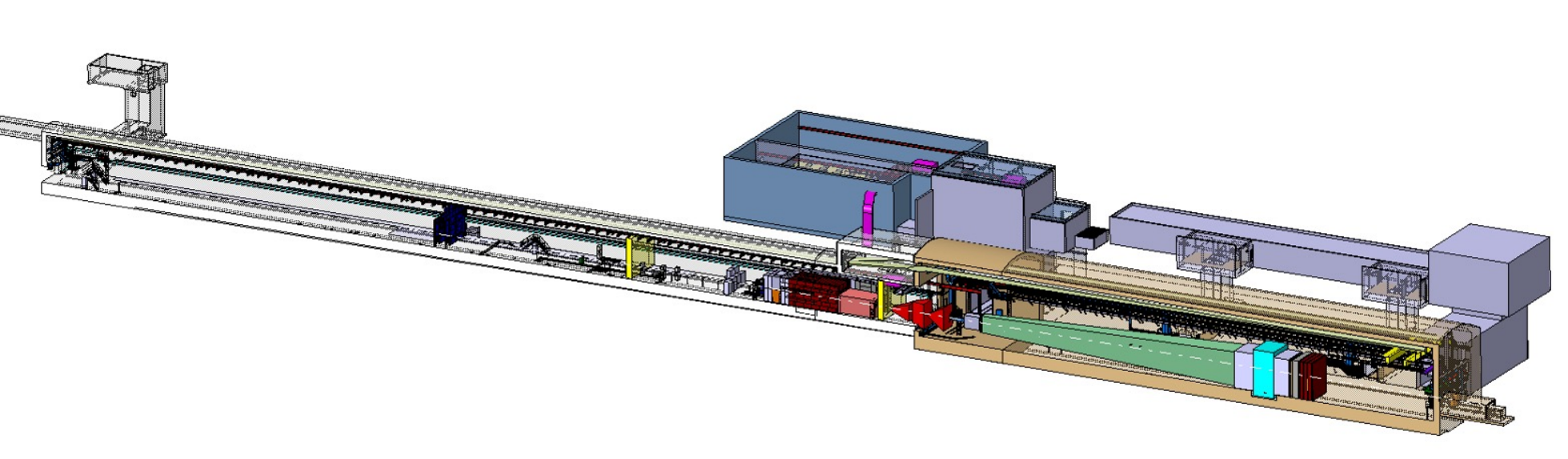}
\caption[BDF/SHiP integration in ECN3]{\acrshort{bdf}/\acrshort{ship} integrated inside the~\acrshort{ecn3} experimental cavern.  Courtesy: CERN.}
\label{f1}
\end{figure}

The consolidation efforts go hand in hand with the recent decision to host a new experimental facility to search for~\acrshort{hs}
particles, for which an~\acrfull{eoi}~\cite{Bonivento:1606085} was submitted to the \acrshort{spsc} in October 2013, identifying underground locations around the~\acrshort{sps} (\acrfull{tcc4},~\acrfull{tnc},~\acrshort{ecn3} and~\acrfull{tt61})~\cite{TaskForce2014}. At that time, all the suitable locations had recently approved programmes (\acrshort{awake}, \acrfull{hiradmat} and~\acrshort{na62}, respectively) and~\acrshort{tt61} was disfavoured for environmental reasons. Studies continued with a focus on the construction of an entirely new facility at a new underground cavern in the~\acrshort{na}, coined~\acrshort{ecn4}. Following a Technical Proposal~\cite{Anelli:2007512,SHiP:2015gkj} submitted to the~\acrshort{spsc} in 2015, a three-year~\acrfull{cds} was carried out under the auspices of~\acrshort{pbc} to present the proposal for the~\acrfull{bdf}~\cite{Ahdida:2703984} and the~\acrshort{ship} experiment~\cite{Ahdida:2654870,Ahdida:2704147} to the 2020~\acrshort{eppsu}. The \acrshort{ship} proposal was recognised as one of the front-runners among the new facilities investigated within the~\acrshort{pbc} studies but for reasons of cost, the project could not, as of 2020, be recommended for construction considering the overall recommendations of the~\acrshort{eppsu}~\cite{European:2720131}. Following a thorough assessment of alternative sitings~\cite{Aberle:2802785}, focus moved to~\acrshort{ecn3} to reduce the investment cost for the facility implementation. Several options were studied within the~\acrshort{pbc} Study Group for the long term use of \acrshort{ecn3} \cite{Ahdida:2867743}. After evaluation by the~\acrshort{spsc}, the~\acrshort{cernrb} selected the~\acrshort{ship} experiment~\cite{Albanese:2878604} for the future exploitation of the~\acrshort{ecn3} experimental facility after~\acrshort{ls3} in conjunction with the implementation of~\acrshort{bdf} and an upgrade of the facility to higher beam intensity. Figure~\ref{f1} shows an overview of \acrshort{ship} in~\acrshort{ecn3}. 

The cost of upgrading and implementing~\acrshort{bdf} in~\acrshort{ecn3} for~\acrshort{ship} came in at approximately 100 MCHF lower than the original~\acrshort{ecn4} proposal by removing the need for major civil engineering works and by re-using existing infrastructure, as well as profiting from the ongoing investment in~\acrshort{nacons}. Alongside the~\acrshort{nacons} project, the upgrade of~\acrshort{ecn3} will secure the long-term future of the~\acrshort{sps}~\acrshort{ft} physics programme at~\acrshort{cern}. 
The~\acrfull{hiecn3} project is presently in a rapid technical design phase with the delivery of the~\acrfull{tdr} expected mid-2026. The project is working in close synergy with the~\acrshort{nacons} project to upgrade the required accelerator infrastructure as part of the foreseen consolidation programme during~\acrshort{ls3}. The~\acrshort{tdr} for~\acrshort{ship} is expected around mid-2027.  The construction plan foresees commissioning of the facility in 2031 and detector in 2032 with at least one year of initial physics data-taking before~\acrshort{ls4} with a programme extending to the mid of the century.\\

\noindent\textit{Proton Sharing}\vspace{0.1cm}

A common feature of most proposals received for the~\acrshort{na}~\acrshort{ft} programme is the request for higher intensities. Proton-sharing scenarios across the~\acrshort{cern} accelerator complex have been analyzed and possible optimizations considering future~\acrshort{pbc} experiments have been studied. 

Since the cessation of~\acrshort{cngs} operations, the~\acrshort{na} experiments have received spills with a duration of \SI{4.8}{\second} at the highest possible repetition rate. The proton sharing scenarios for the high-intensity operation of~\acrshort{ecn3} in parallel to other~\acrshort{na} experiments has been studied also considering the parallel operation of the~\acrshort{lhc}, \acrshort{awake}, \acrshort{hiradmat} and \acrshort{md} sessions~\cite{Bartosik:2650722, Prebibaj:2848908}: $1.2\times10^{19}$ \acrfull{pot}/year will be available to the \acrshort{ehn1} and \acrshort{ehn2} experiments, provided no ion run (1 month) takes place and $0.8\times10^{19}$~~\acrshort{pot}/year can be delivered in case an ion run is included. The integrated intensity to the other \acrshort{na} experiments is maximised by assuming the acceleration of $4.2\times10^{13}$ \acrshort{ppp} with a \SI{4.8}{\s} spill. For some~\acrshort{na} users this might be problematic due to rate limitations and an alternative approach (previously adopted during~\acrshort{cngs}  operation) would be to extend the flat top while proportionally increasing the accelerated beam intensity up to $4.2\times10^{13}$~\acrshort{ppp} (which has been already achieved at the~\acrshort{sps}): $0.8\times10^{19}$~~\acrshort{pot}/year ($0.6\times10^{19}$~~\acrshort{pot}/year with an ion run) can be delivered to the~\acrshort{ehn1} and \acrshort{ehn2} users with \SI{9.6}{\s}-long spills.

Following the decision to launch a \acrshort{tdr} for \acrshort{ship}, additional studies are being performed by the bSAC (Beam Sharing across Complex) and SOX (Spill Optimisation for eXperiments) working groups, mandated by the \acrfull{iefc}, to cope with potentially higher proton requirements for the~\acrshort{na} experiments while minimizing the impact on the users of the upstream~\acrshort{ps} complex~\cite{tirsicham}. The reports from these working groups are expected by the end of Q1 2025. 

\subsubsection{AD complex and ELENA}

The Antimatter Factory at \acrshort{cern} comprises the proton-to-antiproton target area, two synchrotrons [the \acrfull{ad} and the \acrfull{elena}], and several experimental zones housed within the \acrshort{ad} hall.

The \acrshort{ad} ring has been operational since early 2000~\cite{Belochitskii:2001uu}, delivering approximately $3 \times 10^7$ $\bar p$ at \SI{5.3}{\MeV} in a single bunch every two minutes to experiments, which, before~\acrshort{ls2}, operated in 8-hour shifts.
In 2018, the \acrshort{elena} ring was commissioned~\cite{Gamba:2019fcb, Carli:2021ojn} and became fully operational after~\acrshort{ls2} in 2021~\cite{Ponce:2022jbb, Dutheil:2022zco}.
The introduction of \acrshort{elena} has enabled a further deceleration of antiprotons extracted from~\acrshort{ad} to energies as low as~\SI{100}{\kilo\eV}. This advancement significantly improved experimental efficiency, increasing trapping rates by up to two orders of magnitude. In addition, \acrshort{elena} allows the available intensity to be split into four bunches, enabling up to four experiments to run simultaneously.

Since its integration,~\acrshort{elena} has exceeded expectations, doubling the antiproton flux compared to its original design specifications, as shown in Table~\ref{tab:AD_ELENA_perf}.

\begin{table}[h!]
\centering
\begin{tabularx}{\linewidth}{lcccc}
\hline\hline
& \acrshort{pot} [$\times 10^{13}$] & $\bar p$/bunch [$\times 10^6$] & \# $\bar{p}$ bunches & rep. period [s]\\
\hline
\acrshort{ad} pre-\acrshort{ls2} (achieved)        & 1.4  & 30  & 1 & 110 \\ 
\acrshort{elena} (design)       & 1.4  & 4.5 & 4 & $\approx$100 \\ 
\acrshort{elena} (achieved)     & 1.9  & 12  & 4 & 110 \\ 
\hline\hline
\end{tabularx}
\caption{Overall performance of the Antimatter Factory.}
\label{tab:AD_ELENA_perf}
\end{table}

A comprehensive review of the facility’s status is presented in~\cite{indico:1255500}, while a broader overview of potential upgrades can be found in~\cite{indico:1343931}.
These upgrades include essential consolidations of legacy \acrshort{ad} equipment to ensure maintaining high availability and operational reliability, with the assumption of operating the facility till the end of Run 5 (end of 2041).
Short-term performance improvements, such as enhanced beam stability and reduced cycle times, are already underway.

\subsubsection{High Luminosity LHC (HL-LHC)}

The completion of the~\acrshort{liu} project and a number of~\acrshort{hllhc} upgrades already deployed in LS2~\cite{Arduini:2024kdp} have offered to the~\acrshort{lhc} the opportunity and the challenge to operate with up to two times higher beam brightness compared to the design parameters and 50~\% higher as compared to Run~2. The main beam parameters achieved and planned for the~\acrshort{lhc} and the~\acrshort{hllhc}~\cite{HLLHC} at the start of collisions are given in Table~\ref{tab:parameters}. The complete installation of the~\acrshort{hllhc} will take place in~\acrshort{ls3}. In this context, Run~3 is clearly a transition between the~\acrshort{lhc} and the~\acrshort{hllhc}.

 The~\acrshort{hllhc} Project has the following targets:
\begin{itemize}
    \item{} A peak luminosity of~\lumprot{5} with levelling operation (corresponding to an~\acrfull{pu}, i.e.~the number of events per bunch crossing in the detectors, of $\approx$~140) with the possibility to reach a peak levelled luminosity of~\lumprot{7.5} (corresponding to $\left<\mu\right>$~>~200).
    \item{} An integrated luminosity of~\SI{250}{\femto\barn^{-1}} per year, with the goal of achieving a total of~\SI{3000}{\femto\barn^{-1}} in the decade following the upgrade.
\end{itemize}
These ambitious goals will be achieved through a number of important upgrades throughout the accelerator. In particular, a complete renovation of the experimental insertions \acrshort{ir}1 and \acrshort{ir}5 is planned, involving about \SI{1.2}{\km} of the accelerator. The upgrades include the installation of high-field Nb$_3$Sn quadrupole magnets, which will replace the current focusing magnets around the~\acrshort{ip}s to achieve smaller beam sizes. This is complemented by the deployment of crab cavities, which will rotate particle bunches to maximize their overlap at the collision points, thereby enhancing effective luminosity. To support these modifications, additional cryogenic plants, \acrshort{hts}-based superconducting links, new power converters and beam instrumentation will be installed. The project also relies on an improved beam collimation system designed to handle the higher beam intensities.  These upgrades will also ensure the~\acrshort{lhc} efficient operation until the end of the physics programme, presently planned until 2041.

The transition from the~\acrshort{lhc} to~\acrshort{hllhc} in Run~4 foresees a carefully staged performance ramp-up during the initial years of operation. This approach balances machine protection, operational experience with new hardware systems, and the progressive validation of beam dynamics at higher intensities. Luminosity will also be increased incrementally, while the accelerator and experiments gather essential experience with higher pile-up conditions and increased radiation levels in critical areas such as collimation and experimental insertions.

The availability of these upgraded beams at the LHC starting in Run~4 offer further \acrshort{pbc} experimental opportunities presently under consideration. These~\acrshort{pbc} experiments, operating parasitically or semi-parasitically alongside the main collider programme, represent a cost-effective and complementary path to new physics in the~\acrshort{hllhc} era, making full use of the upgraded accelerator infrastructure and physics opportunities presented by the high-luminosity environment.

\begin{table}[t!]
\centering
\begin{tabularx}{\linewidth}{lccc}
\hline\hline
Parameter & Nominal~\acrshort{lhc} & \acrshort{lhc} & \acrshort{hllhc} \\
          & (design report) & (2024) &  \\
\hline
Beam energy in collision [TeV] & 7 & 6.8 & 7 \\ 
Particles per bunch \(N_b\) [$10^{11}$~ppb] & 1.15 &  1.6 & 2.2 \\
\# bunches per beam & 2808 & 2352 & 2760 \\
Half-crossing angle in IP1 and IP5 [$\mu$rad] & 142.5 & 150 & 250 \\
Minimum \bstar [m] & 0.55 & 0.30 &  0.15 \\
Norm. transverse emittance start of collisions $\epsilon_n$ [$\mu$m]  & 3.75 &  2.50 & 2.50 \\
Levelled luminosity in IP1/5 [$10^{34}$~cm$^{-2}$s$^{-1}$] & 1 (peak) & 2.0 & 5.0 \\
Max. average event pile-up $\left<\mu\right>$ [events/crossing] & 27 & 60 & 132 \\ 
\hline\hline
\end{tabularx}
   \caption{\label{tab:parameters}\acrshort{hllhc} nominal parameters compared to the~\acrshort{lhc} design and achieved parameters~\cite{Bruning:782076}.
    }  
\end{table}

\subsection{Accelerator Technology developments}

\acrshort{cern} has developed a number of technologies to support the design and construction of accelerators and detectors for particle physics experiments. The experience and expertise in technological domains such as~\acrfull{sc} and normal conducting magnet and~\acrshort{rf} technology, cryogenics, optics, vacuum and surface technology can benefit and contribute to the development of new detection methods like quantum sensing. It can also support the development and assessment of the feasibility of accelerator and non-accelerator experiment proposals aiming at fundamental~\acrshort{sm} or~\acrshort{bsm} physics measurements. 
As examples of such contributions we can mention the design of the~\acrshort{sc} dipole for the ~\acrfull{babyiaxo} experiment~\cite{IAXO}, the contribution to the~\acrshort{ds20} experiment cryogenic and argon purification system~\cite{aalseth_darkside-20k_2018} (including the support for commissioning the Aria argon distillation plant~\cite{agnes_separating_2021}), or the current ongoing development of~\acrfull{hts} for the~\acrfull{rades} experiment~\cite{Golm:2021ooj}.
To further foster exchanges between technology experts at~\acrshort{cern} and collaborating institutes, the~\acrshort{pbc}~Technology~\acrshort{wg} has organized five workshops on the following topics:~\acrshort{sc}~\acrshort{rf}~\cite{TECH_1st}, lasers and optics~\cite{TECH_2nd}, vacuum, coating and surface technologies~\cite{TECH_3rd}, cryogenic technologies~\cite{TECH_4th} and superconductivity technologies~\cite{TECH_5th}. 

Since its inception~\acrshort{pbc}~\cite{PBC_summary_2020} has supported the feasibility study of several experiments, some of which have now reached maturity, being fully fledged collaborations and funded by the respective agencies. Among the experiments that were part of the~\acrshort{wg} are~\acrfull{alpsii}~\cite{ALPS-II}, which is presently taking data at~\acrfull{desy}, and the already mentioned \acrshort{babyiaxo} and~\acrshort{ds20}, which are now in the construction stage. 
Some other experiments have instead either concluded that their proposal could not reach technical feasibility (\acrfull{vmbcern}), or are running their facilities without further contribution required from the Technology~\acrshort{wg} (Ptolemy,~\acrfull{stax}). The experiments participating in the~\acrshort{wg} at the date of this report are

\begin{itemize}
    \item \acrfull{grahal} \cite{grenet2021grenobleaxionhaloscopeplatform},
    \item \acrfull{akwisp} \cite{cantatore2018akwispinvestigatingshortdistanceinteractions},
    \item \acrfull{flash}~\cite{Alesini_2023},
    \item \acrshort{rades}~\cite{Melcon:2018dba},
    \item Axion heterodyne detection~\cite{Berlin:2020vrk}, and
    \item \acrfull{aion}-100 @~\acrshort{cern}.
\end{itemize}

\acrshort{grahal} and~\acrshort{akwisp} joined~\acrshort{pbc} from its launch~\cite{Siemko:2652165}, while the other experiments have joined more recently. \acrshort{flash} has just started interacting with the Technology~\acrshort{wg} at the date of this report thanks to preliminary discussion established during the workshops. The four new experiments are described in more detail in Section~\ref{subsubsec:non-acc_exp}, and they all are in synergy with the~\acrfull{qti}~\cite{QTI} and the~\acrfull{ecfa}~\acrfull{drd} 5 on quantum sensors~\cite{DRD5}, as they all rely on the support of both programmes, to different extents.

\subsection{Potential future accelerators and facilities at CERN}
\label{sec:future_facilties}

\subsubsection{Short Baseline Neutrino beam proposal}
\label{sec:sbn_acc}

The need for a high-precision neutrino beam for cross-section measurements has been recognized since the 2020~\acrshort{eppsu} and is currently being explored at~\acrshort{cern} within the framework of~\acrshort{pbc}. This effort builds upon~\acrshort{cern}'s investments in neutrino physics, particularly in the development of the proto-DUNE detectors, along with advances in high-precision neutrino-beam design spearheaded by the~\acrfull{enubet} and~\acrfull{nutag} collaborations~\cite{ENUBET:2023hgu,hep-ph_Perrin-Terrin_2022,hep-ph_Baratto-RoldanEtAl_2024}.
The aim of the~\acrfull{sbn} beamline~\cite{SBN@CERN_ESPPU} is to provide a medium-intensity, high-precision neutrino beam that advances our knowledge of neutrino cross sections at the~\SI{}{\GeV} scale in all channels that cannot be addressed by current neutrino cross-section experiments or the upcoming near detectors of~\acrfull{dune} and~\acrfull{hk}. 

\begin{figure}[t!]
\includegraphics[width=\textwidth]{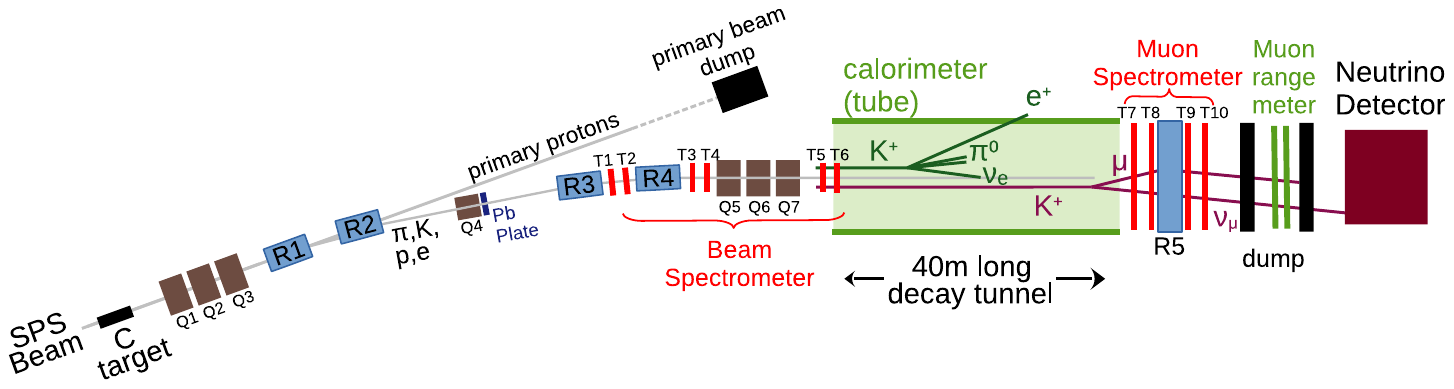}
\caption{Conceptual layout of the~\acrshort{sbn} beamline ~\cite{SBN@CERN_ESPPU}. The secondary mesons produced in a target by the slow-extracted~\acrshort{sps} protons are focused with quadrupoles (in brown) and selected in the~\SI{8.5}{\GeV\per\c} momentum range by a set of dipoles (in blue) before entering the decay tunnel. The secondary beamline, decay tunnel and downstream region are instrumented with calorimeters (in green) and trackers (in red) to monitor the neutrino fluxes and tag individual neutrinos.      
\label{fig:SBNlayout}}
\end{figure}

A conceptual study of the~\acrshort{sbn} beamline has been performed, and an extensive report is under preparation~\cite{jebramcik2024_placement}. The proposed design (Figure~\ref{fig:SBNlayout}) includes a slow extraction of the~\acrshort{sps} protons towards a production target, a 23-meter-long secondary meson beamline and a 40-meter-long decay tunnel, both instrumented with fast radiation-hard detectors for a high-precision monitoring of the neutrino fluxes and individual neutrino tagging (see Section~\ref{sec:sbn_exp}).  Following an extensive optimisation of the target and acceptance of the secondary beamline the expected particle production rates at a particle momentum of~\SI{8.5}{\giga\eV\per\c} are $12.7 \times 10^{-4}\,K^+$ per \acrshort{pot} and $1.9 \times 10^{-2}\, \pi^+$/\acrshort{pot}. In order to meet pile-up constraints, the ideal spill instantaneous intensity should be in the range of $\sim 1.0 \times 10^{12}\,$\acrshort{pot}$/$s. To collect $10^4$ $\nu_e$~\acrshort{cc} interactions at the neutrino detector, roughly $2-3 \times 10^{18}$~\acrshort{pot} per year over five to seven years are required, with the yearly intensity never exceeding 30~\% of the maximum~\acrshort{pot} per year that could be available for~\acrshort{tcc2} during \acrshort{ship} operation~\cite{Prebibaj:2848908}. 

The studies conducted so far indicate that the considered beam line can be integrated neither in~\acrshort{ehn1} nor~\acrshort{ehn2} due to radiation-protection considerations as well as geometric (survey) reasons. Hence, an installation in the~\acrshort{na} would require the excavation of a new underground infrastructure and extensive \acrfull{rd} on the feasibility of the beam delivery. 
A new experimental area close to~\acrshort{tcc6} (where the~\acrshort{hiradmat} facility is installed) could also be considered, given that infrastructure already exists there (like the side tunnel TT61, that is currently empty and not used). This could partly use existing transfer lines and therefore reduce the implementation costs. In this case the extraction system should allow simultaneous slow-extraction to the new experimental area and to the \acrshort{na}, similarly to what was done to serve simultaneously the~\acrshort{sps}~\acrshort{na} and~\acrfull{wa} until the beginning of the 2000's. Such a system would now have to be compatible with the~\acrshort{lhc} extraction. 
A conceptual slow extraction scheme has been proposed~\cite{jebramcik2024_sx} and more studies would be required to demonstrate its feasibility. The installation of a new slow extraction system cannot occur before~\acrshort{ls4} as it would require significant modifications of the~\acrshort{sps}. Significant civil engineering for the construction of the new experimental area would also be necessary.

\subsubsection{Gamma Factory}  

The key~\acrshort{gf} idea~\cite{Krasny:2015ffb,WuChao:2023kci,Krasny:2023ptc} is to produce, accelerate, and store highly relativistic 
atomic (\acrfull{psi}) beams in the~\acrshort{lhc} rings 
and to resonantly excite their atomic degrees of freedom by laser photons to: 
(1) cool and study highly-charged ions,  and (2) produce  
high-energy, polarised photon beams with intensities 
reaching
$10^{17}$ photons/s. 
Emitted photons in the energy range from~\SI{40}{\keV} to~\SI{400}{\MeV} with the above-mentioned intensities can be achieved at the~\acrshort{lhc} with laser photon energies varying from 0.7~\SI{}{\eV} to 12~\SI{}{\eV}
and Fabry-Perot-cavity-stored photon pulses with an average power of up to~\SI{200}{\kW}. The above performance might require a significant upgrade of the~\acrshort{lhc}~\acrshort{rf} system, depending on the specific implementation.

Emitted photons in the above energy range can  be used  as a source of unprecedented-intensity secondary  beams of polarised electrons and positrons [$\approx10^{17}~e^{+}(e^{-})$/s]~\cite{Apyan:2022ysh}, polarised muons [$\approx10^{14} ~\mu ^{+} (\mu^{-})$/s]~\cite{Apyan:2022ysh}, 
quasi-monochromatic neutrons ($\approx10^{16} ~n$/s)~\cite{Baolong:2024ata}, and radioactive ions ($\approx10^{13} ~r.i.$/s)~\cite{Nichita:2021iwa}.

Dedicated simulation tools~\cite{Placzek:2019xpw,Curatolo:2018pza,GF-software-workshop-2021} have been developed to conduct detailed feasibility studies~\cite{Krasny:2021llv,Kruyt:2024sty,Krasny:2020wgx}. 

The technical feasibility of the~\acrshort{gf} photon beam generation and the~\acrshort{psi} laser beam cooling schemes are being investigated and a~\acrfull{gfspspop} experiment has been proposed~\cite{GF-PoP-LoI:2019} for installation in the~\acrshort{sps} tunnel starting from~\acrshort{ls3}. Essential milestones towards the~\acrshort{gfspspop} have already been achieved since the last~\acrshort{eppsu}: 
\begin{enumerate}
    \item demonstration of efficient production, storage and operation of the~\acrshort{psi} beams in the~\acrshort{sps} and ~\acrshort{lhc}~\cite{Hirlaender:2018rvt,Dutheil:2020ekk,Kroger:2019wfh,Gorzawski:2020dgx,Schaumann:2019evk}, including the assessment of the stability and measurement precision in beam position and momentum in the~\acrshort{sps} area where laser pulses would interact with atomic beam bunches~\cite{Rebeca:2022};
    \item design~\cite{Martens:2022ptn} and experimental demonstration of the stable storage of more than~\SI{200}{\kW} of average power in a Fabry-Perot resonator, in fact up to~\SI{700}{\kW} was achieved at the ~\acrfull{ijclab}~\cite{Martens:2022ptn,Lu:2024gwe,Lu:24,Granados:2024sht}.
\end{enumerate}

The successful implementation of the~\acrshort{gf} 
requires the validation of the following additional elements:
\begin{itemize}
    \item stable laser transport and injection of the laser beam into the long Fabry-Perot cavity with fully remote controls and diagnostics. The beam stability is comparable to that of the interferometric systems of the~\acrfull{ligo};
    \item fully remote and continuous operation of the laser source system in a high intensity hadron accelerator tunnel;
    \item operational tools to guarantee the requisite control of the~\acrshort{psi} beam momentum and its spread, and the spatial and temporal overlap of the ion and laser beams in a reproducible fashion;
    \item agreement between simulations and measurements of atomic excitation and ion beam cooling rates; 
    \item atomic and photon beams diagnostic methods to measure and characterise the photon flux from the spontaneous emissions allowing to 
    unfold the photon spectrum and angular distribution.
\end{itemize}

The modelling of the radiation environment at the~\acrshort{gfspspop} experiment site has been performed~\cite{Mazzola:2024xqq} and measurements are ongoing to validate them.
Procurement of an ultra low-phase noise laser and amplification chain is being finalised. The final design of the controls, diagnostics of the laser beam transport system and optical cavity are in progress, with dedicated studies on stabilisation reported in~\cite{Granados:2868829}. 

Scientific opportunities opened by the~\acrshort{gf} across the fields of science are briefly discussed in Section~\ref{sec:PBCatfuturefacilities}. 

\subsubsection{AWAKE and future proton-driven plasma wakefield accelerators}
\label{sec:awake}
\acrshort{awake}, is a proton-driven plasma wakefield acceleration experiment, using self-modulation of a long bunch of~\SI{400}{\GeV} protons in plasma to resonantly excite high amplitude wakefields and accelerate externally injected electrons to high energies [$\BigOSI{}{\GeV}$]. Building on its successful~\acrfull{poc} results~\cite{{bib:batsch-smi-ssm-2021,bib:awake-ssm-2019,bib:turner-ssm-2019,bib:awake-e-acc-2018}} during its first run period (2016 -- 2018), the~\acrshort{awake} collaboration has developed a well-defined program for Run~2~\cite{bib:edda-symmetry-2022}, which started in 2021 and will last until~\acrshort{ls4}. 

The experimental layout for Run~2 incorporates several key changes from Run~1, particularly the introduction of two distinct plasma sources. The first plasma source serves as the ``self-modulator'', where the proton bunch undergoes seeded self-modulation, which was successfully demonstrated~\cite{bib:verra-ssm-2022}. Introducing a density step in the plasma source mantains high-amplitude wakefields~\cite{bib:lotov-densitystep-2015}. Recent measurements are very promising, showing a clear effect, e.g., higher achieved accelerated electron energies.

The second plasma source is the ``accelerator'' source, where electrons are injected and accelerated, while controlling the beam quality. With this layout, which will be ready after~\acrshort{ls3},~\acrshort{awake} Run~2 aims to achieve the next milestone: the acceleration of electrons in the range of~\SIrange[range-units=single]{4}{10}{\GeV} in~\SI{10}{\m} of plasma, a low electron energy spread (5--8~\%) with an accelerated bunch charge of approximately~\SI{100}{\pico\coulomb} and a normalized emittance of~\SIrange[range-units=single,range-phrase=--]{2}{30}{\mm\milli\radian}. To meet these goals, beam loading is required to flatten the wakefield and to control the emittance growth. This necessitates the use of a new electron beam system featuring a~\acrshort{rf} photo-injector, two X-band accelerating structures and a new beamline. A prototype of the new electron source has already been installed and successfully commissioned at~\acrshort{cern}’s~\acrfull{ctf2} facility, validating the new design. 

An additional milestone is the demonstration of the scalability of the acceleration process. In the last phase of Run~2, i.e., before~\acrshort{ls4}, it is planned to replace the second plasma source with a new plasma technology (discharge or helicon plasma source) scalable to lengths of tens to hundreds of meters, which will be crucial for achieving high-energy electron beams. A~\SI{10}{\m}-long prototype of a discharge plasma source was successfully commissioned and operated in~\acrshort{awake} in 2023. 

To accommodate the necessary space for the installation of the second plasma source and the associated electron injection system for achieving the acceleration and scalability milestones, the~\acrshort{awake} facility will undergo major modifications, with works starting in 2025. This includes dismantling the~\acrfull{cngs} target area, which currently occupies a~\SI{100}{\m}-long tunnel section downstream of the~\acrshort{awake} experimental facility. 

Once the goals of Run~2 will be demonstrated, the~\acrshort{awake} acceleration scheme could be used in first particle physics applications, including the production of electron beams with energies between~\SIrange[range-units=single,range-phrase=\text{ and }]{40}{200}{\GeV} for fixed-target experiments (see~Section~\ref{sec:awake_physics}).

\subsubsection{Future colliders and their injectors}\label{sec:futureCollidersAndInjectors}

\begin{figure}[h]
\includegraphics[width=\textwidth]{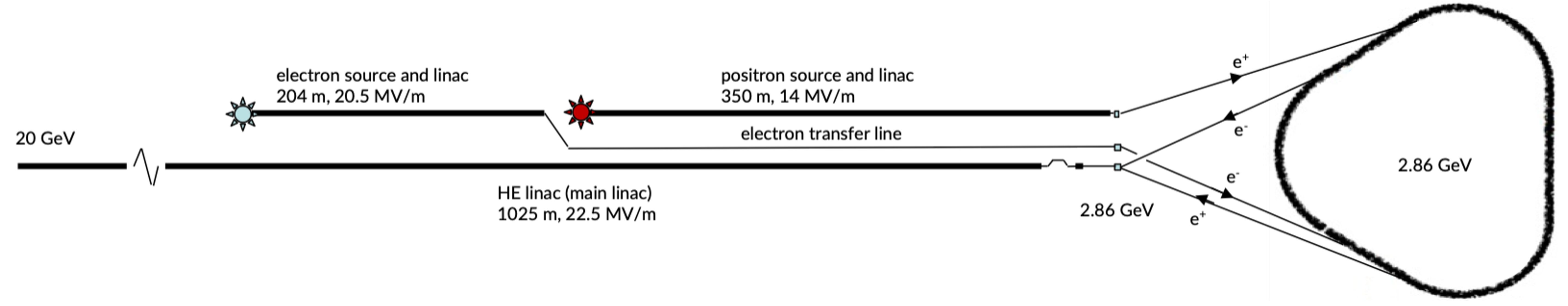}
\caption{Schematic representation of the proposed~\acrshort{fccee} injector complex~\cite{PBCFCCee:2025}.} 
\label{fig:FCCeeinjectors}
\end{figure}

\paragraph{\acrfull{fccee}}
The~\acrshort{fccee} will require a dedicated injector complex for delivering positron and electron beams to the booster for the top-up operation of the collider~\cite{FCCinjectors}. The presently studied injector complex,
as shown in Figure~\ref{fig:FCCeeinjectors}, consists of a low energy electron linac, a positron linac and a damping ring at~\SI{2.86}{\GeV}. For positron production, the electron beam from the low energy linac is sent on a positron target, and the positron linac accelerates the beam to the damping ring injection energy~\SI{2.86}{\GeV}. A high energy linac accelerates either positrons or electrons to the final energy of~\SI{20}{\GeV}. This injector complex will be placed on the~\acrshort{cern} Pr\'evessin site, with the beam exiting the high energy linac close to the~\acrshort{na}, which could provide additional science opportunities beyond collider physics. 

The~\acrshort{fccee} injector complex will provide pulses of 4~bunches with a bunch spacing of \SI{25}{\ns}, with $2.5\times 10^{10}$~\acrfull{ppb} (corresponding to~\SI{4}{\nC}). A summary of relevant beam parameters is given in Table~\ref{tab:FCCeeInjectors}. 
It is important to note that the duty cycle of the injector complex required for providing beam to the booster in top-up mode varies significantly depending on the operation mode of the collider. The injector complex will be used with a duty cycle of 73~\% for the $Z$-pole, 40~\% in $WW$, 19~\% for $ZH$ and 5~\% for $t\bar{t}$.

\begin{table}[h!]
    \centering
    \begin{tabular}{lcc}
    \hline\hline
         & \acrshort{fccee} &\acrshort{pbc} users\\
    \hline
     Beam energy [\SI{}{\GeV}]    & $20$ & $\le 20$\\
     Max. bunch charge [\SI{}{\nC}]    & 4 & 4 \\
     Max. bunch population    & 0.1--2.5 & 2.5\\
     Bunches per pulse   & 2--4 & 1--4 \\
     Linac repetition rate [\SI{}{\Hz}]    & 50-100 & 100\\
     Normalized emittance $x/y$ [\SI{}{\mm\milli\radian}]    & $\le 20/2$ & $\le 20/2$\\
     Physical emittance $x/y$ [\SI{}{\nm\radian}]    & $\le 0.5/0.05$ & $\le 0.5/0.05$\\
     Bunch length rms [\SI{}{\mm}]    & 4 & 1--4\\
     Energy spread [\%]    & 0.1 & 0.1-0.75\\
     Bunch spacing [\SI{}{\ns}]   & 25 & 25 or 50\\
     Pre-injector duty cycle [\%]    & 5--73 & 27--95\\
     \hline\hline
    \end{tabular}
    \caption{Beam parameters of the~\acrshort{fccee} injector complex~\cite{PBCFCCee:2025}.}
    \label{tab:FCCeeInjectors}
\end{table}

Science opportunities offered by the~\acrshort{fccee} injectors have been discussed in a workshop~\cite{FCCotherscienceopportunities} and are summarized in~\cite{PBCFCCee:2025}, those related to~\acrshort{pbc} are outlined in~Section~{\ref{sec:PBCatfuturefacilities}.

The positron beams of the~\acrshort{fccee} booster could be exploited for~\acrshort{dm} searches. These experiments typically require an almost continuous beam with low instantaneous rate~[$\BigOSI{1}{\ns^{-1}}$]. First considerations of producing such beams in between top-up injections of the booster to the collider have been presented in~\cite{Bartosik:Otherscienceopportunities}. The most promising options are to perform a slow extraction directly from the damping ring at~\SI{2.86}{\GeV}, which would result in about $1\times10^{13}$ positrons on target per day with a bunch spacing of~\SI{2.5}{\ns} when considering a duty cycle of 70~\%. Alternatively, a slow extraction at~\SI{20}{\GeV} from the booster, could deliver $2.7\times10^{13}$ positrons on target per day with~\SI{1.25}{\ns} bunch spacing assuming a 40~\% booster duty cycle (available during $H$ and $t\bar{t}$, while only 20~\% booster duty cycle would be available during $WW$). 

\paragraph{~\acrfull{lc}}
A future~\acrfull{lc} at~\acrshort{cern} could host a number of additional detectors, including detectors for~\acrshort{ft} experiments and~\acrshort{bd} experiments making use of the high energy main linac beams or the lower energy injector beams. A~\acrshort{lc} provides very high energy, high intensity, low emittance electron and positron beams. The single pass nature of the~\acrshort{lc} allows us to use the beams even destructively as long as the influence to the collider experiments is not significant. \acrshort{pbc} opportunities at an \acrshort{lc} are outlined in detail in Ref.~\cite{LinearColliderVision:2025hlt}. 

The main purpose of these experiments will be to search for~\acrshort{ds} particles interacting only feebly with the~\acrshort{sm} particles. The intense and high-energy electron and positron beams that the~\acrshort{lc} makes available also have uses in nuclear and hadron physics and in studies of strong-field~\acrfull{qed}. 

The most appropriate locations for using the beams are the beam dumps. There, very high
intensity electron and positron beams interact with thick targets, producing large numbers
of highly penetrating particles. There are several beam dumps distributed over a~\acrshort{lc} facility. Also the injector linacs and damping rings provide opportunities for lower energy beams. In the case of~\acrshort{clic} the very high intensity drive beam surface installation provides another beam available for physics or equipment~\acrshort{rd}. 

A significant numbers of concrete studies have been made for the~\acrfull{ilc} as documented in~\cite{aryshev2023internationallinearcolliderreport}. 
These studies also indicate a not exhaustive list of possible locations for experimental facilities as shown in Figure~\ref{fig:DumpDistribution}. The possibilities offered by injectors and damping rings are not fully shown. For~\acrshort{clic} or other~\acrshort{lc} options similar possibilities exist for providing electron and positron beams at varying energy and intensity.  

\begin{figure}[h!]
\includegraphics[width=\textwidth]{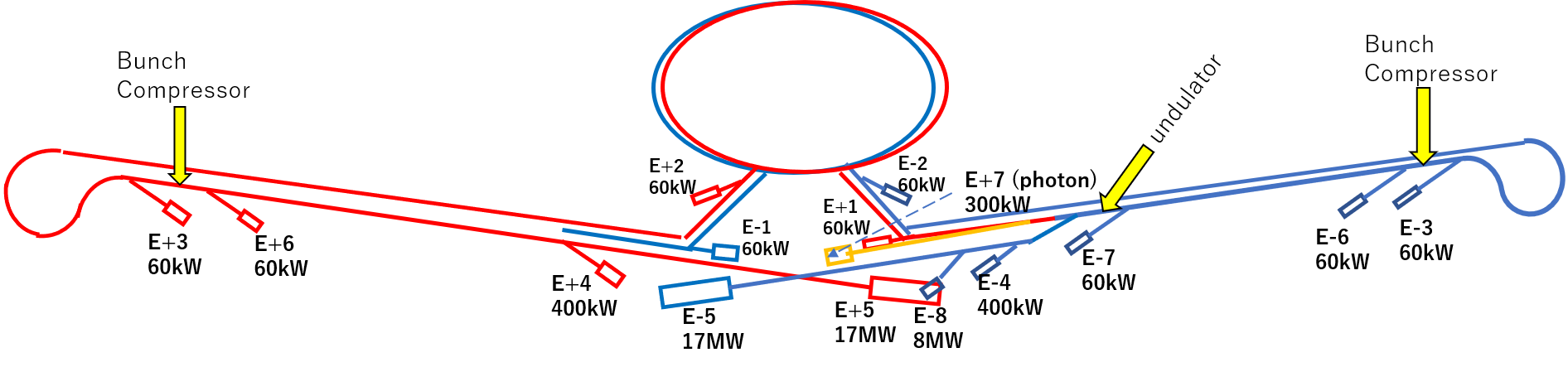}
\caption{Distribution of beam dumps over the~\acrshort{ilc} facility labeled E1-E8~\cite{aryshev2023internationallinearcolliderreport}. The electron, positron and photon beamlines are colored blue, red and yellow, respectively. Similar possibilities exists for other~\acrshort{lc} implementation providing a wide range of energies from injectors to full beam energies, using the opportunities offered by a single pass accelerators to provide beams according to proprieties and needs for collider and~\acrshort{ft} experiments operating at the same time.  
\label{fig:DumpDistribution}}
\end{figure}

\section{PBC proposed experiments and future opportunities at CERN and beyond}
\label{sec:ExpLandscape}
\acrshort{pbc} has been and is currently supporting the development and feasibility study of a wide range of experimental proposals for submission to the relevant scientific committees.
Below, experiments previously supported and now being followed-up by these committees,
are briefly described as well as the opportunities that are currently being scrutinized and studied to support a diverse particle physics programme at \acrshort{cern} and elsewhere. 

\subsection{SPS North Area}
\label{sec:PSPSPS_Opportunities}

\subsubsection{NA61/SHINE-LE}
\label{sec:na61le-intro}

\acrshort{na61/shine} plans to conduct studies of hadron production on nuclear targets using a new very-low-energy (LE) tertiary branch of the H2 beam. These measurements will include both proton- and meson-incident interactions on nitrogen as well as solid materials in order to improve predictions for the flux of neutrinos from atmospheric sources, accelerator beams, and spallation sources. 

Among the highlights of the \acrshort{na61/shine}-LE program \cite{NA61SHINE:SPSC-P-330-ADD-12,NA61SHINE_SPSC_M_793} are:
\begin{itemize}
\item{A comprehensive study of proton-nitrogen interactions at energies below 20~GeV. These will be used to improve models of atmospheric neutrinos, potentially resulting in improvements to the neutrino flux errors of more than a factor of two.}

\item{Measurements of meson production and rescattering from low-energy interactions on aluminium, water, and iron. These will reduce the current large uncertainties on wrong-sign neutrinos (especially $\nu_e$ and $\bar\nu_e$) that result from interactions outside the target at long-baseline beams and are especially significant backgrounds in precise studies of \acrfull{cp} symmetry. These measurements will be geared to the specific needs of the Hyper-K and  \acrshort{dune} experiments.}
\item {A study of 8~GeV $p+{\rm Be}$ interactions geared toward reducing the flux uncertainty of \acrshort{fnal}'s Booster Neutrino Beam from 7--40\% down to 5--6\% across the entire energy spectrum. This will translate directly into improved $\nu+{\rm Ar}$ cross-section measurements.}
\end{itemize}
The \acrshort{na61/shine}-LE beam project has been supported by the leadership of eleven neutrino and muon physics collaborations in a November 2024 letter to \acrshort{cern}'s management. From the  \acrshort{cern} side, a technical study was prepared in the framework of  \acrshort{pbc}-\acrshort{cbwg}~\cite{NA61-lowenergy}. The study is currently on hold waiting for the official approval by the relevant  \acrshort{cern} committees (\acrshort{spsc},  \acrshort{cernrb}). The new beam will be installed parallel to the existing one, which will remain available for \acrshort{na61/shine} and downstream users. A CAD drawing of this proposed tertiary branch is shown in Figure~\ref{fig:NA61le}.

\begin{figure}
    \centering
    \includegraphics[width=0.8\linewidth]{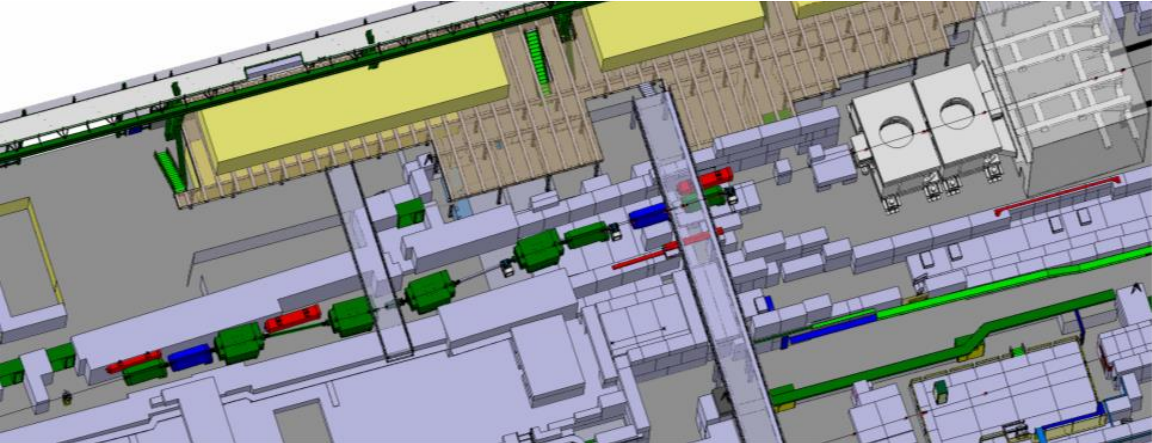}
    \caption{The proposed low-energy branch for  \acrshort{na61/shine}. The new magnets for the low-energy particles appear in green, while a rail system would allow the changeover between this low-energy and the normal high-energy configuration. More details on the proposed implementation can be found in~\cite{NA61-lowenergy}.}
    \label{fig:NA61le}
\end{figure}

\subsubsection{NA61/SHINE with ions} 
The immediate physics goals of the \acrshort{na61/shine} include: (i)
measurements of oxygen-oxygen collisions at $\sqrt{{s}_\mathrm{NN}}=5.1$, 7.6 and~\SI{16.8}{\GeV}, by the end of 
Run~3 and the beginning of Run~4, to elucidate the diagram of high-energy nuclear collisions~\cite{Gazdzicki:2810689,NA61/SHINEaddendum}
and for a precision study of violation of isospin symmetry~\cite{NA61SHINE:2023azp}; 
(ii)
construction and installation of a new \acrfull{last} during \acrshort{ls3};
(iii) 
a new series of measurements of charm and anti-charm hadron correlations in Pb+Pb collisions 
This would be accompanied by extensive measurements of \acrfull{ebye} fluctuations in Pb+Pb collisions as a function of energy and by new data-taking of B+B and/or Mg+Mg collisions~\cite{NA61/SHINEaddendum}.
The realization of oxygen data-taking does not necessitate any additions to the fully operational \acrshort{na61/shine}~experiment.
Implementing the \acrshort{last} for charm and anti-charm hadron correlation studies constitutes a major upgrade of the detector
to be completed at the end of \acrshort{ls3}. These measurements will be fully complementary to the operation of the considered \acrshort{na60+} experiment during Run~4.
Studies of \acrshort{ebye} fluctuations are meant as an independent verification of recent observations made at the \acrfull{rhic}, argued as a possible indication of a Critical Point in the phase diagram of strongly interacting matter.

The proposed large-acceptance measurements are not possible in the foreseeable future at accelerator facilities other than the \acrshort{sps}.

\subsubsection{NA64 ($-e$, $-\mu$, $-h$) -- Phase 2 }
The great advantage of the \acrshort{na64} approach compared e.g. to the \acrshort{bd} approach is that the signal rate in \acrshort{na64} scales as $(coupling)^2$, while for the 
latter one it is $(coupling)^4$. Thus, much less beam particles on target are required for the same sensitivity. In order to use this feature effectively in the future a setup upgrade during \acrshort{ls3} is required which will allow \acrshort{na64} to enter an exciting new phase, leveraging higher particle rates to search for light sub-GeV dark matter by running in background-free mode. After completing the corresponding \acrshort{rd} phase, the upgrades for the electron/positron and muon programs will primarily focus on (i) improving the hermeticity of the detector by installing a new veto hadron calorimeter system; (ii) enhancing \acrfull{pid} capabilities with a new synchrotron radiation detector; and (iii) increasing the beam rate through upgraded read-out electronics. The best location for running in hadron mode is still under study, including the possibility to use the \acrshort{ps} T9 beamline. The target for true muonium production at H4 
is being optimized 
\cite{Gargiulo:2024zyc}.
With the projected $\approx 10^{13}$ electrons, $\approx 10^{11}$ positrons (at 40 and~\SI{60}{\GeV}) and $\approx2\times10^{13}$ muons accumulated on target,  the next stage of the experiment has the potential to probe entirely new regions of canonical \acrshort{ldm} models involving kinetic mixing, offering the possibility to decisively discover, or conclusively disprove, several well motivated models by the end of Run~4.

\subsubsection{NA66/AMBER -- Phase 2}

Phase 2 of the \acrshort{amber} experiment is planned to start in 2031 and will focus on worldwide unique measurements of the internal structure and the excitation spectrum of kaons and other hadrons. 

In the last 20 years, considerable progress has been made in our understanding of the excitation spectrum of light hadrons containing up and down quarks as well as heavy hadrons containing charm and bottom quarks. States beyond the ordinary quark model picture have been firmly established in both the light-quark and heavy-quark sectors. In stark contrast to this, the strange-quark sector has seen very little progress. Most of the experimental data on strange mesons are based on experiments that were performed more than 30 years ago. Only four additional kaon states were included in the \acrfull{pdg} listings since 1990. Many claimed states still need confirmation. In order to progress in our understanding of the internal structure of hadrons, it is extremely important to bridge the gap between the light up-/down-quark and the heavy-quark hadrons. Using a high-intensity kaon beam, \acrshort{amber} will perform unprecedented measurements of the full kaon spectrum up to masses of about $3\,\mathrm{GeV}/c^2$. The goal is to collect at least $2\times 10^7$ exclusive events for the $K^-\pi^-\pi^+$ final state with kaon beam. This will allow for precision measurements of strange mesons and the identification of possible strange exotic states, e.g., multi-quark states or states with gluonic excitations. In addition, the new data will have impact on other fields, e.g., \acrshort{cp}-violation studies in multi-body heavy-meson decays at $B$ factories, where excited strange mesons appear as intermediate states.

Understanding the emergence of hadron sizes is a crucial measurement in hadron physics and beyond. These radius measurement can be performed in inverse kinematics, i.e., scattering of a hadron off an electron in the target material. \acrshort{amber} -- Phase 2 aims at determining the kaon charge radius with a factor 10 better precision than currently known. 
Since pions are predominant in the secondary hadron beam delivered to the experiment, the measurement on the pion can be done in parallel with high statistics. At the same time, this opens the opportunity for a first determination of the radius of an anti-nucleus, namely the antiproton.

The internal structure of pions and kaons is of great theoretical interest due to their dual roles as Goldstone bosons and quark-antiquark bound states~\cite{Chang:2024rbs}. Nowadays, the experimentally available information on the partonic structure of the kaon is extremely limited. It consists of the $K^-$--induced \acrshort{dy} data by the NA3 experiment based on $\sim$700 events~\cite{Saclay-CERN-CollegedeFrance-EcolePoly-Orsay:1980fhh}. The analysis based on these data revealed evidence of the softer behavior of the valence $\bar u$ quark distribution of $K^-$ in comparison to the one in $\pi^-$. This difference was attributed to the breaking of the flavor SU(3) symmetry, resulting in a larger fraction of the kaon’s momentum being carried by the $s$ quark than the lighter $\bar u$ quark~\cite{Chang:2024rbs, Bourrely:2023yzi}, which has triggered intense theoretical investigations of the quark and gluon content of kaons. 
To study the quark and gluon component of the kaon, a measurement of the prompt photon production with a composite hadron beam at \acrshort{amber} is proposed. A system of three electromagnetic calorimeters is able to cover the kinematic region $-0.5 < x_F < 0.7$ and $p_T>3$ GeV/$c$. The $K^+$--induced prompt-photon production via gluon Compton scattering at negative $x_F$  provides information about the gluon content of kaons while the cross-section difference for negative and positive kaons is proportional to the valence $\bar{u}$-quark contribution. 

All three above-mentioned physics programs require the availability of positive and negative hadron beams with medium and high intensity of up to $10^9$ hadrons per spill delivered to the \acrshort{amber} spectrometer in \acrshort{ehn2}.   The hadron beam at the M2 beamline has a natural mix of protons, kaons, and pions requiring efficient and reliable beam particle identification. To allow for the delivery of the high-intensity kaon beams and maximum tagging efficiency, upgrades on the beamline and spectrometer shielding as well as improvements on the beamline vacuum are needed. The full physics program of \acrshort{amber} -- Phase 2 is expected to cover running also post-\acrshort{ls4}.

\subsubsection{SHiP/NA67 in ECN3}
\label{sec:ship-intro}

\acrshort{ship} will be installed in \acrshort{ecn3} and will operate with $4\times 10^{19}$ \acrshort{pot}/year colliding with a thick, high-density target. The target is followed by background-reducing systems -- a hadron absorber and a muon deflector (or muon shield) -- and two \acrshort{ship} detectors. The first one is the \acrfull{snd}. It is designed to detect neutrino interactions and scattering events of hypothetical \acrshort{ldm} particles produced in proton-target collisions. The second detector is the \acrfull{hsds}, aimed at searching for decays of \acrshort{fips}. It is made of a~\SI{50}{\m}-long decay volume, a magnetic spectrometer based on straw technology, and an electromagnetic calorimeter that also serves as a particle identification system. The hadron absorber, the muon shield, veto systems surrounding the \acrshort{hsds}, and the precise straw tracker allow for the reduction of the backgrounds in the \acrshort{hsds} down to a negligible level.

The validity of the predictions of the \acrshort{ship} simulation framework on the particle fluxes has been verified by comparing~\cite{SHiP:2020hyy} them to the data from the \acrfull{charme}~\cite{CHARM:1985anb} and a dedicated experiment performed in 2018 with a replica of the  \acrshort{bdf}/\acrshort{ship} target~\cite{vanHerwijnen:2267770}. With a total of $6\times 10^{20}$ \acrshort{pot}, generating large amounts of heavy hadrons such as $D$ and $B$ mesons, and thanks to efficient background reduction techniques, \acrshort{ship} has a rich physics case that covers neutrino physics and will explore previously inaccessible regions of the \acrshort{fips} parameter space in the~\SI{}{\GeV} mass range, complementing ongoing searches at the \acrshort{lhc} and future searches at \acrshort{fccee}, which mainly focus on the energy frontier and precision physics.

Neutrino physics is explored with the \acrshort{snd}. It comprises measurements of neutrino \acrshort{dis} cross section in the range~\SIrange[range-units=single,range-phrase=--]{5}{100}{\GeV}; the structure functions $F_{4},F_{5}$ parametrizing the differential cross section; tests of the \acrshort{lfu}; precision measurements of the \acrfull{ckm} matrix elements, and neutrino-induced charm production. It especially benefits from the large flux of $\tau$ neutrinos and antineutrinos.

The new physics case may be generically split into exploring scattering signatures, to be studied by \acrshort{snd}, and decay signatures, to be searched for at the \acrshort{hsds} detector. The scattering case covers models of \acrshort{ldm}, both elastic and inelastic~\cite{SHiP:2020noy}, as well as millicharged particles~\cite{Ferrillo:2023hhg}. The decay signatures cover various portal models adding unstable particles~\cite{Beacham:2019nyx}, including (but not restricted to) \acrfull{hnls}, dark photons and other vector mediators, Higgs-like scalars, and \acrshort{alps} ~\cite{Albanese:2878604,SHiP:2018xqw,SHiP:2020vbd}. 

Depending on the new physics model, \acrshort{ship} may explore the parameter space orders of magnitude larger than the currently excluded couplings in the mass range up to $\mathcal{O}(10\text{ GeV})$, being complementary to the main program of \acrshort{atlas}, \acrshort{cms}, and \acrshort{lhcb} experiments. Moreover, the detector will allow for careful reconstruction of the events' kinematics and identification of the particles' type, which makes it possible not only to push the domain of excluded couplings but also to obtain important insights in the case of discovery. For example, in case of a discovery of \acrfull{llps}, \acrshort{ship} will be able to check if they are consistent with the resolution of \acrshort{bsm} problems~\cite{Tastet:2019nqj,Mikulenko:2023iqq}.

\subsubsection{MUonE}
\label{sec:MUonE-intro}

The \acrshort{muone} project~\cite{CarloniCalame:2015obs,MUonE:2016hru} aims to shed light on the potential discrepancy between the theory prediction of the muon's anomalous magnetic moment and its measurement results. Specifically, \acrshort{muone} plans to measure the terms related to the \acrfull{hvp}, which are one of the limiting factors in the precise calculation of the anomalous magnetic moment, by measuring the hadronic contribution to the running of the electromagnetic coupling constant $\alpha$ in elastic high-energy muon-electron scattering. It is thus entirely complementary to experimental efforts using \(e^+e^-\) annihilation into hadrons, including independent systematic uncertainties.

In 2025, \acrshort{muone} will perform a test run with a complete scaled-down setup of the final detector with a possible sensitivity to the hadronic corrections to the running of $\alpha$ \cite{Hall:2896293}.
The \SI{160}{\GeV\per\\c} M2 muon beam, with its present performance and parameters, is adequate for this first measurement.
The apparatus will consist of three silicon-based tracking stations (the final experiment will have 40 of them), the calorimeter of 25 PbWO$_4$ crystals in use now, and a muon identification downstream the calorimeter.
An upgrade of the present hardware of the \acrfull{bms}, which measures the incoming muon energy, is in preparation to improve the accuracy of this muon momentum determination.

Discussions are ongoing within the collaboration about the possibility and the opportunity to run in 2026. The possibility to operate at different beam momenta (e.g., \SI{60}{\GeV\per\c} and \SI{200}{\GeV\per\c}) to improve the control of the systematic uncertainties is being investigated.

If the final proposal will be approved, the experiment will run after \acrshort{ls3}, with the ultimate goal of achieving a 0.3 \% statistical accuracy on the leading-order hadronic vacuum polarization contribution \(a_\mu^\text{HLO}\) to the muon's anomalous magnetic moment. For that purpose, the collaboration is considering two possible upgrades:
\begin{itemize}
 \item increase the beam intensity by at least a factor of two, up to about \((4 - 5) \times 10^8 ~\mu\)/spill;
 \item thermalise the beam tunnel (\(\pm 2^{\circ}\)C or better), for a length of~\SI{50}{\m} and an estimated volume of \SI{800}{\cubic\metre}.    
\end{itemize}

\subsubsection{DICE/NA60+}

The \acrshort{sps} can provide intense beams ($\approx 10^7$/spill) of ions at \acrshort{com} energies 6 GeV $<\sqrt{s_{\rm NN}}<17$ GeV. It represents the ideal location for an experiment aiming at the measurement of rare probes of the \acrshort{qgp}, and in particular of heavy quarks and thermal lepton pairs. Such measurements allow an investigation of the order of the phase transition to the \acrshort{qgp},  restoration of the \acrshort{qcd} chiral symmetry and modification of the hadron spectrum in its vicinity. Studies of various \acrshort{qgp} properties, such as its diffusion coefficients and the modification of the \acrshort{qcd} binding force from large to small distances can also be carried out. 
None of these studies were performed until now below top \acrshort{sps} energy with a decent integrated luminosity, nor will they become possible in the foreseeable future at any other facility in the collision energy range under discussion. On the contrary, this research domain is complementary to that already explored at hadron colliders (\acrshort{rhic}, \acrshort{lhc}), in particular for heavy-quark measurements, and to that expected from lower-energy facilities [\acrfull{fair}], in particular for dilepton-related observables. The use of a heavy-nucleus beam (Pb) is mandatory for these measurements, to ensure the largest size and lifetime of the \acrshort{qgp} droplet that will possibly be formed.

The \acrshort{na60+} project proposes a setup that includes a muon spectrometer coupled to a vertex spectrometer to perform the measurements sketched above~\cite{NA60+LoI}. The need for a high-interaction rate (\SI{e5}{\second^{-1}}) poses significant challenges on (i) the availability of focused high-intensity Pb beams down to low \acrshort{sps} energy, and of proton beams at the same energies, which represent a needed reference; (ii) use of state-of-the-art large area \acrfull{maps} that can be operated in the large charged-hadron multiplicity of Pb--Pb collisions, reaching hundreds of particles.
In addition, an essential requisite for the realization of the experiment are two large-volume dipole magnets needed for the two spectrometers. Both magnets are available at \acrshort{cern}.

After submission of an \acrshort{eoi} in 2019~\cite{Dahms:2673280} and the discussion of an \acrshort{loi} in February 2023~\cite{NA60+LoI}, the collaboration aims to submit a proposal to the \acrshort{spsc} by mid-2025. According to this proposal, data-taking is expected to start in 2029/2030 and to last approximately seven years, with a run with a Pb beam at a different energy each year and periods with proton beams, exploring at least three different energies and with an equivalent integrated luminosity per nucleon-nucleon collision.
The technology challenges are mainly related to the project of the \acrshort{maps} for the vertex detector, with corresponding \acrshort{rd} being carried out in synergy with the \acrshort{alice} \acrfull{its3} project~\cite{The:2890181}, which has the same timeline. For the muon spectrometer, \acrfull{mwpc} detectors will be used, covering a total surface of about~\SIrange[range-units=single,range-phrase=--]{60}{70}{\meter^2}. The collaboration is in the process of strengthening its manpower to ensure that all the items are duly covered. In particular, discussions to extend the participation of groups from the \acrshort{us} are currently in progress.

\subsection{HL-LHC -- Physics in the forward direction (neutrinos and FIPs)}
\label{sec:LHC-Forward}
\subsubsection{Forward Physics Facility}

The \acrfull{fpf} is a proposal to build a new underground cavern at \acrshort{cern} to house a suite of far forward experiments during the \acrshort{hllhc} era~\cite{FPFWorkingGroups:2025rsc,Anchordoqui:2021ghd,Feng:2022inv,Adhikary:2024nlv}.
These experiments will cover the blind spots of the existing \acrshort{lhc} detectors and will maximise the physics potential in the very forward region of the \acrshort{lhc} collisions. 
The physics program of the \acrshort{fpf} covers searches for \acrshort{bsm} particles, neutrino physics, and \acrshort{qcd} with important implications for astro-particle physics.  The \acrshort{fpf} can discover a wide variety of new particles that cannot be discovered at fixed target facilities or other \acrshort{lhc} experiments. 
In the event of a discovery, the \acrshort{fpf}, with other experiments, will play an essential role in determining the precise nature of the new physics and its possible connection to the dark universe.  In addition, the \acrshort{fpf} is the only facility that will be able to detect millions of neutrinos with~\SI{}{\TeV} energies, enabling precision probes of neutrino properties for all three flavors. 
These neutrinos will also sharpen our understanding of proton and nuclear structure, enhancing the power of new particle searches at \acrshort{atlas} and \acrshort{cms}, and enabling \acrfull{icecube}, \acrfull{auger}, \acrfull{km3net} and other astro-particle experiments to make the most of the new era of multi-messenger astronomy. 

An extensive site selection study has been conducted and the resulting site is shown in Figure~\ref{fig:ExecutiveSummaryMap}. The facility dimensions have been optimized following integration studies to demonstrate that it can house the proposed detectors and the main technical infrastructure~\cite{FPFPBCnote2}. The \acrshort{fpf} location is shielded from the \acrshort{atlas} \acrshort{ip} by over~\SI{200}{\m} of rock, providing an ideal location to search for rare processes and very weakly interacting particles.  Vibration~\cite{vibration-note}, radiation~\cite{FPFPBCnote}, and safety studies~\cite{FPFPBCnote} have shown that the \acrshort{fpf} can be constructed independently of the \acrshort{lhc} without interfering with its operation.  A core sample, taken along the location of the~\SI{88}{\m}-deep shaft to provide information about the geological conditions, has confirmed that the site is suitable for construction~\cite{FPFPBCnote2}. 
\acrshort{rp}~studies~\cite{FPFPBCnote} have concluded that the facility can be safely accessed with appropriate controls during beam operations.

The \acrshort{fpf} is uniquely suited to explore physics in the forward region because it will house a diverse set of experiments based on different detector technologies and optimized for particular physics goals. The proposed experiments are shown in Figure~\ref{fig:ExecutiveSummaryMap} and include:
\begin{itemize}
\item \acrfull{faser2}, a magnetic tracking spectrometer, designed to search for light and weakly-interacting states, including new force carriers, sterile neutrinos, \acrshort{alps}, and dark sector particles, and to distinguish $\nu$ and $\bar{\nu}$ charged current scattering in the upstream detectors. 
\item \acrfull{fasernu2}, an on-axis emulsion detector, with pseudorapidity range $\eta > 8.4$, that will detect $\approx 10^6$ neutrinos at~\SI{}{\TeV} energies with unparalleled spatial resolution, including several thousands of tau neutrinos, among the least well-studied of all the known particles.
\item \acrfull{flare}, a 10-ton-scale, noble liquid, fine-grained time projection chamber that will detect neutrinos and search for \acrshort{ldm} with high kinematic resolution, wide dynamic range and good particle-identification capabilities. 
\item \acrfull{formosa}, a detector composed of scintillating bars, with world-leading sensitivity to millicharged particles across a large range of masses~\cite{Foroughi-Abari:2020qar}.
\end{itemize}

\begin{figure}[h!]
\centering
\includegraphics[width=0.82\textwidth]{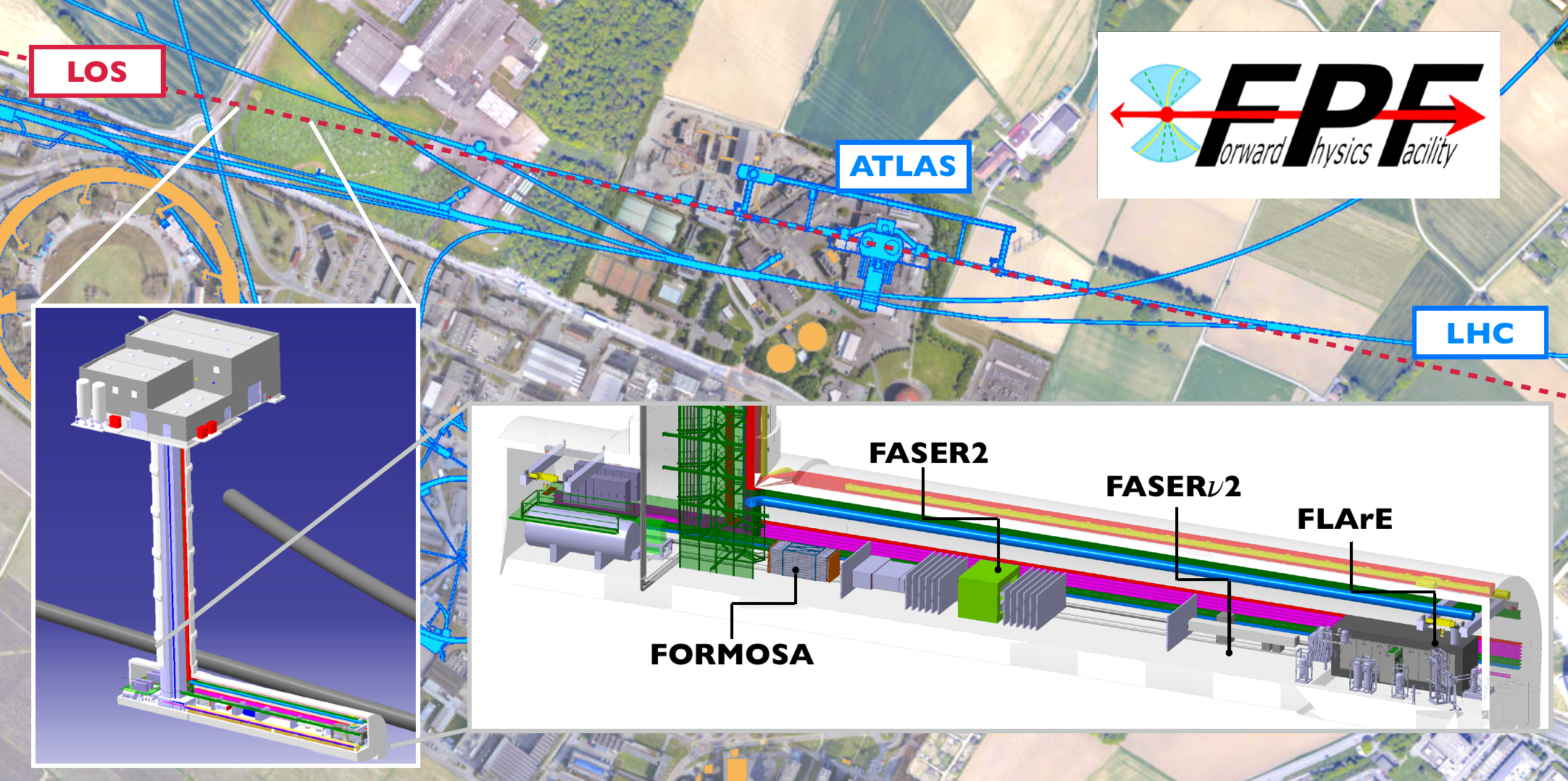}
\caption{The \acrshort{fpf} is located~\SIrange[range-units=single]{627}{702}{\meter} west of the \acrshort{atlas} \acrshort{ip} along the line of sight. The \acrshort{fpf} cavern, \SI{75}{\m} long and~\SI{12}{\m} wide, will house a diverse set of experiments to fully explore the forward region. (Figure taken from \cite{Adhikary:2024nlv}.)
}
\label{fig:ExecutiveSummaryMap}
\end{figure}

Three of the four proposed detectors have existing pathfinder experiments running at the \acrshort{lhc} [\acrshort{faser}, \acrshort{fasernu} and~the \acrfull{milliqan}~\cite{Haas:2014dda} --- see Section~\ref{sec:faser}],
and have released world-leading results, already demonstrating the ability to carry out searches and measurements in the forward region of the \acrshort{lhc}. 

All of the planned experiments are relatively small, low cost, require limited \acrshort{rd}, and can be constructed in a timely way. A Class 4 cost estimate for the Facility by the CERN engineering and technical teams is 35 MCHF for the construction of the new shaft and cavern~\cite{Adhikary:2024nlv} and 10 MCHF for the technical infrastructure.   
An estimate of the core costs for the full suite of experiments (without labour) is around 40 MCHF~\cite{Adhikary:2024nlv}. 
The \acrshort{fpf} requires no modifications to the \acrshort{lhc} and will support a sustainable experimental program, without additional power consumption for the beam beyond the existing \acrshort{lhc} program.  

To fully exploit the forward physics opportunities, the \acrshort{fpf} and its experiments should be ready for physics as early as possible in Run~4.  A possible timeline is for the \acrshort{fpf} cavern to be built during \acrshort{ls3}, the support services and experiments to be installed starting in 2030, and the experiments to begin taking data during Run~4. All of the experiments will be supported by international collaborations, and they will attract a large and diverse global community.
The \acrshort{fpf} is a mid-scale project composed of smaller experiments that can be realized on short and flexible timescales, which are well aligned with the \acrfull{astae} programme recommended from the \acrshort{us} \acrfull{p5} report~\cite{ASTAE}. In addition, it will provide a multitude of scientific and leadership opportunities for junior researchers, who can make important contributions from construction to data analysis in a single graduate student lifetime.

\subsubsection{Scattering and Neutrino detector at the HL-LHC (SND@HL-LHC)}

First results~\cite{SNDLHC:2023pun,SNDLHC:2024qqb} provide a clear picture of the signal and background environment for further exploitation of the neutrino physics potential at the \acrshort{lhc}. The physics program in Run~3 will be statistically limited in most  channels. The exploitation of the potential of the \acrshort{hllhc} with some key improvements will largely extend the physics reach of the experiment both in neutrino physics and in BSM searches.

The \acrshort{sndlhc} detector upgrade for the \acrshort{hllhc} is expected to observe over 10$^{4}$ neutrino interactions. This high statistics will allow for reducing the uncertainties in all of the main physics measurements using neutrinos: forward charm production, lepton flavour universality tests, and high energy neutrino interactions. The experiment is expected to observe a few hundred tau neutrino interactions and will possibly observe tau antineutrino interactions for the first time~\cite{Abbaneo:2926288}. 

The experiment's relatively large cross-sectional area of 40 by 40 cm$^2$ allows for a wide coverage in pseudo-rapidity ($6.9 < \eta  < 7.7$), which is complementary to all other HL-LHC experiments, including \acrshort{faser}. This wide $\eta$ coverage enhances the fraction of neutrinos originating in charm decays and allows for measurements differential in $\eta$ which are sensitive to the gluon \acrshort{pdf} in unexplored low-$x$ regions. In addition to the different $\eta$ coverage compared to \acrshort{faser}, the experiments also differ significantly in their implementation, with the \acrshort{sndhllhc} design including precise instrumentation of the neutrino target and of the hadronic calorimeter and muon spectrometer~\cite{Abbaneo:2926288}. This design, which also foresees high precision timing layers, provides the unique opportunity to correlate neutrino-like events with \acrshort{atlas} data via a dedicated trigger.

The detector will be located in the same \acrshort{ti}18 tunnel where the \acrshort{sndlhc} experiment currently operates.  The experiment will use silicon strip detector modules inherited from the \acrshort{cms} outer barrel tracker and is divided into two sections. The upstream section includes tungsten as the neutrino interaction target, and denser instrumentation. The downstream section includes magnetized iron as an absorber and has sparser instrumentation. The latter section of the experiment acts as a hadronic calorimeter and muon spectrometer. Fast detector planes using either plastic scintillator or resistive-plate chamber technology will be used to trigger the read out.

The large yield of neutrino events offers a unique opportunity to identify sub-samples of neutrino events in coincidence with charm hadrons detected in \acrshort{atlas}. While the charm hadron that decays into the neutrino detected in \acrshort{sndhllhc} is always outside of the \acrshort{atlas} detector acceptance, in about 11\% of these events the charm hadron produced in association with the neutrino parent is emitted within the acceptance of \acrshort{atlas}, and therefore it can, potentially, be detected.

A detailed design exists for the components of the detector and their integration. A prototype of a silicon strip layer has been assembled with \acrshort{cms} spare components, and timing detector components have been exposed to a hadron test-beam in 2024.  Significant synergy exists in detector \acrshort{rd} for the \acrshort{ship} experiment, which has a highly complementary physics case.

It is worth noting that a modest civil engineering work, with an excavation of~\SI{4.5}{\m^3}, would  improve by a factor $\approx7$ the accumulated statistics and hence the physics reach, in particular in the tau neutrino sector.

\subsection{HL-LHC -- FIPs Physics at large angle}
\label{sec:LHC-LargeAng}

The programme to search for \acrshort{llps} at \acrshort{atlas}, \acrshort{cms}, and \acrshort{lhcb} is vibrant and draws on the expertise of data analysts, instrumentation specialists, and theory experts~\cite{Alimena:2019zri}. The sensitivity of both \acrshort{atlas} and \acrshort{cms} to the decay-in-flight of \acrshort{llps} is greatest when these particles are relatively heavy (\mbox{$m_{llp} \gtrsim 10$ GeV}), though there are some important exceptions (e.g.,~\cite{CMS:2021sch,CMS:2021juv}). This is due to the presence of large backgrounds when exploring the low-mass region (below a few GeV) in a high-energy hadron collider, where data acquisition systems are overwhelmed with primarily hadronic activity. 
Transverse detectors that can probe \acrshort{fips} physics at large angles are imperative as they provide complementary sensitivity to \acrshort{ft} and forward detectors discussed in Sections \ref{sec:PSPSPS_Opportunities} and \ref{sec:LHC-Forward}, which kinematically cannot probe heavy \acrshort{llps} produced at the \acrfull{ew} scale and above. Furthermore, transverse detectors provide complementary sensitivity to the main \acrshort{lhc} detectors \acrshort{atlas}, \acrshort{cms}, and \acrshort{lhcb}, which cannot probe \acrshort{llps} with large decay lengths of 10 m and above. To achieve comprehensive coverage of the full \acrshort{llps} parameter landscape, and in particular, to have a large sensitivity to \acrshort{llps} produced from Higgs-boson decays, one or more high-volume, transverse \acrshort{llps} detectors are needed.

\subsubsection{ANUBIS}

The primary physics motivation of \acrfull{anubis} is providing unique sensitivity to massive \acrshort{llps} ($m_{llp}>$\SI{1}{\GeV}) with significant lifetimes ($c\tau>\BigOSI{e2}{\m}$) produced in high-$Q^2$ collisions at the \acrshort{ew} scale or above~\cite{Bauer:2019vqk,Satterthwaite:2839063}. 
Such models cannot be probed either at existing general purpose detectors like \acrshort{atlas} and \acrshort{cms} due to their finite geometrical acceptance nor at dedicated forward detectors like \acrshort{faser} or beam dump experiments like \acrshort{ship} as they cannot probe phenomena produced at an effective \acrshort{com} energy $\sqrt{s}\gtrsim 10~$~\SI{}{\GeV}. 

The \acrshort{anubis} and \acrshort{atlas} trigger and data acquisition system should be fully integrated allowing the full information about \acrshort{llps} candidate events registered by the two detectors to be exploited. This would provide unique sensitivity to the associated production of new particles together with a Higgs boson or $W$ or $Z$ bosons, enabling new searches for \acrshort{hnls}~\cite{Hirsch:2020klk}, \acrshort{alps}~\cite{Bauer:2017ris,Brivio:2017ije} and hidden photons~\cite{Argyropoulos:2021sav} that could not be performed otherwise. 
\acrshort{anubis} would extend the reach of all current \acrshort{lhc} detectors to \acrshort{llps} produced via the Higgs portal by 3 orders of magnitude~\cite{Satterthwaite:2839063}.

The detector requirements are (i)~high efficiency of $>98\%$ per detector layer, (ii)~angular resolution of $\lesssim~\SI{0.01}{\radian}$, (iii)~spatial resolution of $\lesssim~\SI{0.5}{\cm}$, and (iv)~time resolution of $\lesssim~\SI{0.5}{\ns}$. 
The large instrumented detector area of $\approx\SI{1600}{\m^2}$ imposes stringent constraints on the detector costs. 
\acrfull{rpc} is the optimal choice of technology given the requirements above.
For cost optimization and to benefit from a mature technology, a technology based on the \acrshort{atlas} Phase~2 \acrshort{rpc}s~\cite{CERN-LHCC-2017-017} is adopted.

The proposed implementation of the \acrshort{anubis} detector takes a staged approach:
\begin{itemize}
\item
First, a small-scale $\approx1\times2\times\SI{2}{\m^3}$ prototype, \acrfull{proanubis}, was commissioned in 2024 and is currently taking physics data in a representative location close to the ceiling of the \acrshort{atlas} cavern. 
The primary physics goals of \acrshort{proanubis} are: (i) directly measuring the expected background levels in combination with \acrshort{atlas}, and (ii) validating the experimental setup and its technology. 
\item
\acrshort{proanubis} is scheduled to run during Run~3, while \acrshort{rd} of the full \acrshort{anubis} detector electronics is performed and the civil engineering related to its installation is finalized. 
\item It is foreseen to partially deploy \acrshort{anubis} in \acrshort{ls3}, allowing for partial data-taking in Run~4 before the full deployment in \acrshort{ls4} for data-taking in Run~5.
\end{itemize}

Currently, \acrshort{proanubis} has collected~\SI{104}{\femto\barn^{-1}} of $pp$ collisions. 
This is being used to develop analysis techniques and detector calibrations that would allow for easier commissioning and early data-taking for the full detector. 
\acrshort{atlas} can be used as an active veto against high-energy charged tracks and ‘punch-through jets’ (jets that are not fully contained in the calorimeter) that are likely to produce displaced vertex candidates from hadronic interactions. Hence, displaced vertices aligned with punch-through jets and energetic charged tracks are vetoed.
Synchronization with \acrshort{atlas} would allow for a strong \acrshort{poc} for the active-veto concept and the expected background level, \acrshort{anubis} would operate under.

The \acrshort{rd} efforts focus on the detector geometry and sensor layout to optimize its performance, the identification of an eco-friendly gas mixture for operating the \acrshort{rpc}s, and on the serialization and daisy-chaining of the data acquisition system and its integration with the \acrshort{atlas} detector.

\subsubsection{CODEX-b}

The \acrfull{codexb} proposal is a low-cost transverse \acrshort{llps} detector option that uses existing technology and infrastructure.
\acrshort{codexb} is a special-purpose detector to be installed near the \acrshort{lhcb} \acrshort{ip}8 to search for displaced decays of exotic \acrshort{llps}~\cite{Gligorov:2017nwh,Aielli:2019ivi,Aielli:2022awh}. A recent \acrshort{eoi} presented the physics case with extensive experimental and simulation studies for the proposal~\cite{Aielli:2019ivi}.
The proposed \acrshort{codexb} detector would be located roughly 25 meters from \acrshort{ip}8 and have a nominal fiducial volume of $10\times10\times 10$~\SI{}{\m^3}. Backgrounds would be controlled by passive shielding provided by the existing concrete shielding wall in \acrfull{ux}85, combined with an array of active vetoes and passive shielding to be installed nearer to the interaction point. The installation of the detector and the passive shielding would not impact \acrshort{lhcb} operation.
The core advantages of \acrshort{codexb} are
\begin{enumerate}
\item competitive sensitivity to a wide range of \acrshort{bsm} \acrshort{llps} scenarios, exceeding or complementing the sensitivity of other dedicated \acrshort{llps} or general purpose detectors, both existing and proposed~\cite{Aielli:2019ivi};
\item a near-zero background environment, as well as an accessible experimental location with many of the necessary services already in place;
\item the ability to tag events of interest within the existing \acrshort{lhcb} detector, independently from the \acrshort{lhcb} physics program;
\item a relatively compact size, instrumented with low-cost  \acrshort{rpc}s, results in a modest cost. There is also the realistic possibility of extending detector capabilities for final-state neutral particles using technologies like scintillating tiles.
\end{enumerate}

A smaller proof-of-concept demonstrator detector, \mbox{CODEX-$\beta$}, consisting of 42 \acrshort{rpc} singlets, is currently being installed and will be integrated with \acrshort{lhcb} and operated during a fraction of Run~3~\cite{CODEX-b:2024tdl}. This demonstrator has been approved as an \acrshort{lhcb} \acrshort{rd} project and will be placed close to the proposed location of \acrshort{codexb}, shielded only by the existing concrete wall. The assembly and first commissioning of the \acrshort{rpc}s for \mbox{CODEX-$\beta$} is complete, and the first steps towards installation have been taken. Data-taking with \mbox{CODEX-$\beta$} will commence in 2025 and continue to the end of Run~3. A \acrshort{rpc} gas mixture compliant with  \acrshort{cern} regulations in terms of \acrfull{ghg} emissions will be used during the production of the \acrshort{codexb} detector.

\subsubsection{MAPP-2}

The physics programme of \acrfull{mapp2} focuses on searching for \acrshort{llps} and weakly interacting particle avatars of new \acrshort{bsm} physics at the \acrshort{hllhc}. \acrshort{mapp2} targets a wide range of exotic particles, including dark photons, new scalars, \acrshort{alps}, \acrshort{hnls}, and supersymmetric particles. By searching for \acrshort{llps} and weakly interacting particles, \acrshort{mapp2} addresses fundamental questions related to dark matter, new physics, and \acrshort{hs}s.

The \acrshort{mapp2} detector would be located in the UGC1 gallery near \acrshort{ip}8 and it would cover an intermediate pseudorapidity range (\(1.4 \le \eta \le 3\)) not addressed by other detectors dedicated to \acrshort{llps} searches. Additionally, \acrshort{mapp2}'s \SI{1000}{\m^3}, fully open, monitored decay zone would allow to detect charged particles with kinetic energies as low as~\SI{50}{\MeV} and photons with energies down to~\SI{100}{\MeV}, enabling the observation of low-mass \acrshort{llps} decays.
\acrshort{mapp2} would use position-sensitive scintillator technology instead of \acrshort{rpc}s, removing the need of using gases. The estimated cost of the detector is 6 to 7 MCHF.

\acrshort{mapp2} would consist of three position-sensitive scintillator layers arranged in 14 “houses” along UGC1's length, following its cross section and leaving the central decay region free of material. \acrshort{mapp2} covers a pseudorapidity range from ~1.3 to about 3. 
The detector is shielded from \acrshort{sm} backgrounds by at least \SI{24}{\m} of rock and concrete and from cosmic rays by about \SI{110}{\m} of rock.
Each \acrshort{mapp2} “house” contains \SI{1}{\m} $\times$ \SI{1}{\m} scintillator panels with embedded \acrfull{wls} fibers on both sides. These fibers form an X-Y grid with a \SI{2}{\cm} pitch, read out by \acrfull{sipm}. 
The electronic readout, software trigger, and calibration systems mirror those of the MAPP-1 detector presently installed in the UA83 gallery near \acrshort{ip}8.
\acrshort{mapp2} is expecting to integrate \SI{300}{\femto\barn^{-1}}during the \acrshort{hllhc} phase. 

A \acrshort{loi} for the \acrshort{lhcc} is in preparation and a \acrshort{tdr} will be submitted for approval in 2026.  
the MoEDAL-MAPP collaboration plans to test a \acrshort{mapp2} demonstrator unit in 2026/27 and to apply for funding to construct the full \acrshort{mapp2} detector during \acrshort{ls3}. In this scenario,  \acrshort{mapp2} data-taking would take place during Run~4 and beyond.

\subsubsection{MATHUSLA40}
The proposed \acrfull{mathusla40} experiment is a dedicated \acrshort{llps} detector for the \acrshort{hllhc}, situated on the surface near \acrshort{cms}. 
\acrshort{llps} that are produced at \acrshort{cms} \acrshort{ip}5, travel to the surface, and decaying in the \acrshort{mathusla40} decay volume would be reconstructed as displaced vertices by six plastic scintillator tracking layers installed in the ceiling and rear wall of the decay volume~(see Figure~\ref{fig:mathusla}). The primary physics target are \acrshort{llps} in the $\BigOSI{10}{\GeV}$ -- $\BigOSI{100}{\GeV}$ mass range with lifetimes $\gtrsim\SI{100}{\m}$ that decay hadronically.
This highly motivated~\cite{Curtin:2018mvb} new physics signal, including exotic Higgs decays to \acrshort{llps}, arises in many \acrshort{bsm} theories, including the Neutral Naturalness solutions to the little Hierarchy Problem~\cite{Chacko:2005pe, Craig:2015pha}. 
The \acrshort{lhc} is the only collider that can produce these \acrshort{llps} today, but they are in a blind spot for \acrshort{atlas} and \acrshort{cms} due to high backgrounds and severe trigger limitations \cite{Alimena:2021mdu}. 
\acrshort{mathusla40} could search for these \acrshort{llps} with near-zero background, with an ultimate sensitivity that surpasses the main detectors by orders of magnitude in \acrshort{llps} cross section and long lifetime.

\begin{figure}[b!]
\centering
\includegraphics[width=0.79\textwidth]{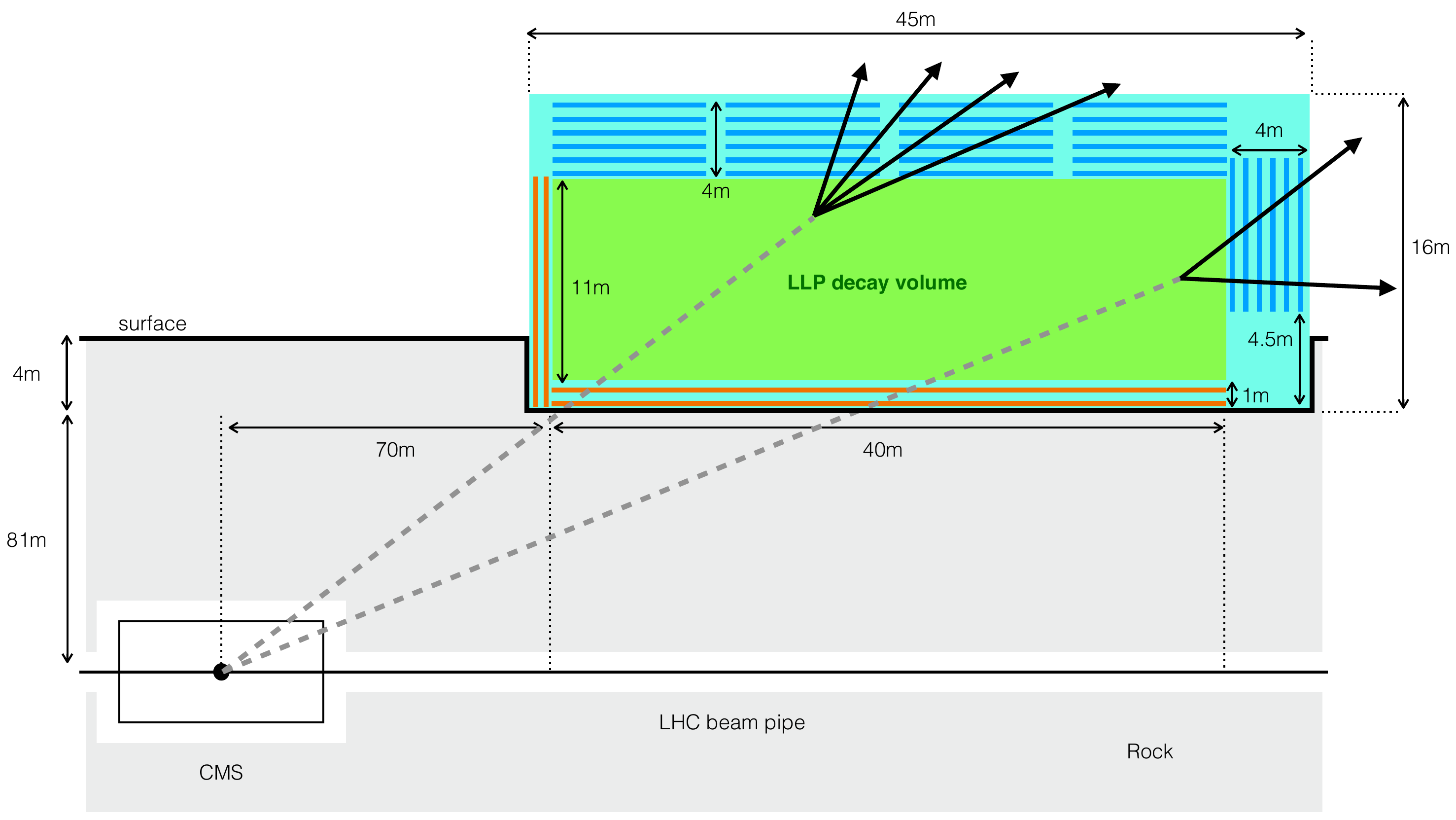}
\caption{
\acrshort{llps} (gray dashed lines) can decay into \acrshort{sm} charged states (black arrows) in the \acrshort{mathusla40} decay volume (green),
and be reconstructed as displaced vertices by the $9\times9\times\SI{4}{\m^3}$ tracking modules (thin blue rectangles), with veto layers (orange) aiding in background rejection. (Figure taken from \cite{Aitken:2025cjq}.) 
}
\label{fig:mathusla}
\end{figure}

The current proposal consists of an \acrshort{llps} decay volume with area $\approx\SI{40}{\m^2}$ and height $\approx\SI{11}{\m}$ (see Figure~\ref{fig:mathusla}). Veto detectors in the floor and front wall help reject cosmic, \acrshort{lhc} muon, and atmospheric backgrounds. A trigger system identifies upward traveling tracks, prompting the detector to be read out for permanent storage and offline analysis. The \acrshort{mathusla40} trigger system can also provide a \acrfull{l1} trigger signal to \acrshort{cms}, enabling correlated offline analyses to determine the underlying new physics model in the event of a discovery \cite{Barron:2020kfo}.

The \acrshort{mathusla40} collaboration presented an \acrshort{loi} to the \acrshort{lhcc} in 2018~\cite{alpigiani2018letter} and an update in 2020~\cite{MATHUSLA:2020uve}, and operated a test-stand on the surface above \acrshort{atlas} in 2018~\cite{Alidra:2020thg}.
Canadian \acrshort{mathusla40} groups constructed and operate \acrshort{rd} test-stands at the University of Toronto and the University of Victoria.
The collaboration developed a proposal for a~\SI{100}{\m}-scale detector towards the conceptual design stage, but over the last year, downscaled its proposal to the current~\SI{40}{\m} version in response to recent developments in the \acrshort{us} funding landscape~\cite{P5:2023wyd}, which made explicitly clear that the original proposal was too large to attract \acrshort{us} (and international) support. 
\acrshort{mathusla40}'s inherent modularity facilitates this rescaling, and also suggests a clear progression towards implementation of the full detector, by first building a~\SI{10}{\m} prototype which then becomes the first module of \acrshort{mathusla40}. 
This redesign is now complete, and reflected in the recently released \acrshort{mathusla40} \acrshort{cdr} \cite{Aitken:2025cjq} for an achievable 40~m geometry, covering the design, fabrication, and installation at \acrshort{cern} Point~5 \cite{MATHUSLACDR}. The collaboration aims to submit the proposal to the \acrshort{lhcc} in 2025.

\subsection{HL-LHC -- Fixed Target}
\label{sec:LHC-FT}

\subsubsection{ALADDIN}

\acrfull{aladdin} is a proposed fixed-target experiment at the \acrfull{ir}3 of the \acrshort{lhc} aiming to measure the \acrfull{edm} of charm baryons, such as $\Lambda_c^+$ and $\Xi_c^+$~\cite{Akiba:2905467}. These measurements are highly sensitive probes of the \acrshort{sm} and to potential new physics but remain unexplored due to the challenges imposed by the short lifetimes of these particles. \acrshort{aladdin} employs a technique using two bent crystals to channel forward-produced charm baryons from proton collisions, as sketched in Figure~\ref{fig:IR3layout}, inducing spin precession to enable the determination of their \acrfull{mdm} and \acrshort{edm}. The first crystal, with \SI{50}{\micro\radian} bending, splits part of the halo from the 7~TeV circulating proton beam, and steers it onto the target, itself paired to the second crystal with \SIrange[range-units=single,range-phrase=--]{5}{10}{\milli\radian} bending. It is followed by a compact detector including a spectrometer and a \acrfull{rich} detector for particle identification.
\acrshort{aladdin} not only provides a unique opportunity to probe uncharted charm baryon properties but also paves the way for innovative advancements in experimental techniques and detector technologies.
Its deployment during~Run~4 builds upon validation tests from machine experiment called \acrfull{twocryst} scheduled for 2025 to demonstrate the feasibility of the technique~\cite{Hermes:2025peb}. 

\begin{figure}[htb]
\centering
\includegraphics[width=0.99\textwidth]{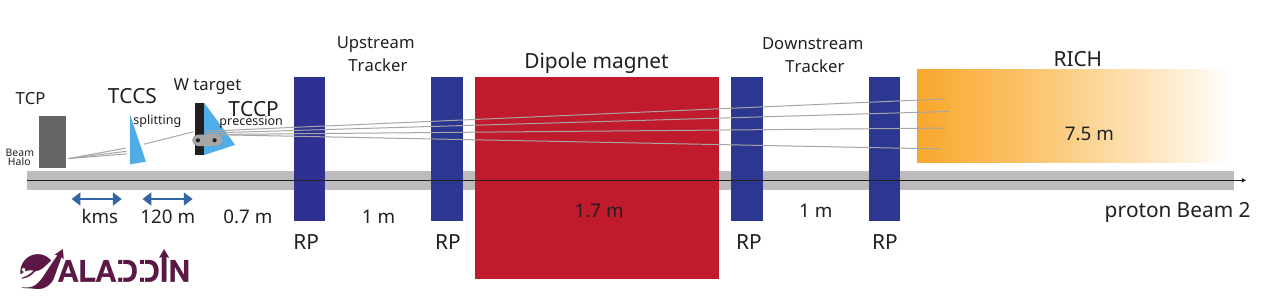}
\caption{Sketch of the \acrshort{aladdin} experiment (not to scale), illustrating the primary collimator (TCP), the splitting (TCCS) and precession (TCCP) crystals, the tungsten target just upstream of the TCCP, the spectrometer including the dipole magnet, and the upstream and downstream trackers, housed in Roman Pots, and the \acrshort{rich} detector. (Figure adapted from \cite{Akiba:2905467}.)} 
\label{fig:IR3layout}
\end{figure}

The operation of a double-crystal setup, parasitically to standard proton-proton collisions \cite{Mirarchi:2019vqi}, is both complex and challenging. A number of critical issues were identified early on by the \acrshort{lhcft} study team~\cite{Barschel:2653780}, which call for a direct demonstration of this scheme. 
The machine experiment \acrshort{twocryst} was conceived to validate the feasibility of this setup and to characterize precisely the performance reach in the \acrshort{hllhc} era. The \acrshort{twocryst} setup is installed in the \acrshort{lhc} \acrshort{ir}3 to address the performance in the multi-TeV range of the new long-crystals required for the charmed baryons precession experiments. The double-crystal setup has been installed in \acrshort{ir}3 together with two Roman Pots that will enable transverse profile monitoring of single- and double-channeled beams. The specifications of these crystals are identical to those planned for \acrshort{aladdin}. Two detector technologies will be applied: a fibre tracker system recovered from the \acrshort{atlas}-ALFA experiment and a pixel detector derived from the \acrshort{lhcb} \acrshort{velo}, which is close to the technology for the final experiment.  

The goal is to exploit the last two years of operation of the \acrshort{lhc} before the \acrshort{ls3} to collect important data as an input to the future \acrshort{mdm}/\acrshort{edm} experiment: performance of the long \SI{7}{\milli\radian} crystal in the multi-TeV energy range; operational aspects of the double-crystal setup; background checks with the final sensors. An ambitious \acrshort{md} plan is expected in 2025 and 2026, with the goal of performing the relevant measurements to establish the achievable \acrshort{pot} and the rate of channeled $\Lambda_c^+$ that the machine can deliver. 

\subsubsection{LHCspin}
Building on the experience obtained with \acrshort{smog2} (see Section~\ref{sec:SMOG2}), a proposal called \acrshort{lhcspin}~\cite{lhcspinproject, DiNezza:2022V7} is being studied which envisages to install a transversely 
polarized hydrogen (and deuterium) target inside the \acrshort{lhc}. This setup would facilitate unique
spin-dependent measurements of \acrshort{dy} and heavy-flavour production processes, by using both proton and lead beams, providing insight into details of the nucleonic transverse momentum-dependent parton 
distributions~\cite{TMDHavakian, Hadjidakis_2021, Gross_2023}.
The setup would use an \acrfull{abs} to inject a nuclear-polarized H or D beam into a storage cell 
just in front of the \acrshort{lhcb} \acrshort{velo}.
The main challenge lies in the ability to maintain a high nuclear target polarization inside the storage cell,
which requires finding an adequate coating of the inner cell wall compatible with \acrshort{lhc} operation. 
Measurements performed with an amorphous carbon coating produced promising results which revealed that a 
high degree of nuclear polarization is 
maintained despite a substantial hydrogen recombination rate~\cite{ELKORDY2024169707}, thus providing an 
almost ideal storage cell target (molecular hydrogen, being slower than atomic hydrogen, produces a denser 
target). However, if a molecule-dominated nuclear-polarized hydrogen target is used, a technique must be 
developed to measure its polarization in situ. This could be achieved by using a calibrated single-spin 
asymmetry, such as in elastic scattering near the Coulomb-nuclear interference region~\cite{Buttimore_2001,ZELENSKI2005248,Poblaguev_2020,p-Carbon-Huang}. Therefore, an auxiliary experiment 
is also proposed~\cite{IR4-LHCspin} which would be installed in \acrshort{ir}4 and which 
would measure the asymmetry of a free atomic beam of known polarization. The details of this setup are being 
developed and include the integration with the \acrshort{lhc} beam pipe of an \acrshort{abs}, 
a Breit-Rabi--type polarimeter and large-angle (\SI{}{\MeV} range) proton recoil detectors.

\subsection{ISOLDE}
\subsubsection{Symmetry violating effects in radioactive atoms and molecules}
\label{sec:radiomolecules}
The unique environments within heavy, radioactive atoms and molecules provide opportunities to probe \acrshort{bsm} physics
~\cite{Safronova2018}. \acrshort{cp}-violating interactions manifest in these systems as observable \acrfull{em} moments such as an\acrshort{edm}. These measurements of \acrshort{cp}-violation in \acrshort{em} moments can provide sensitivity to new particles and interactions on energy scales comparable to those achievable at the \acrshort{lhc} and the \acrshort{fcc}~\cite{Roussy2023}, enabling discovery of new physics or imposing extremely stringent constraints on models of new interactions~\cite{radmols2024}.

To disentangle possible sources of \acrshort{cp}-violation, multiple searches are required in complementary probe systems. The aim of the radioactive molecules initiative is firstly to develop a catalogue of probe systems at \acrshort{isolde} and enable preliminary characterization at the existing experimental facility. Secondly, the initiative aims to make a selection of long-lived radioactive probes available to offline experiments. Finally, investigations are underway towards the feasibility of adapting the high-precision techniques used for stable molecules in offline symmetry-violating experiments towards online experiments with radioactive probes.

\subsection{AD and ELENA Complex}

The Antimatter Factory at \acrshort{cern} is unmatched worldwide in producing low-energy antiprotons ($\bar p$). It supports over 60~institutes and about 350~scientists organised into six collaborations, which conduct precise comparisons of the fundamental properties of stable matter-antimatter conjugates at low energies, and study antimatter gravity~\cite{Carli:2022htc}.

With increased $\bar p$ flux and improved beam availability, \acrshort{elena} has supported groundbreaking studies in fundamental physics.
Highlights include the most precise comparisons of fundamental properties of protons and antiprotons~\cite{BASE:2022yvh,BASE:2016yuo}, the first observation of the ballistic behaviour of antihydrogen in the gravitational field of the earth~\cite{ALPHA:2023dec}, the 1S-2S spectroscopy of laser-cooled antihydrogen atoms~\cite{Ahmadi:2018eca,ALPHA:2021wcn}, and the antiproton-to-electron mass ratio measurements in buffer-gas cooling of antiprotonic helium~\cite{ASACUSA:2016xeq}. 
Some of the collaborations are progressing towards production of an anti-hydrogen beam~\cite{ASACUSA:2014mse} and anti-hydrogen production via charge-exchange~\cite{AEgIS:2023xnq,Adrich:2023tua}. The recently reported observation of positronium laser cooling~\cite{AEgIS:2023lpw} is a crucial step towards this goal.
Two collaborations, \acrfull{base} and \acrfull{puma}, are working toward transporting antiprotons out of the \acrshort{elena} facility, and \acrshort{base}-\acrfull{step} reported very recently on first transport of protons~\cite{BASE:2025_p_transport}. \acrshort{puma} will soon move the antiprotons from \acrshort{elena} to \acrshort{isolde} aiming at probing the surface properties of stable and rare isotopes using low-energy antiprotons~\cite{PUMA:2022ngr}.

Looking ahead, physicists are interested in studying the properties of the antineutron and strongly interacting antimatter with high precision. 
In another attempt, some \acrshort{ad} physicists are currently looking into the production and spectroscopy of anti-hydrogen molecular ions. These techniques have potential to study fundamental antimatter properties with resolutions on the $10^{-16}$-level and below~\cite{Myers:2018ntk}, which could pave the way for novel experiments in the long-term future~\cite{AD_ELENA_Users_Contributions_ESG}.

In response to those needs, the feasibility of producing anti-deuterons at the \acrshort{ad} target is being explored. 
Further upgrade studies being considered include the ability to vary the $\bar p$ extraction energy and the introduction of slow extraction, aiming at antineutron-related studies by the current users as well as to broaden the facility’s scientific reach beyond the present user community.

\subsection{Non-accelerator experiments}
\label{subsubsec:non-acc_exp}
\subsubsection{Atomic Interferometer@CERN}
\acrfull{ai} is a promising quantum sensing technology~\cite{Buchmueller:2023nll} with great promise for searching for \acrfull{uldm} and measuring \acrfull{gws}.
The sensitivity of \acrshort{ai} is maximised in experiments manipulating with the same laser two or more atom clouds that are as widely separated as possible. Two geometric configurations are being considered, using either a vertical shaft or a horizontal gallery, with the former being pursued by the \acrfull{magis} experiment~\cite{MAGIS-100:2021etm} in the \acrshort{us} and the \acrfull{aion} project~\cite{Badurina:2019hst} in the UK. The \acrshort{aion} Collaboration has already demonstrated quantum interference between superposed clouds of atoms and the cancellation of laser noise between the interference patterns in a pair of $^{87}$Sr clouds manipulated by the same laser in a vertical gradiometer configuration. The \acrshort{aion} collaboration has proposed to build a \SI{10}{\m} prototype in the UK, to be followed by full-scale~\SI{100}{\m} and~\SI{1}{\km} experiments~\cite{Badurina:2019hst}. 

One of the \acrshort{cern} access shafts, PX46, has been identified as a possible site for a~\SI{100}{\m} \acrshort{ai} detector based on \acrshort{aion} technology, see Figure~\ref{fig:PX46situation}, and a conceptual feasibility study~\cite{Arduini:2023wce} has established its suitability as far as infrastructure, seismic stability and electromagnetic interference are concerned. The installation in PX46 of an international experiment is currently being discussed within the framework of the \acrfull{tvlbai} proto-collaboration~\cite{Abdalla:2024sst}. Assuming successful evolution of the current technical \acrshort{rd} programme and the availability of funding, it is suggested to undertake preparatory work at Point 4 during \acrshort{ls3} to enable the subsequent installation and operation of the \acrshort{ai} detector in PX46 without compromising \acrshort{lhc} operation. The preparatory work would consist primarily of installing radiation shielding at the base of PX46 and an access control system.

\begin{figure}[h!]
\centering

\includegraphics[width=1.0\textwidth]{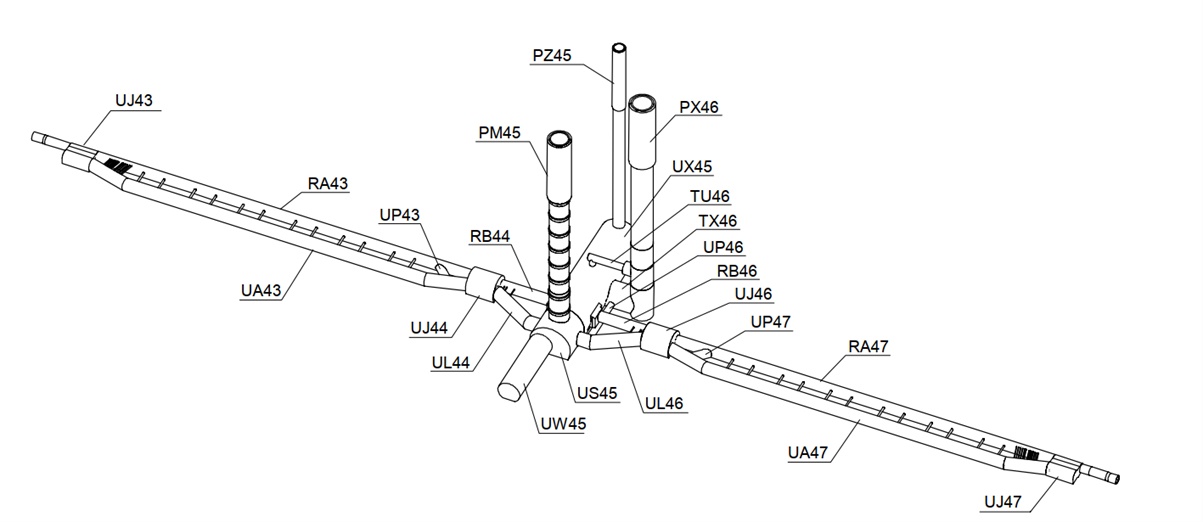}

\caption{Schematic drawing of the civil engineering infrastructure at Point~4 on the \acrshort{lhc}, including the \acrshort{lhc} tunnel, the PX46 shaft and other underground locations~\cite{Arduini:2023wce}.}
\label{fig:PX46situation}
\end{figure}

No equally suitable European site for a \SI{100}{\m} vertical \acrshort{ai} experiment has yet been identified, and this project would maximise the exploitation of unique \acrshort{cern} infrastructure to make world-leading probes of \acrshort{uldm} and \acrshort{gws}.

\subsubsection{Axion Heterodyne Detection}
The proposal in~\cite{Berlin:2019ahk, Berlin:2020vrk} for the heterodyne detection of axion \acrshort{dm} in microwave cavities was realized in the form of two prototypes that are currently taking data at \acrfull{slac}~\cite{SLACLDRD} and within the \acrfull{sqms} initiative at \acrshort{fnal}~\cite{Giaccone:2022pke}. A third prototype has been funded by the \acrfull{qti}, but has not been built, yet.

The basic idea consists in designing a cavity with two modes that have a small frequency splitting compared to their frequencies. Their geometry is designed to maximize photon transitions between them when an axion \acrshort{dm} background is present and one of the two modes is loaded. This setup gives a large parametric advantage, of order $1/(m_a L)^2$ in signal power compared to traditional resonant searches, when exploring small axion masses $m_a < L^{-1}$, where $L$ is the size of the cavity.

Preliminary results of the prototype at \acrshort{slac} have demonstrated a good mode separation in the readout of the signal mode and the possibility to tune the frequency splitting between the modes by about a~\SI{}{\MHz}. The recorded mode separation is better than a part in $10^{7}$, i.e., if the loaded mode is readout through the signal mode port, less than $10^{-7}$ of its input power is measured. These preliminary results are still unpublished but demonstrate the key features needed to realize a full future experiment~\cite{Berlin:2019ahk, Berlin:2020vrk}. The \acrshort{slac} prototype is a~\SI{50}{\liter}-copper cavity, while the \acrshort{qti} prototype will demonstrate the same features in a \acrshort{sc} cavity of similar size. It is projected to have sensitivity to unexplored axion \acrshort{dm} parameter space. 

The long-term goal of this initiative is to build a full scale experiment as proposed in~\cite{Berlin:2019ahk, Berlin:2020vrk}, based on a~\SI{}{\m^3} \acrshort{sc} cavity cooled to~\SI{2}{\kelvin}. A future experiment based on this idea can probe about two orders of magnitude in axion mass on the \acrshort{qcd} line~\cite{Berlin:2019ahk}, the prime target of all axion searches. Additionally, it is the only existing proposal that can probe in the laboratory the lightest viable \acrshort{dm} candidates~\cite{ Berlin:2020vrk}, whose De Broglie wavelength is comparable to the size of dwarf galaxies and whose coherence time is about ten times longer than recorded human history.

\subsubsection{FLASH}
\acrfull{flash}~\cite{Alesini:2023qed} is a large resonant-cavity haloscope in a high static magnetic field which is planned to probe new physics in the form of \acrshort{dm} axions, scalar fields, chameleons, hidden photons, as well as \acrfull{hfgw}. Concerning the \acrshort{qcd} axion, \acrshort{flash} will search for these particles as the \acrshort{dm} in the mass range~\SIrange[range-units=single,range-phrase=--]{0.49}{1.49}{\micro\eV}, thus filling the mass gap between the ranges covered by other planned searches. A dedicated microstrip \acrfull{squid} operating at ultra-cryogenic temperatures will amplify the signal. The frequency range accessible overlaps with the \acrfull{vhf} \acrshort{rf} spectrum and allows for a search of \acrshort{gws} in the frequency range~\SIrange[range-units=single,range-phrase=--]{100}{300}{\MHz}. The experiment will make use of the cryogenic plant and magnet of the \acrfull{finuda} experiment at \acrfull{infn} -- \acrfull{lnf} (Italy). Recently, an \acrfull{erc} Synergy Grant has been awarded to the project \textit{GravNet}, which has the ambition to search for \acrshort{hfgw} with a network of haloscopes, including \acrfull{quaxlnf}~\cite{PhysRevD.110.022008}, already operating at \acrshort{lnf}, and \acrshort{flash}.

\subsubsection{RADES}
\acrshort{rades} is a haloscope for axion \acrshort{dm} searches using \acrshort{rf} cavities in dipole magnets, with physics results obtained at around~\SIrange[range-units=single,range-phrase=--]{8}{9}{\GHz}~\cite{CAST:2020rlf,Ahyoune:2024klt}. The use of \acrshort{hts} tapes on the cavity walls, gives also a unique synergy with studies that explore \acrshort{hts} coating for the \acrfull{fcchh} beam-pipes~\cite{Dhar:2024xrx,Golm:2021ooj}. In addition, it has resulted in the development of a novel type of cavity tuning mechanism through a `split-cavity'~\cite{Golm:2023iwe}.

The \acrshort{erc} Synergy Grant \textit{Dark Quantum} and the \acrfull{qrades} Grant within QuantERA~\cite{quantera}, which have been awarded to \acrshort{rades} collaborators for the development of single photon detection in \acrshort{rades}, have started in 2024 with relevant contracts in place. In addition, many \acrshort{rades} collaborators are members of the \acrshort{iaxo} collaboration.
In 2024, a \acrshort{loi} has been submitted by \acrshort{rades} to the Canfranc Underground Laboratory (Spain) in order to run the experiment shielded from cosmic radiation.

The long-term goal of \acrshort{rades} will be the installation of a haloscope in the \acrshort{babyiaxo} magnet~\cite{Ahyoune:2023gfw}. In terms of axion-mass-range to be probed, there is a unique experimental reach in axion masses between \SIrange[range-units=single,range-phrase=\text{ and }]{1}{2}{\micro\eV} (except for \acrshort{flash}~\cite{Alesini:2023qed}). 
Within the \acrshort{erc} Synergy Grant there is also the plan to explore quantum-limited detection with magnetic-field-resilient transmons for axion masses corresponding to frequencies of~\SIrange[range-units=single,range-phrase=--]{10}{20}{\GHz}.

\subsection{PBC Opportunities at new Facilities}\label{sec:PBCatfuturefacilities}

\subsubsection{SBN beam performance}
\label{sec:sbn_exp}
The proposed novel \acrshort{sbn} beam facility introduced in Section~\ref{sec:sbn_acc} would provide an outstanding playground for precision neutrino physics. In order to evaluate its experimental performance, a reference experimental setup has been defined with realistic ingredients as detailed below (see Figure~\ref{fig:SBNlayout}).

The \acrshort{sbn} reference beam consists of \SI{9.6}{s} spills of $1.0\times 10^{13}$ protons and an \SI{8.5}{GeV} energy tune of the secondary meson narrow-band beam. A possible lower secondary energy tune at, e.g., \SI{4}{GeV}, optimized for \acrshort{hk}, is being evaluated \cite{SBN@CERN_ESPPU}. 
The calorimetric instrumentation of the decay tunnel monitors the $\nu_e$ flux by measuring the continuous yield of electrons produced in the kaon $K_{3e}$ decays. The instrumentation on the tunnel walls also monitors muons from $K_{2\mu}$, while a muon range-meter based on PICOSEC micromegas \cite{Kallitsopoulou:2024dmk} (instrumented hadron dump) can access muons from pion decays, thus monitoring the $\nu_\mu$ fluxes of the facility \cite{ENUBET:2023hgu}. Rate capability, radiation hardness, and ability to discriminate electrons from photons and pions are key specifications of these devices. The  \acrshort{sbn} reference calorimeter is based on the results of the prototype built and beam-tested within the \acrfull{np06}/\acrshort{enubet}~\cite{ENUBET:2023hgu}.
The tracking instrumentation of the beam and muon spectrometers is designed to measure individual 2-body decays of charged pions and kaons to tag each $\nu_\mu$ of the beam, similarly to what the current  \acrshort{na62} \acrshort{gtk} tracker performs. The peak meson rate at the entrance of the decay tunnel is \SI{20}{MHz/mm^2}, 10 times higher than the present \acrshort{na62} conditions. Rate capability, radiation hardness, time resolution, and material budget are key specifications of these devices. The \acrshort{sbn} reference trackers anticipate the expected performance of Si detectors developed for the \acrshort{hllhc} era~\cite{hep-ph_LHCb_2021}. 
The \acrshort{sbn} reference neutrino detector is a
ProtoDUNE-like LAr \acrfull{tpc} of \SI{500}{t} with a fiducial transverse size of $4\times$\SI{4}{m^2} and an entry plane located \SI{25}{m} from the end of the decay tunnel. With this detector statistical samples of about $10^4$ $\nu_e$ and $5\times 10^5$ $\nu_\mu$ \acrshort{cc} events are expected for $1.4\times 10^{19}$ \acrshort{pot}, corresponding to $\approx5$ years of data-taking. An additional water Cherenkov detector of interest for \acrshort{hk} is being considered \cite{SBN@CERN_ESPPU}.

The quantitative response of this reference setup has been simulated based on a detailed propagation of secondary particles in the optimized beam optics. Some key results are listed below:
\begin{itemize}

\item $\nu_e$ and $\nu_{\mu}$ flux: the simulation confirms that the flux can be controlled within 1~\%, in line with the expected statistical precision of the $\nu_e$ \acrshort{cc} total sample.

\item $\nu_\mu$ tagging: the $\nu_\mu$ \acrshort{cc} events in the detector are associated to the tagged $\pi$ and $K$ decays using time and space coincidence. The simulation shows that most of the observed events can be associated to a single decay, with mis-associations at the level of a few \%. In addition, $\pi$ and $K$ parents can be discriminated using the decay kinematics, which should allow for providing complementary control of the $K$ flux and hence the $\nu_e$ flux. A proof of principle of neutrino tagging was recently provided by  \acrshort{na62} with the first detection of a tagged neutrino~\cite{hep-ph_NA62Collaboration_2024}.  

\item $\nu_\mu$ energy resolution: with a detector close to the source, the transverse position of interacting vertices is correlated with the neutrino energy by the two-body kinematics of a Narrow-Band beam. The \acrfull{nboa} technique exploits this correlation, similarly to the approach currently used in experiments such as \acrfull{sbnd} at \acrshort{fnal}. \acrshort{nboa} provides a $\nu_\mu$ energy resolution of approximately 10~\%. For the tagged $\nu_\mu$ \acrshort{cc} sample the energy resolution is improved to the sub-\% level thanks to full reconstruction of the individual decays.

\end{itemize}

The physics impact of the \acrshort{sbn} expected from these experimental performances is discussed in Section~\ref{subsec:NU}.

\subsubsection{Gamma Factory research opportunities}

The \acrshort{gf} facility, introduced in  Section~\ref{sec:future_facilties},
proposes to extend the \acrshort{cern} scientific programme in the domains of particle, nuclear, atomic, fundamental, accelerator, and applied physics, delivering  technological leaps opening new research opportunities and creating potential for unexpected discoveries  in many  branches of physics.
Its  multidisciplinary research could re-use 
the existing \acrshort{cern} accelerator infrastructure, including \acrshort{lhc}. 
It could be conducted over the time interval between the end of \acrshort{hllhc} proton and ion collision programme, and the start of a  new, high-energy-frontier collider operation.

The \acrshort{gf} primary goal is to create novel research tools and novel research methods for the high-intensity frontier research. 
Several application domains of the \acrshort{gf} tools have already been studied: 
 
\begin{itemize}
\item 
{\bf Particle physics} \cite{Apyan:2022ysh,Krasny:2004ue,Placzek:2024bkv} -- precision \acrshort{qed} and \acrshort{ew} studies,  
Higgs boson studies in the $\gamma\gamma$ collision mode,  searches 
for very rare muon decays, precision studies in the neutrino sector, electron-proton collision physics at the \acrshort{lhc}, 
 \acrshort{qcd}-confinement studies;
\item
{\bf Nuclear physics} \cite{Budker:2021fts,Nichita:2021iwa,Budker:2022kwg} --  nuclear spectroscopy, studies of cross-talk of nuclear and atomic processes, 
nuclear photo-physics studies, photo-fission research, precision studies of rare radioactive nuclides;
\item
{\bf Atomic physics} \cite{Budker:2020zer,Bieron:2021ojp,Serbo:2021cps,Flambaum:2020bqi,Budker:2022kwg} -- studies of   highly charged atoms, high precision studies  of muonic and pionic atoms; 
\item
{\bf Fundamental   physics} \cite{Wojtsekhowski:2021xlh,Karbstein:2021otv,Balkin:2021jdr,Chakraborti:2021hfm,Budker:2022kwg} --  studies of the basic symmetries of the universe, vacuum birefringence studies, dark matter searches, atomic interferometry; 
\item
{\bf Accelerator physics} \cite{Krasny:2021llv,Zimmermann:2022xbv,Krasny:2020wgx,Cooke:2020arc, Zimmermann:2022svn,Granados:2024sht, Zimmermann:2024cnr,Zimmermann:2023rqo,Apyan:2022ysh} -- laser beam cooling techniques, high-intensity sources of polarised 
positrons and muons, beams of radioactive ions and neutrons, very narrow band, \acrshort{cp} and flavour-tagged neutrino beams;
\item
{\bf Applied physics} \cite{Baolong:2024ata} -- beam-driven energy sources,  fusion research, medical isotopes and isomers, precision lithography.
\end{itemize}

\subsubsection{Experiments at AWAKE}\label{sec:awake_physics}

The demonstration of the \acrshort{awake} programme technology (see Section~\ref{sec:awake}) would enable the upgrade and extension of the accelerator facility and the acceleration of electrons to~$\BigOSI{50}{\GeV}$
for use in particle physics experiments.  Some possibilities have already been discussed 
elsewhere~\cite{bib:edda-symmetry-2022,Caldwell:2018atq,Wing:2019rsta}, although this list is by no means exhaustive and other possibilities could 
provide a compelling particle physics motivation.  A brief summary of some of the experiments is given here, principally the possibility to search 
for dark photons in a \acrshort{bd} experiment or the investigation of strong-field \acrshort{qed} in electron--laser interactions.

Using protons from the \acrshort{sps}, experiments could be performed under the assumption of electrons delivered at \SI{50}{\GeV} in bunches with 
$5\times10^9$\,particles and a total running time of, e.g., 3 months giving 10$^{16}$ \acrfull{eot}~\cite{awake++}.  As this would be a \acrshort{bd} experiment, 
searches for dark photons would be performed in the $e^+e^-$ decay channel.  Given these parameters, the search can be extended to higher dark 
photon masses ($m_{A^\prime} \approx\SI{0.1}{\GeV}$) in the coupling range, $10^{-5} < \epsilon < 10^{-3}$, a region not covered by current searches.  If 
protons from the \acrshort{lhc} could be used to drive the wakefields,~\SI{}{\TeV} energy electrons could be produced and a similar search for dark photons would extend to even higher masses ($m_{A^\prime} \approx\SI{1}{\GeV}$) and for the same $\epsilon$ region, beyond any currently planned experiment.

The \acrfull{e144}~\cite{Bamber:1999zt} at \acrshort{slac} in the 1990s pioneered studies in strong-field \acrshort{qed}~\cite{Fedotov:2022ely} through investigation 
of electron--laser interactions.  Understanding how \acrshort{qed} becomes non-linear is relevant for a number of physical systems, such as in astrophysics, 
and can be investigated in the laboratory through experiments in which high-energy electrons and a high-power laser interact.  These studies have 
undergone a resurgence with the improvement in laser technology and the consequent increase in laser power.  Experiments could be performed with 
electrons of energy~\SI{50}{\GeV}, as in the \acrshort{e144}~experiment, and beyond other current experiments~\cite{e320,LUXE:2023crk}. As the rate will be limited by the~\SI{}{\Hz}-level laser system, high bunch rates for electrons are not required and so this is an ideal first experiment for the \acrshort{awake} facility.

\subsubsection{PBC opportunities at FCC-ee and LC}

Future colliders, such as the \acrshort{fccee} or a \acrshort{lc}, and their injectors, as described in Section~\ref{sec:futureCollidersAndInjectors}, will offer various options for \acrshort{pbc}. Both types of facilities have the potential to test \acrshort{qed} in the strong-field regime. This is possible due to the very high particle energy available of \SI{100}{\GeV} and beyond and the developments in the available peak intensity of high-power laser systems. Combining both in an electron-laser collision allows the investigation of the harmonic structure of the non-linear Breit--Wheeler pair production and eventually also delve into the fully non-perturbative domain of \acrshort{qed} where there is no reliable theory at all~\cite{Fedotov:2022ely}. A viable and sustainable approach to testing strong-field \acrshort{qed} is the use of oriented crystals. At \acrshort{fccee} energies, electrons and positrons traversing a crystal along its main lattice axes experience an average electric field in their rest frame that is amplified by a factor $\gamma$ due to Lorentz contraction. This enhancement brings the field to values comparable to the Schwinger critical field of \acrshort{qed}, beyond which electrodynamics becomes non-linear \cite{baier-1998,uggerhj-2005,baryshevskii-1989}. A linear collider additionally enables to explore strong-field \acrshort{qed} effects in the beam-beam interaction, which becomes particularly relevant in possible energy upgrades above the \SI{1}{\TeV} scale~\cite{Schulte:1999xb, Filipovic:2021xge}. Since strong-field \acrshort{qed} processes produce a high flux of energetic photons, they can also be further used in a \acrshort{ft} experiment to search for new physics~\cite{Bai:2021gbm,Schulthess:2025tct}. 

\paragraph{\acrshort{pbc} opportunities at \acrshort{fccee}}

The science opportunities of \acrshort{fccee} are detailed in~\cite{PBCFCCee:2025}. 
The high-intensity positron source from the \acrshort{fccee} booster, operating at~\SI{20}{\GeV} and possibly up to~\SI{46}{\GeV}, offers opportunities for dark sector searches and studies of true muonium. If positrons slow extraction can be realized, an experiment akin to \acrshort{na64} could significantly extend the sensitivity to \acrshort{ldm}, leveraging resonant annihilation to enhance reach in the dark photon parameter space.
True muonium, the yet-unobserved bound state of a muon and antimuon, presents a unique test for bound-state \acrshort{qed} and potential New Physics. Previous proposals suggested production at the \acrshort{sps} H4 beam line with \SI{43.7}{\GeV} positrons, requiring precise energy resolution. The \acrshort{fccee} booster could dramatically increase positron availability, enabling true-muonium production with a controlled energy spread and with expected production rates of $10^3-10^4$ true muonium atoms per day. This would allow studies of its decay properties, hyperfine structure, and possibly the Lamb shift.
The \acrshort{fccee} accelerator complex will provide high-energy, high-quality positron and electron beams, reaching energies up to \SI{183}{\GeV}, with an intense primary beam and low divergence. This unprecedented beam availability would significantly enhance the experimental performance and increase statistics, which is essential for advancing strong-field \acrshort{qed} studies in crystals. Indeed, it could enable the observation of spin dynamics effects in strong-field \acrshort{qed}. At \acrshort{fccee}, spin rotation via channeling in bent crystals would offer a unique opportunity to investigate strong-field \acrshort{qed} spin dynamics, particularly the significant reduction of the anomalous magnetic moment $\mu'$ in strong fields at high particle energies \cite{baryshevsky-1990,osti_6944261}. Thus, measuring the spin precession angle of positrons in a bent crystal at \acrshort{fccee} energies would provide a direct test of strong-field \acrshort{qed}. Bent crystals would not only facilitate the measurement of the drastic decrease in the positron (electron) magnetic moment, but also enable the observation of radiative self-polarization in strong fields \cite{tikhomirov-1993}, observing circularly polarized gamma-radiation by positrons, polarized electron-positron pair production by gamma-quanta and spin rotation in the circularly polarized crystal field harmonics \cite{VTikhomirov-1996,tikhomirov-1996,tikhomirov-2001}.

\paragraph{\acrshort{pbc} opportunities at the \acrshort{lc}}

In addition to the capabilities to test strong-field \acrshort{qed}, which are similar for \acrshort{fccee} and an \acrshort{lc}, a future \acrshort{lc} facility offers other options for \acrshort{pbc}, which are elaborated in~\cite{LinearColliderVision:2025hlt}. The main difference is that a \acrshort{lc} will continuously dump its beam. As an example, the main beam dump of \acrshort{ilc} is designed to handle \SI{17}{MW} of power~\cite{Satyamurthy:2012zz}, indicated in Figure~\ref{fig:DumpDistribution}. The extremely large quantity of electrons and positrons on target is unique, and high production rates of heavy mesons and tau leptons are expected~\cite{Nojiri:2022xqn}. It enables the exploration of uncharted regions of parameter space for dark-sector particles, like \acrshort{alps}, light scalar particles, \acrshort{hnls}, and \acrshort{ldm}. The electrons are highly polarized, above 80\%, which enhances certain mechanisms of production of new physics. On the other hand, positrons enhance the production via resonant annihilation. 
Besides the main beam dumps, an \acrshort{lc} has other beam dumps, such as the tune-up dump. At such a location, single bunches can be extracted from each bunch train at full energy after the linear accelerator. They can be used, for example, for \acrshort{ft} searches with thin targets, where lower rates are desired~\cite{Ishikawa:2021qna} or for the \acrshort{rd} of electron-driven plasma-wakefield acceleration at \SI{100}{GeV}~\cite{Litos:2014yqa}.

\subsubsection{cpEDM measurements}

  Proposals to build a synchrotron to measure a possible \acrshort{edm} of charged particles \cite{PhysRevLett.93.052001, EDM:CPEDM:2019nwp, PhysRevD.105.032001} are motivated by the physics case described in Section~\ref{subsec:EDM}. Whereas the \acrshort{edm} of electrically neutral particles such as neutrons can be measured with particles at rest, \acrfull{cpedm} measurements require to confine the particles in a precision storage ring. Most proposals to measure static \acrshort{edm}s are based on circulating polarized bunches satisfying the ``frozen spin'' condition as sketched in Figure~\ref{fig:cpEDM_Principle}. In a perfect machine and particles without \acrshort{edm} and the well known \acrshort{mdm} of the particles, an initially longitudinal polarization is maintained by appropriate choice of the magnetic and electric field bending the beam. This means that the angular frequencies $\omega_M$ describing the rotation of the spin and $\omega_p$ describing the rotation of the particle momentum or direction are identical. A finite \acrshort{edm} generates a rotation of the polarization out of the horizontal plane into the vertical direction. A common proposal by an international community for proton \acrshort{edm} measurements is a ring with only \acrfull{es} bendings operated at the ``magic energy'' ($\approx\SI{232.8}{\MeV}$), where the magnetic fields to satisfy the frozen spin condition vanish. Focusing can be done with electric~\cite{EDM:CPEDM:2019nwp} or magnetic~\cite{PhysRevD.105.032001} quadrupolar components and a typical circumference is~\SI{500}{\m} to keep the required electric fields at reasonable values.

  \begin{figure}[!h]
   \centering
     \includegraphics*[width=7cm]{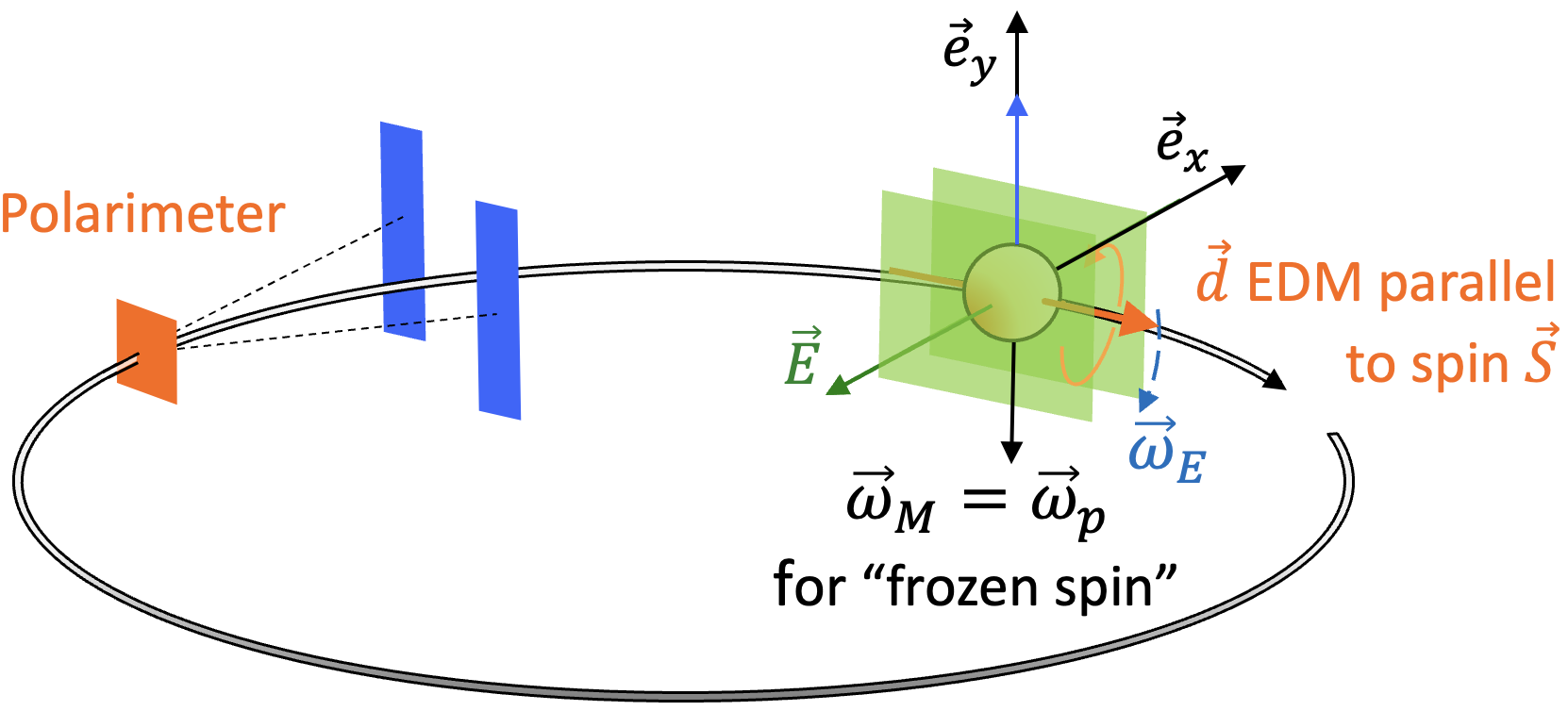}
   \caption{Sketch of a frozen-spin \acrshort{cpedm} measurement ring.}
   \label{fig:cpEDM_Principle}
\end{figure}

  The main challenge of \acrshort{cpedm} measurement proposals is to show that the achievable sensitivity, i.e., the smallest detectable \acrshort{edm}, is superior to current limits, e.g., for the neutron~\cite{EDM:PhysRevLett.124.081803}. This requires that a sufficient spin coherence time can be reached and that systematic effects, i.e., spin rotations caused by machine imperfections for particles with a \acrshort{mdm} only, can be understood and sufficiently suppressed. The former requires that the beam is bunched and sextupole families are used to suppress contributions to spin decoherence related to betatron and synchrotron oscillations. Some systematic effects are mitigated for ``magic energy'' rings by circulating counter-rotating beams, state-of-the-art magnetic shielding (in particular for fully \acrfull{es} rings), magnetic focusing and high symmetry lattices (in particular for the ``hybrid'' ring with magnetic focusing). However, many possible systematic effects remain and require thorough and careful evaluations~\cite{hacıomeroglu2017systematicerrorsrelatedquadrupole, PhysRevAccelBeams.24.034003, PhysRevAccelBeams.25.064001, Cilento:2902257, Cilento:2907231}.

Many experimental studies relevant for the design of any \acrshort{cpedm} ring have been done with the \acrfull{cosy} ring \cite{PhysRevLett.117.054801, PhysRevLett.115.094801, PhysRevAccelBeams.21.024201, PhysRevX.13.031004} in J\"ulich using a different method not relying on frozen spin concept.

    A proposal to design and construct a smaller $\approx\SI{100}{\m}$-circumference prototype \acrshort{edm} ring \cite{EDM:CPEDM:2019nwp}  is motivated by the observation that achieving the required \acrshort{cpedm} measurement sensitivity is very challenging without thorough in-depth investigations. Such a prototype ring would be an intermediate step to gain expertise in operating a large scale \acrshort{es} facility, to develop key technologies (as very sensitive beam position pick-ups and high electric field equipment), and to study in-depth systematic effects and mitigation measures.

\subsection{Other Early Stage Proposals Discussed within PBC}

Below a brief summary of ideas presently discussed within the PBC (in alphabetical order):
\begin{itemize}
    \item{} \textbf{AQN@LHC}: Search for axion-antiquark-nuggets at the \acrshort{lhc} through energy deposition from their strong interaction with the \acrshort{lhc} environment leading to a measurable warm up (cf.~\cite{Zioutas:2024cmy}).
    \item{} \textbf{FAMU}: Measurement of the Zemach proton radius with an accuracy better than 1\% via high-precision spectroscopy of the 1S hyperfine splitting in muonic hydrogen~\cite{Pizzolotto:2020fue}.
    \item{} \textbf{FLOUNDER, SINE \& UNDINE}: \acrshort{dm} searches and TeV neutrino physics studies with under-water and ground detectors from very forward production in \acrshort{lhc} collisions \cite{Ariga:2025jgv,Kamp:2025phs}.
    \item{} \textbf{Gravitational effects of the LHC beam}: Measurement of the gravitational near-field of the \acrshort{lhc} beam with optomechanical detectors testing gravity in a completely different domain from previous test of general relativity \cite{Spengler:2021rlg}. 
    \item{} \textbf{PAX}: Test of strong-field quantum electrodynamics via high-precision X-ray spectroscopy of antiprotonic atoms at the \acrshort{cern} \acrshort{ad} facility~\cite{Paul:2020cnx}.
    \item{} \textbf{SHIFT@LHC}: Search for long-lived low-mass \acrshort{bsm} particles with a fixed-target setup upstream of the \acrshort{cms} experiment at the \acrshort{lhc} \cite{Niedziela:2020kgq}.
\end{itemize}

\subsection{FPC -- interfacing experiment and phenomenology}

As the spectrum of \acrshort{pbc} experiments targeting \acrshort{fips} and other types of \acrshort{hs}s broadens, the role of the \acrshort{fpc} continues to grow. 

With data from the \acrshort{cern} experiments \acrshort{na62}, \acrshort{na64}, and \acrshort{faser} coming in, and with the construction of \acrshort{ship} on the horizon, maintaining the benchmark models that facilitate a comparison of different results is an important indicator of the achieved progress -- both within \acrshort{pbc} and in the worldwide context. The improved sensitivity and sophistication of experimental analyses also require more accurate signal calculations, especially in difficult hadronic channels (cf., e.g., \cite{NA62:2025yzs} for some recent measurements of \acrshort{na62}) as well as the development of tools facilitating sensitivity calculations and reinterpretations, e.g., \cite{Jerhot:2022chi,Ovchynnikov:2023cry}. Further updates of the benchmark models may be needed to improve overall consistency and agreement with data from a broader range of experiments, such as, e.g., neutrino experiments. Many of these updates are driven by contributions from individual members of the \acrshort{fpc}, but the hub that the \acrshort{fpc} provides is central for disseminating these developments and identifying the requirements of the different experiments and the wider community.

This exchange will become even more important with the broadening landscape of fixed target experiments, \acrshort{lhc} searches for long-lived particles and forward physics, as well as other types of experiments and facilities. These developments stimulate new discussions on interesting new physics targets that are within reach of \acrshort{pbc} experiments, such as exotic bound states called quirks~\cite{Feng:2024zgp}. In exchange with the \acrshort{bsm} working group, the \acrshort{fpc} also aims to identify experimental capabilities that are not yet captured by \acrshort{pbc} benchmark models and new ways to present them. Some of these new developments will be discussed in Section~\ref{subsec:HiddenSector}.

Maintaining and expanding this lively, collaborative and adaptive effort between theory and experiment, but also promoting individual investigations will be central to continuing the successful and broad physics program supported by \acrshort{pbc}.

\subsection{Summary of the proposals at CERN}

 \begin{table}[h!]
 \scriptsize
    \centering
    \begin{tabularx}{\linewidth}{lllcc}
        \hline \hline
        \textbf{Location} & \textbf{Proposal} & \textbf{Status} & \textbf{Cost category} & \textbf{Earliest operation} \\
        \midrule
        & \acrshort{na61/shine} ions  & Addendum submitted to \acrshort{spsc} in 2024 & C0 & Run~4   \\
   & \acrshort{na61/shine}-LE & Proposal submitted to \acrshort{spsc} in 2024 & C1 & Run~4 \\
\acrshort{ehn1}  & \acrshort{na64} - Phase~2 -- e   & Proposal to be submitted to \acrshort{spsc}    & C0    & Run~4   \\
        & \acrshort{na64} - Phase~2 -- e$^+$ & Conceptual Design & N/A & Run~4 \\
        & \acrshort{na64} - Phase~2 -- h & Conceptual Design & N/A & Run~4 \\ 
        & \acrshort{na60+} & Proposal to be submitted to \acrshort{spsc} in 2025 & C2 & Run~4    \\\hline        
        & \acrshort{na64} - Phase~2 -- $\upmu$  & Proposal to be submitted to \acrshort{spsc} & C0    &  Run~4 \\
\acrshort{ehn2}        & \acrshort{amber} - Phase 2 & Conceptual Design & C1/C2  & Run~4  \\ 
& \acrshort{muone}   & Conceptual Design      & C1    &   Run~4      \\
         \hline
To be defined    & \acrshort{sbn}  & Conceptual Study & N/A   &   Run~5 \\ \hline
        & \acrshort{fpf} & \acrshort{loi} to be submitted to \acrshort{lhcc} in 2025 & C3 & Run~4 \\
        & \acrshort{sndhllhc} & Proposal submitted to \acrshort{lhcc} in 2024 & C1 & Run~4 \\
        & \acrshort{anubis} & Data-taking with prototype detector (\acrshort{proanubis}) & N/A & Run~4 \\
        & \acrshort{codexb} & Data-taking with prototype detector (\acrshort{codexbeta})& N/A & Run~4 \\
\acrshort{lhc}     & \acrshort{mapp2} & \acrshort{loi} to be submitted to \acrshort{lhcc} in 2025 & C1 & Run~4 \\
        & \acrshort{mathusla40} & Data-taking with prototype detectors & N/A & Run~4 \\
        & \acrshort{aladdin} & Preparation of proposal to \acrshort{lhcc} & C2 & Run~4 \\
        & \acrshort{lhcspin} & Conceptual Study & N/A & Run~5 \\ 
        & \acrshort{aion} & \acrshort{eoi} to be submitted to \acrshort{lhcc} in 2025 & C1 & Run~4 \\
        & \acrshort{gf} & Conceptual Study & N/A & After Run~5 \\
      
        \hline \hline
    \end{tabularx}
        \caption{Preliminary cost estimates for beam and infrastructure upgrades as obtained by a \acrshort{pbc} working group for the initial proposals where available. The definition of cost categories is explained in the text. N/A = Not Available.}
    \label{tab:proposals}
 \end{table}

The list of proposed experiments and facilities with an overview of the present status and indicative cost are shown in Table~\ref{tab:proposals}.
The cost estimates are limited to the beam and infrastructure upgrades and do not account for the costs of the experiments themselves. The cost categories are defined as follows:
\begin{itemize}
  \item   C0:	Up to 300~kCHF
  \item	C1:	From 300~kCHF to 2~MCHF
  \item	C2:	From 2~MCHF to 10~MCHF
  \item	C3: From 10~MCHF to 50~MCHF
  \item   C4: Above 50~MCHF
\end{itemize}

\section{Physics reach of PBC projects in the global context}

\subsection{Hadronic vacuum polarisation and the muon anomalous magnetic moment}
\label{subsec:HVP}

The apparent discrepancy between the experimental
value and theoretical prediction of the muon anomalous magnetic moment $(g-2)_\mu$ has
triggered enormous activities in improving both the
experimental results and theoretical predictions~\cite{Aoyama:2020ynm},
but also in formulating extensions to the Standard Model.
Recent experimental results on $(g-2)_\mu$ confirm
the earlier measurements~\cite{Muong-2:2006rrc} with reduced uncertainties of
0.19 ppm when combined, and a final goal of 0.10 ppm~\cite{Muong-2:2023cdq,Muong-2:2024hpx}.
Additional data is also expected from the planned \acrfull{jparc} E34 (g-2/EDM) experiment~\cite{Abe:2019thb}.

The experimental determination is significantly more precise than
phenomenology, which is above 0.3 ppm.
To improve on that it is of utter importance to better understand the contributions
from the leading-order \acrfull{hvp} and from the hadronic contribution to light-by-light scattering,
which are main drivers of current theory uncertainties~\cite{Aoyama:2020ynm}.

A traditionally promising way of calculating the leading hadronic
contribution is the use of time-like dispersion relation, which relies
on the knowledge of the $e^+ e^-$ annihilation cross section into
hadrons from the $\pi^0$ mass threshold upwards. The low-energy range
is affected by large non-perturbative effects, and experimental data
from $e^+e^-$ machines are needed. At present, large amounts of data
have been collected at different machines for many exclusive and
inclusive channels, through direct energy scan or radiative return
methods, which -- in principle -- allow for a rather precise
determination of the leading hadronic contribution. However, from
Figure~\ref{fig:g-2} it also becomes clear that there is a clear tension
between results that use different experimental data sets.

More recently, lattice \acrshort{qcd} has become more competitive in
the calculation of the leading-order \acrshort{hvp}.
Not only that, they suggest much smaller discrepancies between
the experimental results of $(g-2)_\mu$ and its theoretical determination,
reducing them to less than a one-sigma effect~\cite{Boccaletti:2024guq}.

An overview of current experimental and theoretical results on
the muon anomalous magnetic moment is given in Figure~\ref{fig:g-2},
highlighting a large spread in the theory predictions, which indicates
a lack of a full understanding of the systematics involved.
In this situation, corroboration from independent methods, like the space-like
approach~\cite{CarloniCalame:2015obs} with muon-electron scattering data of
the \acrshort{muone} experiment~\cite{MUonE:2016hru} would be extremely valuable.
\acrshort{muone} aims to measure the leading-order hadronic contribution to $(g-2)_\mu$
with a statistical precision of 0.3\%, in line with current determinations.
The goal in accuracy and the novel technique will make this measurement an
important contribution to the test of the Standard Model prediction of $(g-2)_\mu$.

\begin{figure}[t!]
    \centering
    \includegraphics[width=.6\linewidth]{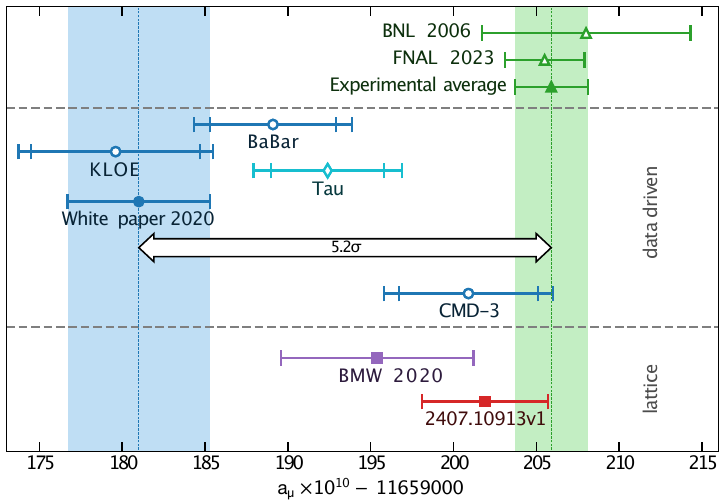}
    \caption{Overview of current experimental~\cite{Muong-2:2006rrc,Muong-2:2023cdq,Muong-2:2024hpx} and data-driven \cite{BaBar:2009wpw,BaBar:2012bdw,CMD-3:2023alj,CMD-3:2023rfe,KLOE:2008fmq,KLOE:2010qei,KLOE:2012anl,KLOE-2:2017fda,Davier:2010fmf,Davier:2013sfa} as well as theoretical ~\cite{Boccaletti:2024guq,Borsanyi:2020mff,Aoyama:2020ynm} results related to the muon anomalous magnetic moment a\(_\mu\). 
    (Figure adapted from Ref.~\cite{Boccaletti:2024guq}.)
    }
    \label{fig:g-2}
\end{figure}

\subsection{Hadron structure and properties
}
\label{subsec:HadronStructure}

Hadron structure studies remain an important part of high-energy physics, addressing such questions as the origin of mass of ordinary matter, partons with a very large fraction of the parent hadron's momentum, or the onset of saturation at very low Bjorken-$x$. 
Although the physics program of related \acrshort{pbc} projects partially overlaps with that of ongoing and future experiments, like at the \acrshort{eic} at \acrshort{bnl}, which is scheduled to begin data-taking in the middle of 2030s, it is important to note that the covered kinematical regions are often different and complementary (see, e.g., Figure~\ref{fig:PDF-Landscape}). While the \acrshort{eic} will mainly cover the low-$x$ region with good coverage also of intermediate-to-high $x$ at high-$Q^2$, fixed-target collisions at the \acrshort{lhc}---such as those achievable with \acrshort{smog2} and \acrshort{lhcspin}---mainly cover the intermediate-to-high $x$ region at intermediate $Q^2$. Furthermore, while \acrshort{eic} exploits \acrshort{dis} measurements, \acrshort{lhcspin} (and \acrshort{smog2}) will allow one to explore the nucleon structure by means of hadronic collisions, providing a fully complementary approach. This is of crucial importance, especially to constrain poorly explored aspects of the nucleon structure, such as the essentially unknown gluon \acrfull{tmd}. In addition, comparisons between \acrshort{eic} and \acrshort{lhcspin} results will be extremely useful to test fundamental \acrshort{qcd} predictions, such as evolution, universality and factorization.
Likewise, the \acrshort{fpf} will probe high-$x$ but also very low-$x$ \acrshort{pdf}s at low scales, virtually unreachable at current and upcoming  experiments. 
\acrshort{amber}'s focus is on pion and kaon \acrshort{pdf}s using the theoretically clean \acrshort{dy} process, typical \acrshort{dis} experiments can contribute to this avenue only in a model-dependent approach.

\begin{figure}[tb]
\centering
\includegraphics[width=0.75\textwidth]{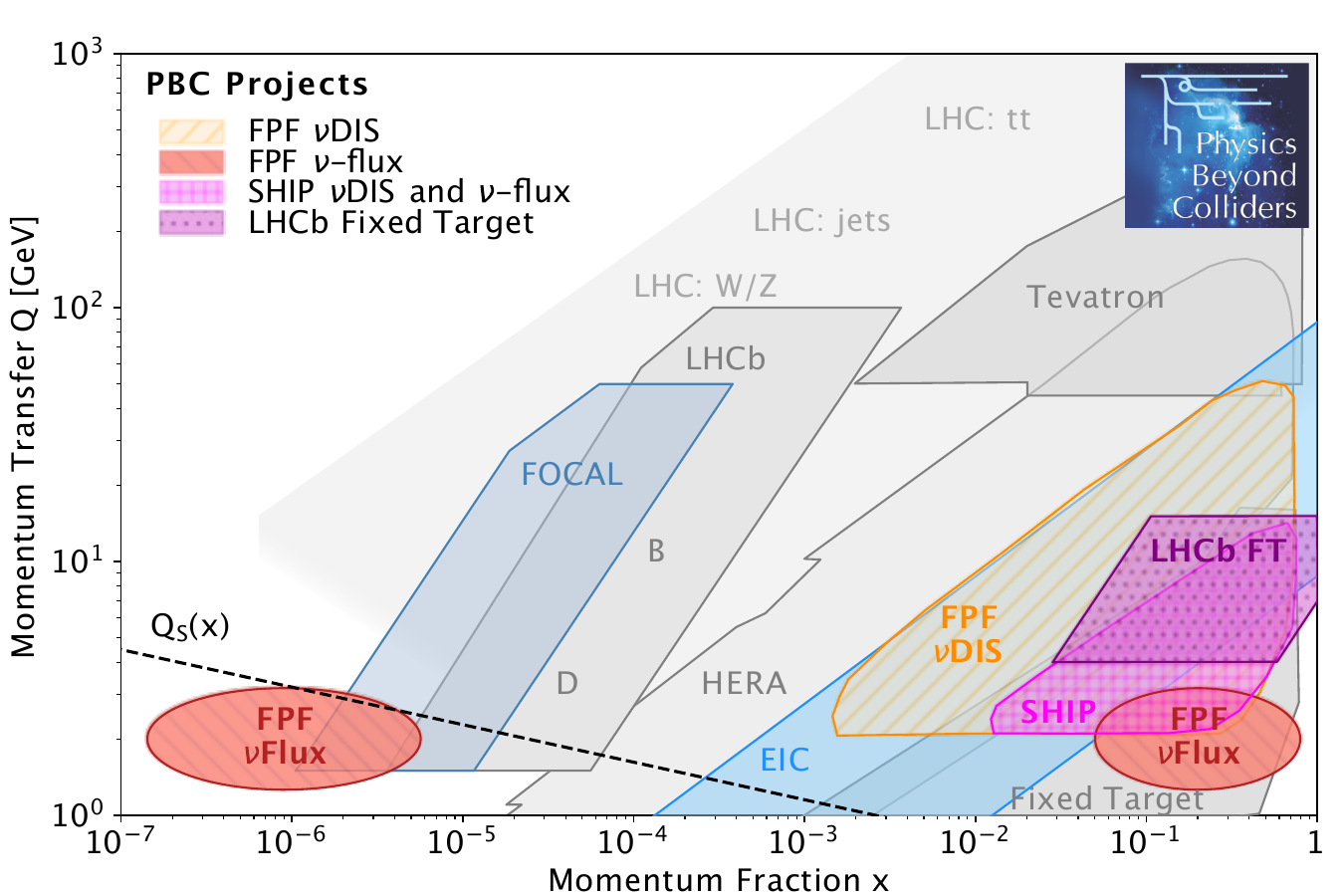}
\caption{Kinematic coverage of proposed \acrshort{pbc} experiments (as labelled), past/existing experiments (gray), and other proposed experiments (blue) in the momentum transfer $Q$ versus momentum fraction $x$ plane. Also indicated is the expected saturation scale $Q_s$ in protons (dashed line)~\cite{ALICE:2023fov}.}
\label{fig:PDF-Landscape}
\end{figure}

\paragraph{LHCb-FT}

{\bf \acrshort{smog2}}: Unpolarized fixed-target collisions at \acrshort{lhcb} open a new chapter in \acrshort{lhc} physics, offering novel opportunities that complement the traditional beam-beam studies. These opportunities notably include: (i) access to nucleon and nuclear \acrshort{pdf}s at large Bjorken-$x$, including the unpolarized \acrshort{tmd}s, and (ii) studies of effects arising from (non-perturbative) intrinsic-charm contributions, expected at large $x$. Opportunities regarding nuclear matter and dark matter studies with \acrshort{smog2} will be addressed further below.

{\bf \acrshort{lhcspin}}: The physics case of \acrshort{lhcspin} is mainly focused on the investigation of the nucleon spin structure, with a special emphasis on the \acrshort{tmd}s and the \acrshort{gpd}s of quarks and gluons. Polarized quark and gluon distributions can be studied at \acrshort{lhc} with \acrshort{lhcspin} through high-energy collisions of proton beams on transversely polarized hydrogen or deuterium targets. The golden process for accessing quark \acrshort{tmd}s is \acrshort{dy}, where a quark and an anti-quark annihilate to produce a charged lepton pair (e.g., $\mu^+\mu^-$) in the final state. Another key aspect of this project is the possibility to access heavy quarks, which are primarily produced via gluon-gluon fusion. The production of quarkonia and open heavy-flavour states provides an efficient way to study the gluon dynamics inside nucleons and probe gluon \acrshort{tmd}s. This is achieved by measuring inclusive production of heavy mesons such as $J/\Psi$, D$^0$, $\eta_c$, $\chi_c$, $\chi_b$, etc., for which the \acrshort{lhcb} detector is well suited and optimized. Furthermore, isospin effects can be studied by comparing $pp^{\uparrow}$ and $pd^{\uparrow}$ collisions. Finally, by studying collisions of heavy-ion beams with polarized p or d targets, \acrshort{lhcspin} enables innovative investigations of collectivity in small systems as well as access to higher-twist contribution to fragmentation functions.

\paragraph{\acrlong{gf} possibilities/opportunities}
The secondary photons of the \acrshort{gf} at \acrshort{lhc} would have an energy between 40 keV and 400 MeV. Unprecedentedly high intensity of the polarized photon beam will allow to study a number of low-energy observables relevant to nuclear and nucleon structure and to tests of fundamental symmetries, hitherto inaccessible with the available photon sources. If the secondary photons of the \acrshort{gf} can collide with a polarized fixed target at the \acrshort{lhc}, then further opportunities would arise. The \acrshort{com}~energy would be at most $\sqrt{s}=1$ GeV at a photon energy of 400 MeV, hence insufficient to probe nucleon partonic structure, but polarized structure functions $g_1$ and $g_5$ would be accessible in case of longitudinally polarized proton or nuclear targets. The polarized structure function $g_1$ is of relevance to the Gerasimov-Drell-Hearn sum rule. The parity-violating polarized structure function $g_5$ has never been measured with real photons before.

The further possibility of colliding these secondary photons with the \acrshort{lhc} proton/ion beam would yield a \acrshort{com}~energy $\sqrt{s}$ up to 100 GeV, allowing one to perform studies of nucleon structure at (sub)asymptotic photon energies. 
This would allow to study with high statistics the parity violating structure function $F_3$ with real photons that are circularly polarized, arising from loop contributions, rather than through $\gamma$-$Z$ interference. This would be complementary to $F_3$ studies at the future \acrshort{eic}, which  will require high $Q^2$ (electroproduction rather than photoproduction). 
Parity-violation studies with circularly polarized real photons moreover will provide information on odderon exchange without the need for comparison to antiparticle beams. Further opportunities are offered by exclusive and semi-inclusive vector-meson photoproduction, which would allow for studies of \acrshort{gpd}s and \acrshort{tmd}s. 

\paragraph{\acrshort{ship} possibilities/opportunities}
The \acrshort{ship} experiment will provide information about the internal structure of the nucleon using neutrino interactions  as a probe (see a discussion in~\cite{Alekhin:2015byh} and in section 1.3 and 4.4.4 of~\cite{Albanese:2878604}). First, it can measure the structure functions $F_{4,5}$ with the precision of around $5\%$, corresponding to the systematic uncertainty in neutrino production flux. This estimate is reasonable assuming that the charm hadron production will be well measured; this may be done via combining the charm production measurements at the thin target NA65/DsTau experiment and $J/\psi$ production in the thick \acrshort{ship}-like target (section 4.5 of~\cite{Albanese:2878604}). Second, the neutrino-induced charm production will allow probing the strange content of the nucleon. Last but not least, the neutrino cross-section measurements in the range of energies accessible at \acrshort{ship} (mainly within the range $E_{\nu}<100\text{ GeV}$) can be used in fitting nuclear \acrshort{pdf}s. These questions will be studied in detail once the collaboration decides on the technology, dimensions, and placement of the \acrshort{snd}@\acrshort{ship}.

\paragraph{\acrshort{fpf} possibilities/opportunities}
The \acrshort{fpf} will also enable unique studies of the strong interaction and proton structure both via measurements of neutrino fluxes and interactions. Both high-energy electron neutrinos and tau neutrinos primarily originate from charm hadrons. These are mainly produced via gluon fusion with one gluon carrying a momentum fraction $x\sim 1$ and the other $x \sim 4 m_c / s \sim 10^{-7}$. Neutrino flux measurements at the \acrshort{fpf} will therefore constrain the gluon \acrshort{pdf}s in the low-$x$ region, beyond the coverage of all other existing and proposed experiments, as shown in the left panel of Figure~\ref{fig:PDF}. This will allow to probe BFKL dynamics~\cite{Ball:2017otu} and gluon saturation~\cite{Bhattacharya:2023zei}. Stringent constraints on models of $D$-meson fragmentation will also be provided.

The \acrshort{fpf} also augments the \acrshort{lhc} program with a \textit{neutrino-ion collider} to probe charged current \acrshort{dis}. This complements the planned \acrshort{eic}, which will probe neutral current \acrshort{dis} in a similar energy range. The large neutrino sample enables differential cross-section measurements to constrain \acrshort{pdf}s. An example is shown in the right panel of Figure~\ref{fig:PDF}. These improvements will benefit key measurements at the \acrshort{lhc} central detectors, such as Higgs or weak boson production~\cite{Cruz-Martinez:2023sdv}, and help to break the degeneracy between \acrshort{qcd} and new physics effects in \acrshort{lhc} data interpretations~\cite{Hammou:2024xuj}. 

\begin{figure}[tbh]
\centering
\includegraphics[width=0.95\textwidth]{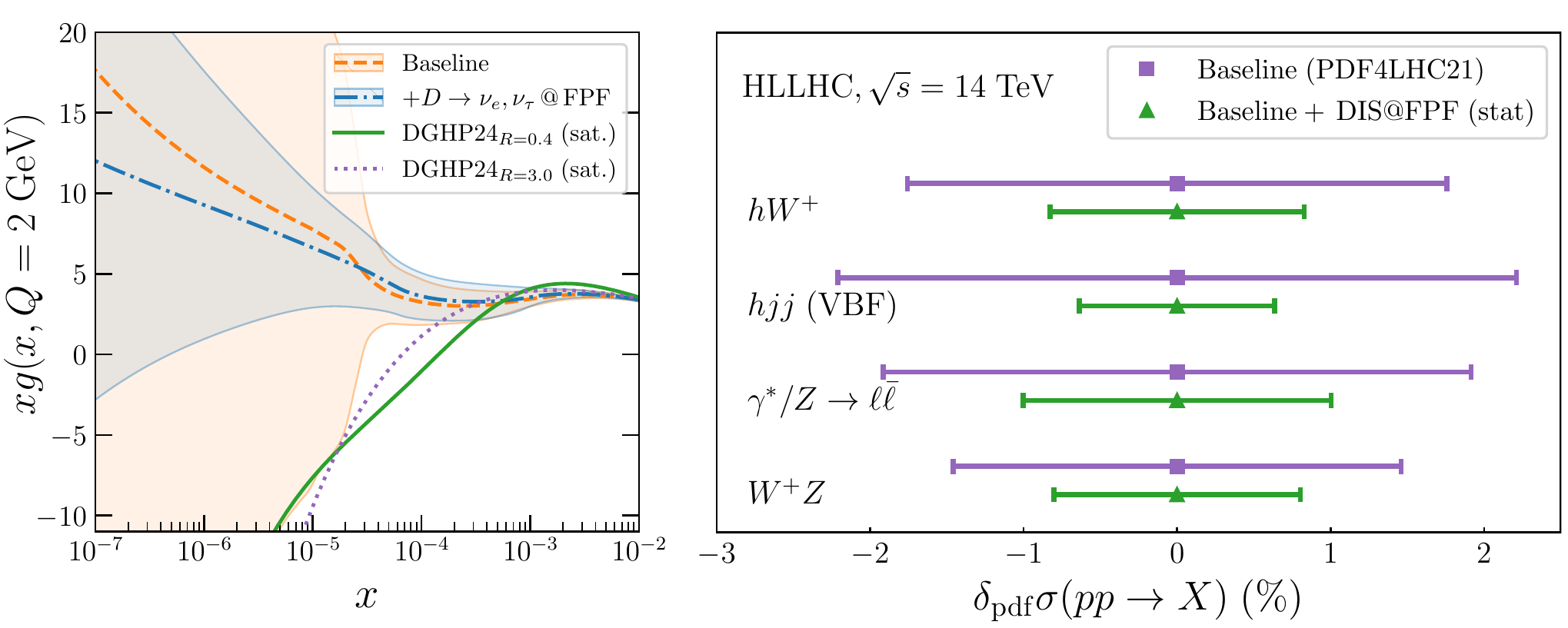}
\caption{Left: Impact of neutrino flux measurements at the \acrshort{fpf} on the small-$x$ gluon \acrshort{pdf}. Right: Reduction of the \acrshort{pdf} uncertainties on Higgs and weak gauge-boson cross sections at the \acrshort{hllhc}, enabled by neutrino \acrshort{dis} measurements at the \acrshort{fpf}. 
(Figures adapted from Ref.~\cite{Adhikary:2024nlv}.)
}
\label{fig:PDF}
\end{figure}

\paragraph{\acrshort{amber}}
The partonic structure of light mesons (e.g., pions and kaons) has garnered significant interest recently, driven by both phenomenological analyses of existing data and advancements in lattice \acrshort{qcd} calculations~\cite{Barry:2018ort,Novikov:2020snp,Barry:2021osv,JeffersonLabAngularMomentumJAM:2022aix,Kotz:2023pbu,ExtendedTwistedMass:2024kjf}. The observed differences in the relative momentum fractions carried by quarks and gluons in mesons compared to protons may provide valuable insights into fundamental questions, such as the mechanisms of baryon mass generation in \acrshort{qcd}. However, the existing pion and especially kaon \acrshort{dy} cross-section data, collected primarily in the 1980s, partially suffer from limited precision and notable tensions between experiments. This underscores the importance of the \acrshort{amber} experiment's physics program, which aims to measure the structures and charge radii of pions and kaons with unprecedented precision. \acrshort{amber}'s efforts will provide the first comprehensive data on kaon \acrshort{pdf}s while significantly enhancing the accuracy of pion structure determinations.

The kaon spectroscopy program of \acrshort{amber} offers worldwide unique
possibilities to deliver much-awaited new data for the strange-meson
sector, allowing for a detailed comparison to theory predictions and for
the search of strange exotic mesons like hybrids and tetraquarks.
The envisaged $2\times 10^7$ exclusive events for the $K^-\pi^-\pi^+$ 
final state with kaon beam would correspond to 100 times more data 
than that of the ACCMOR experiment \cite{ACCMOR:1981yww}, and 30 times more than the (acceptance-limited) \acrshort{compass} data.
Kaon spectroscopy with kaon beams could, in principle, be pursued with 
\acrshort{jlab} Hall D's tertiary $K_L$ beam, but with a limited flux of about $10^4~\text{s}^{-1}$
in the momentum range 0--11 GeV/c \cite{KLF:2020gai}. Besides the low flux, there is an additional uncertainty in
the production mechanism due to the rather low beam energy. 
Likewise, the new hall and beamline in planning at 
\acrshort{jparc} (funding and timelines unclear), with a  
very high ($10^7~\text{s}^{-1}$) flux and reach of about 10 GeV \cite{Fujioka:2017gzp},
could, in principle, be used for kaon spectroscopy. However, there is no program 
for strange meson spectroscopy so far and the low energy, again,
makes the separation between beam and target excitation very
challenging. 
\acrshort{besiii}, \acrshort{belleii}, as well as \acrshort{lhcb} can study strange mesons in $D$ or $B$-meson
decays. $D$-meson decays are limited by the mass of the $D$-meson, and both
channels need the knowledge of strong phases and final-state
interactions, where input from \acrshort{amber}  is actually crucial.

\subsection{Heavy-ion physics} 
\label{subsec:HIP}

\begin{figure}[tb]
\centering 
\includegraphics[width=.7\textwidth]{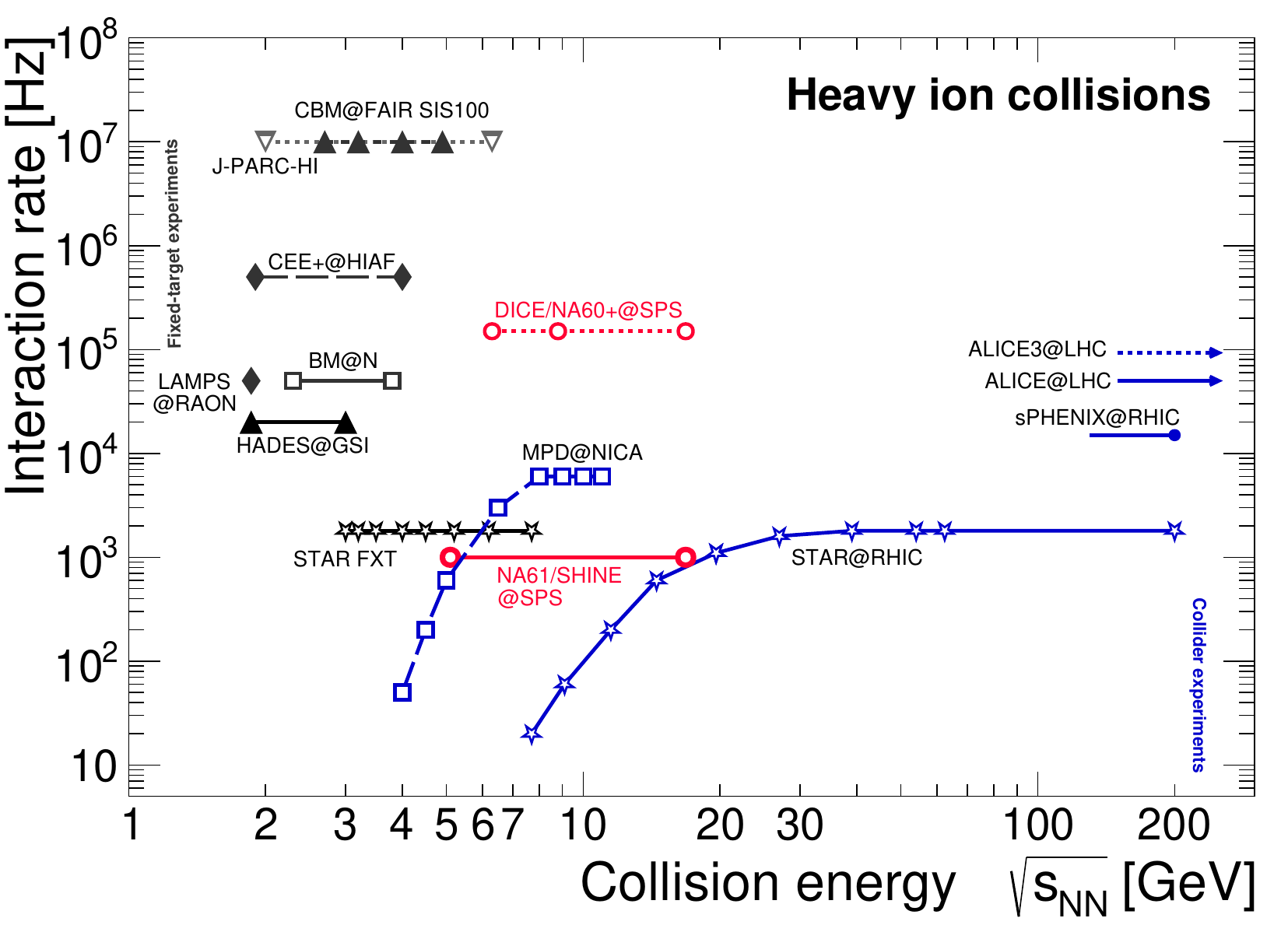}
\footnotesize\caption{
Interaction rate for existing (solid line), under construction (long-dashed line) and planned (short-dashed line) heavy-ion (Au--Au or Pb--Pb) experiments. Figure and caption from~\cite{PBC_ionreport}, which is adopted and updated from~\cite{high_mub:IRplot,Galatyuk:2019lcf,CBM:2016kpk,QCDWorkingGroup:2019dyv}. Highlighted in red are the \acrshort{sps} \acrshort{na60+} and \acrshort{na61/shine} projects.
}
\label{fig:HIC-Landscape} 
\end{figure}

Within \acrshort{pbc}, heavy-ion physics is pursued both at the \acrshort{sps}  (\acrshort{na60+} and \acrshort{na61/shine}) and \acrshort{lhc}  (\acrshort{smog2}) fixed-target setups. The \acrshort{sps} experiments occupy a unique and crucially important energy range from $ 5.1 - 16.8$\,GeV $\sqrt{s_{NN}}$ in the global landscape of heavy-ion experiments, see Figure~\ref{fig:HIC-Landscape} and Table~\ref{tab:synergy}. In this regime we expect a change of the dynamics of heavy-ion collisions. This includes the approach to the potential critical end point of \acrshort{qcd} and the emergence of inhomogeneous patterns such as a moat regime. In combination, \acrshort{na60+} and \acrshort{na61/shine} offer a rather complete cover of interesting observables, see Table~\ref{tab:synergy}. Furthermore, \acrshort{sps} and \acrshort{lhcft}  measurements will also impact astroparticle physics via  production studies relevant to cosmic-ray physics and \acrshort{dm} searches. Collisions on hydrogen and helium targets reproduce primary cosmic-ray collisions in the interstellar medium, while data from nitrogen or oxygen targets (or proxies like neon) contribute to modeling extensive air showers from Ultra-High-Energy (UHE) cosmic rays in the atmosphere.

\begin{table}[t]
\scriptsize
\centering
\begin{tabularx}{\linewidth}{  l | c    c || c c  c c ||c c | c } 
\hline\hline
                                  & \multicolumn{2}{c || }{ SIS18~/~SIS100 } &   \multicolumn{4}{c|| }{ \acrshort{sps}   }    & \multicolumn{2}{c| }{\acrshort{rhic}}  &  \acrshort{lhc} \\
                                  & \acrshort{hades}      &  \acrshort{cbm}                       &   \multicolumn{3}{c   }{\acrshort{na61/shine}}  &  \acrshort{na60+}      & \multicolumn{2}{c| }{\acrshort{star}}        &  \acrshort{alice}~/~\acrshort{alice3} \\
\hline
\hline
     $\sqrt{s_{NN}}$ [GeV]      & $1.9-4.9$  & $2.7-4.9$ &  \multicolumn{2}{c   }{$5.1-16.8$}      & & {$6.3-16.8$}      & \multicolumn{2}{c| }{$3-200$}  &  $2760-5440$ \\
     Start date                &   running  & 2028                         &  \multicolumn{2}{c   }{running}    &    & $>$$2029$~~       &\multicolumn{2}{c| }{             running} &  running / 2035 \\
                               &            &                              &light ions&\multicolumn{1}{c   }{Pb+Pb}&  &                &                                           &                 \\
\hline
     Bulk properties           &$+$         &$+$                           & $+$     &   $+$     &   & $+$                   &\multicolumn{2}{c| }{$+$}        &$+$\\
     E-by-E fluctuations       &$+$         &$+$                           & $+$     &  $+$      &   &                   &\multicolumn{2}{c| }{$+$ }       &$+$\\
     Resonances                &$+$         &$+$                           & $+$     &  $+$      &   & $+$               &\multicolumn{2}{c| }{  $+$}      &$+$\\
     Open charm                &            &{higher energies}                           & $+$     &  $+$      &   & $+$               &\multicolumn{2}{c| }{only top energy}        &$+$\\
     $c-\bar{c}$ correlations  &            &{higher energies}                           &         &    $+$    &   & $+$               &\multicolumn{2}{c| }{   }        &$+$\\
     Quarkonium                &$+$         &{higher energies}                          &         &           &   & $+$               &\multicolumn{2}{c| }{$+$}        &$+$\\
     Dileptons                 &$+$         &$+$                           &         &           &   & $+$               &\multicolumn{2}{c| }{$+$}        &$+$\\
\hline\hline
\end{tabularx}
\caption{Synergy between \acrshort{sps} and \emph{selected} other heavy-ion physics programs, covering a broad range of topics with a gradual transition from (left to right) properties of dense hadronic matter over onset of deconfinement and the location of the critical endpoint to the properties of quark-gluon plasma without always clear boundaries between them at the various experiments. Table and caption taken from~\cite{PBC_ionreport}.}
\label{tab:synergy}
\end{table} 

\acrshort{na60+} aims at measuring rare probes of the \acrshort{qgp}, and in particular heavy quarks and lepton pairs. None of these studies were performed until now below top \acrshort{sps} energy with a decent integrated luminosity, nor will they become possible in the foreseeable future at any other facility in the collision energy range under discussion. On the contrary, this research domain is complementary to that already explored at hadron colliders (\acrshort{rhic}, \acrshort{lhc}), in particular for heavy-quark measurements, and to that expected from lower-energy facilities (\acrshort{fair}), in particular for dilepton-related observables.

More in detail, an energy scan in the region 6~GeV $<\sqrt{s_{\rm NN}}<17$ GeV is planned. With the foreseen experimental setup, which includes a muon spectrometer coupled to a vertex spectrometer, the dimuon spectrum from threshold up to the charmonium region will be addressed. Lepton pairs are a unique tool to determine the temperatures and lifetime of the strongly interacting medium created in the collisions. Those are key quantities to investigate a possible  anomalous behavior related to the onset of a first-order transition. In addition, dileptons provide direct information about hadron spectral functions, in particular the $\rho$-meson and its mixing with the chiral partner $a_{\rm 1}$, sensitive to chiral symmetry restoration which occurs in vicinity of the transition from hadrons to a \acrshort{qgp}.  

The investigation of charmonium states, and in particular J/$\psi$, has always been of paramount importance in \acrshort{qgp} studies. Anomalous suppression effects are likely due to color screening in the \acrshort{qgp}. They were observed at top \acrshort{sps} energies in the 1990s (NA50/NA60 experiments). They should disappear when moving toward lower energies. It would be particularly interesting to correlate them with a precise measurement of the temperature of the fireball, and perform comparisons with the temperature dependence of the modifications of the inter-quark potential, which can be computed starting from lattice \acrshort{qcd} results.

Open heavy flavor  hadrons  (D$^{\rm 0}$, D$^{\rm +}$, $\Lambda_{\rm c}$)  represent excellent probes of the strongly interacting matter created in nuclear collisions, as demonstrated by extensive studies at \acrshort{rhic}/\acrshort{lhc} energy.  The initial temperature of system will be lower and more time will be spent in the hadronic phase, allowing stringent tests of theory predictions for the diffusion coefficient in this region. The degree of thermalization of charm quarks may also be accurately measured, through studies of the elliptic flow of charmed hadrons. Finally, the hadronization of a baryon-rich \acrshort{qgp} may lead to even larger enhancement of baryon/meson ratios with respect to \acrshort{lhc} energies. 

\acrshort{na61/shine} is a fixed-target hadron spectrometer operating
at the \acrshort{sps}, with the unique capability for large-acceptance hadron measurements over a versatile set of beams and targets in the specific regime of collision energy: 5~GeV~$<\sqrt{s}_\mathrm{NN}<17$~GeV. 
Its accumulated data set includes $p$+$p$, $p$+C, $\pi$+C, $K$+C, $p$+Pb, Be+Be, C+C, Ar+Sc, Xe+La, and Pb+Pb reactions, recorded at up to six beam momenta per reaction.

The primary objectives of the Collaboration are {\bf (a)} to further explore \acrshort{qgp}-related signatures in light-ion systems (oxygen-oxygen collisions) at $\sqrt{s}_\mathrm{NN}=5.1$, 7.6, and 16.8~GeV
by
the end of 
Run~3 (2026) and the beginning of Run 4~\cite{Gazdzicki:2810689,NA61/SHINEaddendum},
{\bf (b)} a precision study of the  violation of isospin symmetry in the O+O system,
{\bf (c)}
   the construction and installation of a new \acrshort{last} during \acrshort{ls3} (2027-2029), and
  {\bf (d)}
  a new series of measurements of charm and anti-charm hadron correlations in Pb+Pb collisions in Run 4, to investigate the space correlation (locality) of the $c\bar{c}$  pair \cite{Gazdzicki:2023niq}. 
    The studies performed in O+O reactions
    will be continued in B+B and/or Mg+Mg collisions~\cite{NA61/SHINEaddendum}. Being charge-symmetric systems ($Z$$=$$A$$-$$Z$), O+O and Mg+Mg reactions give a unique experimental opportunity to address the issue of the isospin--symmetry-violating charged-kaon excess~\cite{NA61SHINE:2023azp}. On the
other hand, the expected multiplicity of about one $c\bar{c}$ pair in central Pb+Pb collisions at top
\acrshort{sps} energy makes studies of $c\bar{c}$ hadron correlations very promising for testing heavy-quark
production locality and transport properties \cite{Gazdzicki:2023niq}. Also, extensive measurements of even-by-event (EbyE) fluctuations in Pb+Pb collisions are planned as a function of energy.
    They are meant as an independent verification of recent observations made at \acrshort{rhic}, argued as a possible indication of a Critical Point 
    in the phase diagram of strongly interacting matter.
  The realization of oxygen data-taking does not necessitate any additions to the fully operational \acrshort{na61/shine} 
experiment. Implementing the \acrshort{last} for $c\bar{c}$ hadron correlation studies constitutes a major upgrade of the
detector
to be completed at the end of \acrshort{ls3}. 

The proposed large-acceptance measurements are not possible in the foreseeable future at accelerator facilities other than the \acrshort{sps}.

\acrshort{smog2}: Studies of nuclear-matter effects can be performed using hydrogen as a reference system for comparison with larger nuclear targets such as argon, krypton, and xenon.
Quarkonia production is a key tool to study the high temperature and density \acrshort{qcd} regime. In these collisions, where on average only one $c\bar{c}$ bound state is produced, one can probe charmonium suppression in a regime where contributions from charm quark recombination are expected to be negligible. Jet measurements represent an additional tool to explore cold nuclear matter in this unique phase space and collision system. Using $p\text{H}_2$ collisions as a pp reference, measurements of nuclear modification factors ($R_{pA}$) can be performed with jets produced in \acrshort{ft} $pA$ collisions, enabling precise studies of parton energy loss in cold nuclear matter as a function of the nuclear size $A$ and of the relevant kinematic variables. Measurements with photon-tagged jets and charm jets will further allow flavour-dependent studies of parton energy loss with increasing nuclear size.

Precise antiproton and anti-nuclei production measurements in $p\text{H}$ and $p\text{He}$ collisions in the relevant energy regime of $\sqrt{s_{NN}}\sim 100$ GeV allow for significant reduction of the present uncertainties on the production cross sections, which constitute the main limitation to the interpretation of data collected by space-borne experiments (PAMELA, AMS) searching for signals of \acrshort{dm}.

\subsection{Neutrino physics} 
\label{subsec:NU}

\paragraph{The \acrfull{sbn} beam}

The novel short-baseline neutrino beam facility introduced in Sections \ref{sec:sbn_acc} and \ref{sec:sbn_exp} would represent a paradigm shift in the field by enabling the determination of neutrino-nucleus scattering amplitudes with a precision comparable to that achieved in electron-nucleus scattering. In particular, achieving sub-percent precision in cross-section measurements within the neutrino energy range of~\SIrange[range-units=single]{0.6}{5}{\GeV} would be invaluable for  \acrshort{dune} and \acrshort{hk} since their current uncertainties, $\mathcal{O}(4\%)$, exceed the expected experiment statistical uncertainties and would be the limiting factor in their systematics budgets.

The physics potential of the \acrshort{sbn} experimental reference setup presented in Section \ref{sec:sbn_exp} has been quantified for an integrated luminosity of $1.4\times 10^{19}$ \acrshort{pot}s, corresponding to $\approx5$ years of data-taking. The main highlights are:

\begin{itemize}

\item An absolute measurement of the inclusive 
 $\nu_\mu$~\acrshort{cc} cross section in argon (and possibly water) with a precision of 1~\% as function of
 energy from~\SIrange[range-units=single]{0.6}{5}{\GeV}. Unlike current experiments, the neutrino energy is determined a priori, independent of final state particles reconstruction in the neutrino detector (which is biased by nuclear effects), with precisions of 10--20~\% using the \acrshort{nboa} technique and below 1\% for the tagged events. Figure~\ref{fig:SBN_xsect} shows the expected energy distributions and the corresponding cross-section precision using both methods.

\item A measurement of differential $\nu_\mu$ cross sections where the momentum transfer is known a priori with a $\approx$10~\% precision ($\approx$1\% for tagging). This precision can be achieved in exclusive channels such as single-pion production in \acrshort{cc} and \acrshort{nc} events, thanks to the unprecedented resolution of the liquid argon \acrshort{tpc}.

\item An absolute measurement of the inclusive 
$\nu_e$~\acrshort{cc} cross section in argon (and possibly water) with a precision of 2~\% (5~\%) in the region of interest for \acrshort{dune} (\acrshort{hk}). By comparing measurements of $\nu_e$ and $\nu_\mu$, the ratio of the cross sections, which is currently responsible for the largest projected source of systematic uncertainties on measurements of the \acrshort{cp}-violating phase $\delta$ with \acrshort{dune} and \acrshort{hk}, can also be measured to a similar level of precision.

 \item A direct determination from data of the energy smearing function (an essential input to neutrino oscillation studies but notoriously challenging to model) by comparing the reconstructed neutrino energy in the neutrino detector (as used in oscillation analyses) to the true neutrino energy estimated with neutrino tagging. 

\end{itemize}

\begin{figure}[h!]
    \centering
    \includegraphics[width=0.483\linewidth]{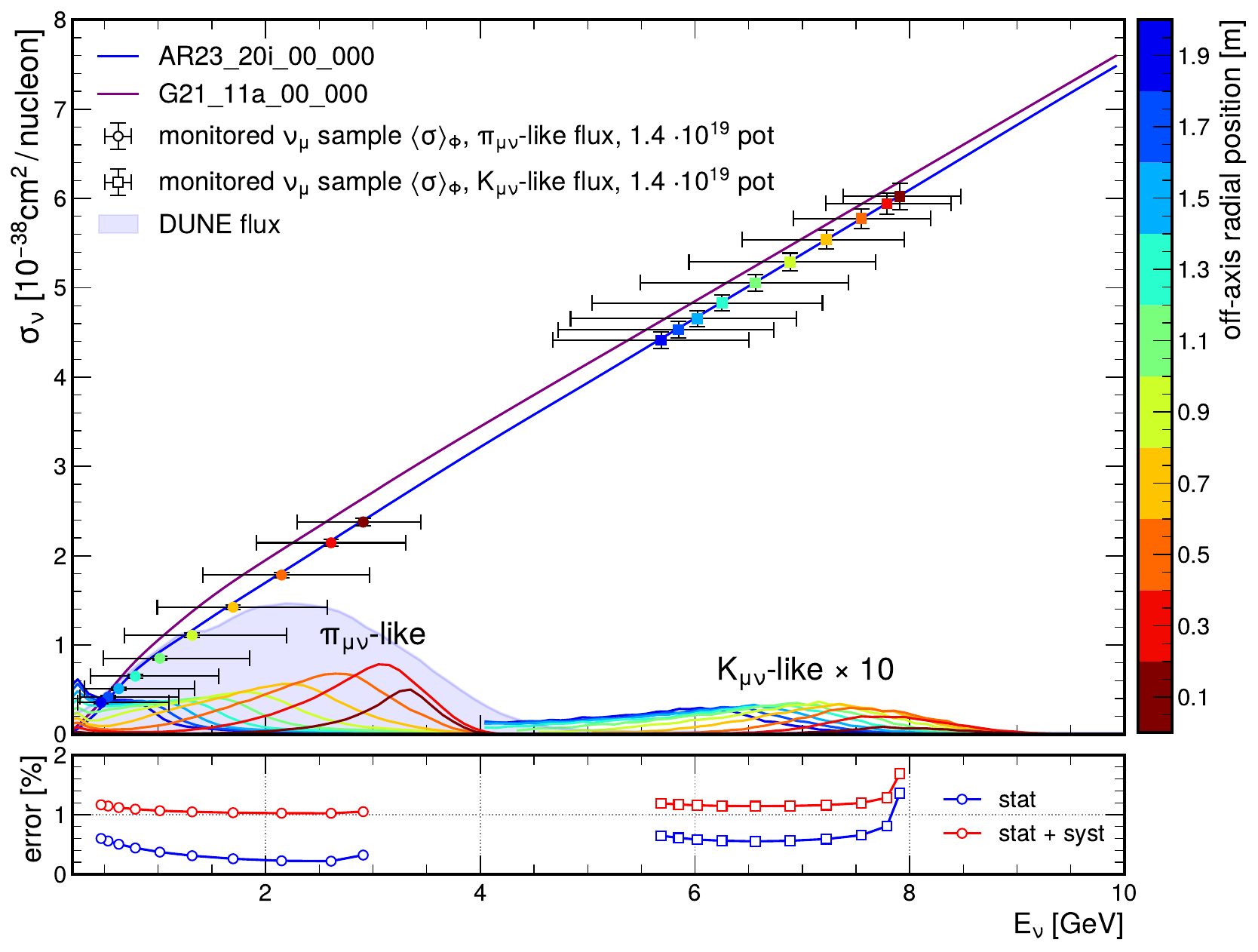}
    \includegraphics[width=0.458\linewidth]{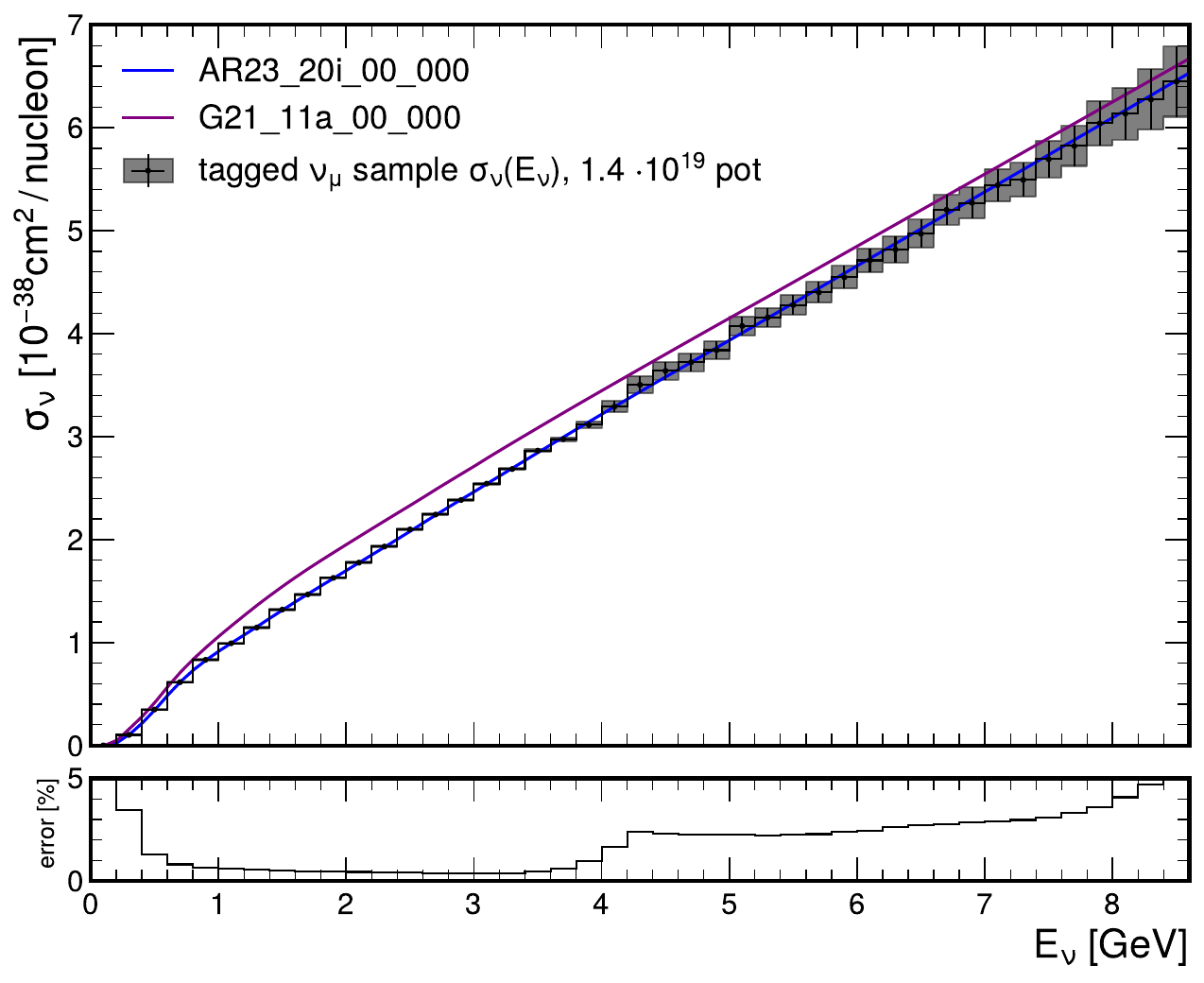}
    \caption{
Determinations of the $\nu_\mu$ \acrshort{cc} inclusive cross section with the \acrshort{sbn} facility. The measurements are compared to two models (blue and red lines).
Left: Flux averaged cross section as a function of neutrino energy using the \acrshort{nboa} technique. The colored lines correspond to the fluxes at different radial distances, given by the colored scale to the right of the figure. The $K_{\mu\nu}$ decay component of each flux has been artificially inflated by a factor of 10 for illustration purposes. Each \acrshort{nboa} flux has a corresponding predicted measurement point of the same color. Horizontal error bars encase the 68\% percentiles with respect to the mean energy for the \acrshort{nboa} fluxes. The underlying figure shows the size of uncertainties due to available statistics (blue) and considering also systematic uncertainties related to the monitored flux prediction, assumed to be $\sim$1\%, (red). The \acrshort{dune} near-detector flux is shown for reference using an arbitrary normalization. Right: Projected measurement of the cross section as a function of neutrino energy using tagged neutrinos. In this case, the individual neutrino energies are measured with a sub-percent precision by reconstructing the kinematics of the parent meson 2-body decays. The error bars represent the statistical error expected on the measurement, also shown below the main figure. (Figures taken from \cite{SBN@CERN_ESPPU}.)}
    \label{fig:SBN_xsect}
\end{figure}

Given the outstanding precision in source characterization, \acrshort{sbn} could also explore \acrshort{bsm} processes, such as \acrfull{nsi}, the oscillation of sterile neutrinos at the~\SI{1}{\eV^2} mass scale, and the production of neutral \acrshort{llps}. Additionally, the facility's performance in anti-neutrino mode for the study of $\bar{\nu}_\mu$ cross sections and the first direct measurement of $\bar{\nu}_e$ interactions are under evaluation. 

Last but not least, the neutrino tagging technique may pave the way to a post \acrshort{dune}/\acrshort{hk} sustainable high-precision Long Baseline Program in synergy with large deep-sea detectors developed for neutrino astronomy~\cite{hep-ph_Perrin-Terrin_2022,hep-ph_Baratto-RoldanEtAl_2024}.

\paragraph{Collider Neutrino Experiments}

\begin{figure}[htb]
\centering
\includegraphics[width=0.99\textwidth]{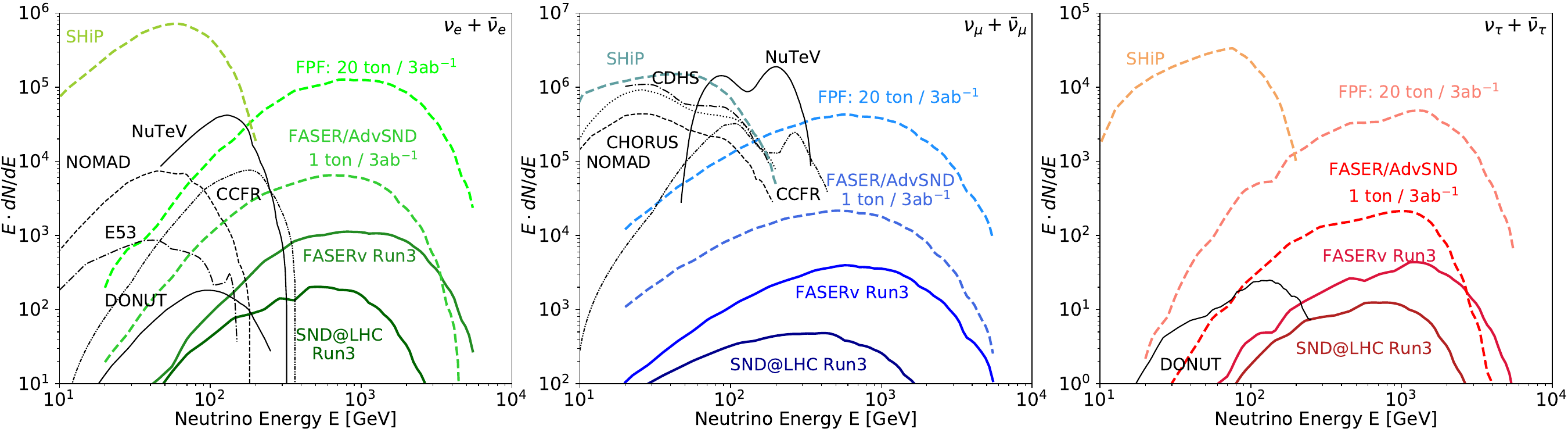}
\caption{\textbf{Neutrino yields at existing and proposed \acrshort{lhc} neutrino experiments.} The plots show the expected event rate and energy spectrum of neutrinos interacting at collider neutrino experiments for all three neutrinos flavors, following Ref.~\cite{FASER:2024ykc}. Spectra from previous accelerator experiments and the planned \acrshort{ship} experiment are shown for comparison~\cite{Ahdida:2867743}. (Figure adapted from Ref.~\cite{Ariga:2025qup}.)
}
\label{fig:nu_fluxes}
\end{figure}

As the highest-energy particle collider built to date, the \acrshort{lhc} is also the source of the most energetic human-made neutrinos. Indeed, it generates an intense, collimated beam of neutrinos of all three flavors with TeV energies in the forward direction. These neutrinos have been detected for the first time by the \acrshort{lhc} far-forward \acrshort{faser} and \acrshort{sndlhc} experiments in 2023~\cite{FASER:2023zcr, SNDLHC:2023pun}, a breakthrough observation making the \textit{dawn of collider neutrino physics}~\cite{Worcester:2023njy}, and first measurements of the neutrino flux and cross section have already been performed~\cite{FASER:2024hoe, FASER:2024ref}. Both \acrshort{faser}~\cite{FASER:2025myb} and \acrshort{sndlhc} will operate until the end of \acrshort{lhc} Run~3, where they are expected to collect about $10^4$ neutrino interactions, and plan to continue operation with upgraded neutrino detectors during the \acrshort{hllhc} era, where they could collect and study about $10^5$ interactions. Additional neutrino detectors, \acrshort{fasernu2} and \acrshort{flare}, with significantly larger target mass have been proposed as part of the \acrshort{fpf}~\cite{Anchordoqui:2021ghd, Feng:2022inv, Adhikary:2024nlv, FPFWorkingGroups:2025rsc}, and could collect more than $10^6$ neutrino interactions. Additional detector places at the surface exit points of the neutrino beam were also considered, but would require 1000 times larger detectors and correspondingly cheaper detector technologies to collect a comparable event rate~\cite{Ariga:2025jgv, Kamp:2025phs}. 

Figure~\ref{fig:nu_fluxes} shows the event rates and energy spectra for these experiments, which will probe neutrino energies from hundreds of GeV to several TeV, i.e., well beyond current accelerator experiments, and with far higher event rates than \acrshort{faser} and \acrshort{sndlhc} in Run~3. The high statistics enable precision studies, including measurements of neutrino cross sections for all three flavors and tests of lepton flavor universality in neutrino scattering, complementing results from flavor physics and main \acrshort{lhc} experiments. Notably, the large number of expected tau neutrino interactions, identifiable in the \acrshort{fasernu2} emulsion detectors, will allow for detailed studies of tau neutrinos, the least known of all \acrshort{sm} particles. Measurements at \acrshort{fasernu2} will also allow for detecting the anti-tau neutrino for the first time. 

\begin{figure}[htb]
\centering
\includegraphics[width=0.99\textwidth]{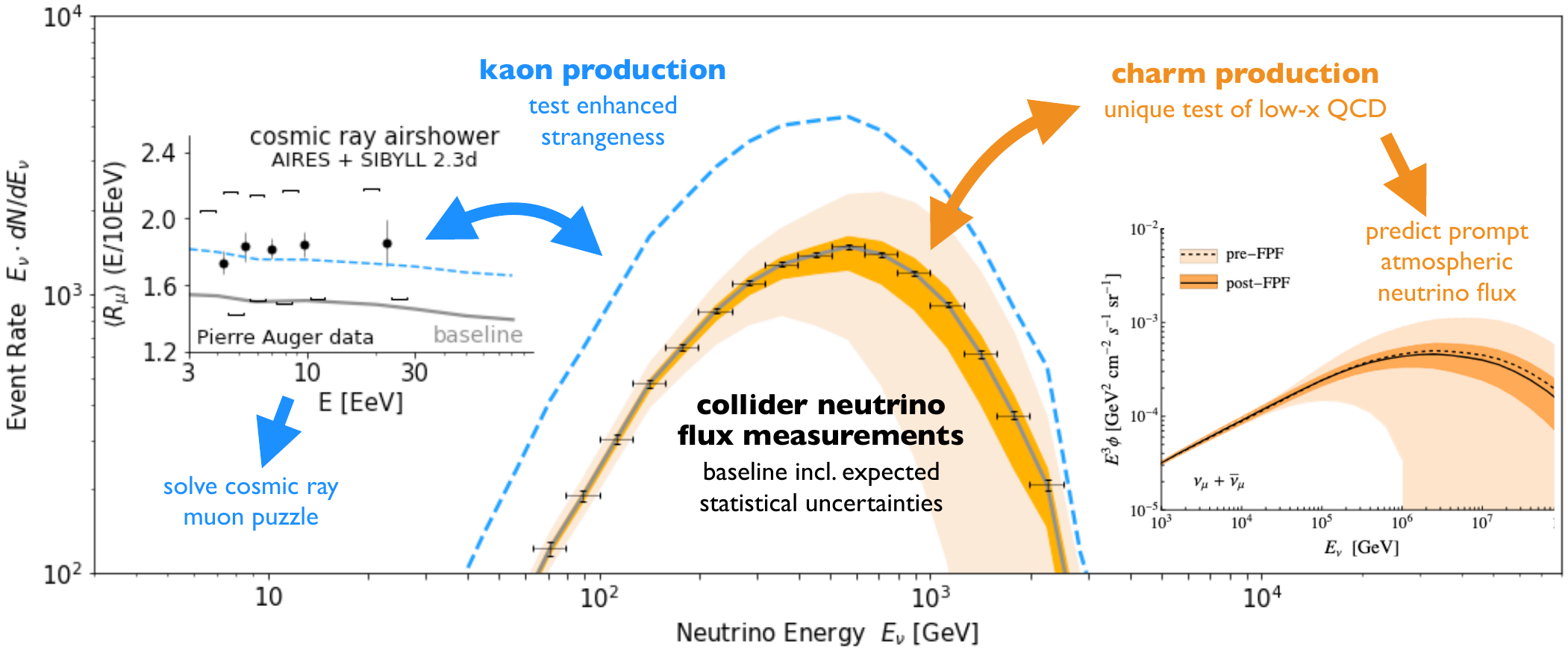}
\caption{\textbf{Astroparticle physics at collider neutrino experiments.} The central part of the figure show expected energy spectrum of interacting electron neutrinos in the \acrshort{flare} detector at the \acrshort{fpf} (solid gray curve) and expected statistical uncertainties (black error bars). The left part considers a model of strangeness enhancement introduced in Ref.~\cite{Anchordoqui:2022fpn} which can resolve the discrepancy in the dimensionless muon shower content $R_\mu$ for air shower data and would lead to sizable changes of the neutrino energy spectrum at the \acrshort{fpf} and can therefore be tested. The right side of the figure shows how \acrshort{fpf} data will reduce \acrshort{pdf} uncertainties (orange band) on the prompt neutrino flux $\Phi$ as a function of $E_\nu$~\cite{Gauld:2015kvh, Rojo:2024tho, Bai:2022xad}. 
(Figure from Ref.~\cite{Adhikary:2024nlv}.)
} 
\label{fig:Spectrum}
\end{figure}

\acrshort{lhc} neutrino flux measurements will significantly impact astroparticle physics by validating and improving models of forward particle production. These models are essential for understanding particle production in extreme astrophysical environments and cosmic-ray interactions with Earth’s atmosphere. Collider neutrino experiments offer a unique opportunity to test these models at comparable energies under controlled conditions. For instance, they will shed light on the cosmic ray muon puzzle---a long-standing tension between measured and predicted muon counts in high-energy cosmic-ray air showers~\cite{Soldin:2021wyv}. Thorough analyses suggest that an enhanced forward strangeness production could resolve this discrepancy~\cite{Albrecht:2021cxw}: this scenario can be tested using the \acrshort{lhc} neutrino flux as shown in Figure~\ref{fig:Spectrum}. Additionally, the figure shows how collider data will reduce uncertainties in the prompt atmospheric neutrino flux~\cite{Gauld:2015kvh}---arising from charmed hadron decays in cosmic-ray collisions---which is the dominating background for astrophysical neutrino searches above a few 100~TeV~\cite{Bai:2022xad}. Collider neutrino data will therefore enhance astrophysical neutrino studies and multi-messenger astronomy~\cite{Bai:2022xad}.

\paragraph{Neutrinos at \acrshort{ship}} 

Proton interactions on the tungsten target produce a very intense neutrino flux, of all three flavors. Indeed, the beam dump is the ideal instrument to enrich the neutrino yield of tau neutrinos, since the thick and high-density target absorbs most of the hadrons rather than letting them decay, thus enhancing neutrinos from prompt, i.e., charmed hadron, decays. This feature enhances enormously the tau as well as the electron neutrino components when compared to muon neutrinos, thus allowing for unique studies of their behavior when interacting with matter. More than 90\% of the $\nu_e$s interacting in the neutrino detector target are expected to originate from charmed hadron decays. Therefore, systematic uncertainties from the neutrino flux cancel out when, e.g., forming the $\nu_e/\nu_{\tau}$ ratio, thus enabling \acrshort{lfu} studies with neutrino interactions with unprecedented accuracy. 
The collected dataset will enable a rich program of neutrino physics measurements, providing valuable \acrshort{sm} tests and serving as an essential benchmark for future neutrino experiments. \acrshort{ship} will measure nucleon/nucleus structure functions, probe the strange quark content of the nucleons, study neutrino-induced charm production and $V_{cd}$ matrix elements with unprecedented accuracy. Notably, this aspect of \acrshort{ship}’s program offers a set of guaranteed measurements, ensuring high-impact physics results even in the absence of \acrshort{bsm} discoveries.

\subsection{Feebly-interacting particles and dark sectors}
\label{subsec:HiddenSector}

In spite of the many successes of the \acrshort{sm} of particle physics, there are strong reasons for expecting new physics beyond the \acrshort{sm}. From the theoretical viewpoint, it would be desirable to explain the large hierarchies of masses and apparent fine-tuning of parameters present in the \acrshort{sm}. On the experimental side, there are various observations that cannot be explained within the \acrshort{sm}, such as the amount of particle-antiparticle asymmetry in the universe, non-zero neutrino masses and the existence of a new form of matter called \acrlong{dm}. All these problems may be addressed by postulating the existence of new particles at or below the GeV scale that have thus far evaded detection because of their feeble couplings to known matter. Examples include \acrshort{qcd} axions~\cite{Agrawal:2017ksf,Hook:2019qoh}, relaxions~\cite{Flacke:2016szy,Banerjee:2020kww}, right-handed neutrinos~\cite{Asaka:2005pn,Bondarenko:2018ptm} and mediators of \acrshort{dm} interactions~\cite{Feng:2008ya}, such as new scalar or gauge bosons arising from a spontaneously broken symmetry. All these particles, collectively referred to as \acrfull{fips}, have been the focus of great experimental and theoretical efforts~\cite{PBC_summary_2020,Beacham:2019nyx,Agrawal:2021dbo,Antel:2023hkf}. 

To manage the large number of potential \acrshort{fips} models with different coupling structures, \acrshort{pbc} pursues a model-agnostic bottom-up approach~\cite{Batell:2009di}, considering a small number of benchmark scenarios with coupling structures that resemble interactions known from the \acrshort{sm}~\cite{Beacham:2019nyx}. These benchmark scenarios highlight the complementarity of different \acrshort{fips} production modes (including for example Higgs decays, meson decays and bremsstrahlung) and decay channels (including both charged and fully neutral final states). A key feature of \acrshort{fips} is that, due to their small masses and feeble couplings, their proper decay length can be comparable to typical detector dimensions, and they can be produced with substantial boost factors. In large regions of parameter space, \acrshort{fips} are therefore predicted to escape from the \acrshort{lhc} main detectors before decaying, but they can be targeted with lower-energy accelerators or new \acrshort{lhc} detectors placed further away from the production point. Depending on the specific processes under consideration, these dedicated detectors can be situated either in the forward direction or at large angles from the beam axis. 

In the following, we present the sensitivity of various \acrshort{pbc} experiments for a variety of different \acrshort{fips} models. The sensitivities corresponds to an integrated luminosity of $300~\text{fb}^{-1}$ for projects using collisions at \acrshort{lhcb} and $3000~\text{fb}^{-1}$ for projects using collisions at \acrshort{cms} or \acrshort{atlas}, and $6 \times 10^{20}$ \acrshort{pot} for \acrshort{ship}.

\subsubsection{Benchmark models}

The \acrshort{pbc} \acrfull{bcs} were proposed originally in~\cite{Beacham:2019nyx}, then developed further in~\cite{Agrawal:2021dbo,Antel:2023hkf}, and have by now become a community standard. While several of these benchmarks are of primary relevance for high-intensity beam-dump experiments, others can be targeted by a wider range of complementary search strategies. The former \acrshort{bcs} have been presented in detail in~\cite{Ahdida:2867743}; therefore we focus here on the latter. We emphasize that in spite of substantial efforts by the community, the various production and decay modes of \acrshort{fips} are still subject to sizable theoretical uncertainties. 
The sensitivity projections of the various experiments treat these uncertainties in different ways. Hence,  care must be taken when comparing different curves.

\acrshort{bc}3 considers particles $\chi$ with a fractional electric charge (so-called millicharge). In Figure~\ref{fig:bc3_gray}, we show the sensitivity projections and existing limits for the parameter space of this model in the plane of the effective charge $Q_{\chi}$ in units of elementary charge $e$ versus mass $m_{\chi}$.

\begin{figure}[t]
\centering
\includegraphics[width=0.73\textwidth]{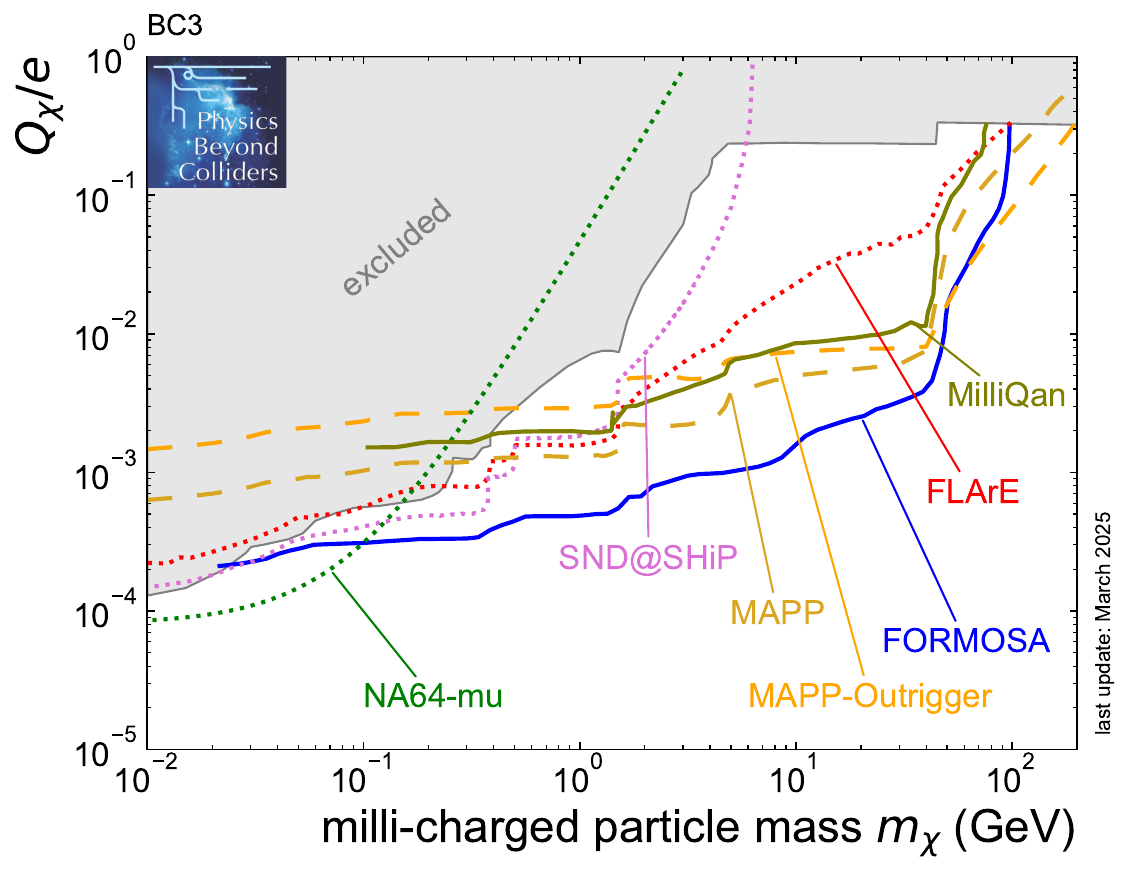}%
\caption{\label{fig:bc3_gray} Sensitivity projections for the \acrshort{fpc} benchmark \acrshort{bc}3 for millicharged particles $\chi$ in the plane of the effective charge $Q_{\chi}$ in units of elementary charge $e$ versus $\chi$ mass $m_{\chi}$. 
For the various projections, the line style reflects the maturity of the background estimates: solid lines correspond to background estimates based on the extrapolation of existing data sets, dashed lines indicate background estimates based on full Monte Carlo simulations, and dotted lines represent projections based on toy Monte Carlo simulations or on the assumption that backgrounds are negligible.
Upper limits on the effective charge $Q_{\chi}/e$ as function of the milli-charged particle mass $m_\chi$ stem from 
ArgoNeuT\,\cite{ArgoNeuT:2019ckq}, MilliQan-Prototype\,\cite{milliQan:2021lne}, 
MilliQ/SLAC\,\cite{Prinz:1998ua}, LSND\,\cite{Magill:2018tbb,LSND:2001akn}, SuperK\,\cite{Wu:2024iqm},
SENSEI\,\cite{SENSEI:2023gie}, and \acrshort{cms}\,\cite{CMS:2012xi} as well as from reinterpretations of BEBC\,\cite{Marocco:2020dqu} and a combination of \acrshort{lep}-based results\,\cite{milliQan:2021lne}.
Note that these reinterpretations were not performed by the experimental collaborations and without access to raw data.
Relevant sensitivity is expected also for a proposed LAr detector at \acrshort{ship}, but no detailed estimates are available~\cite{Ferrillo:2023hhg}.}
\end{figure}

\acrshort{bc}5 considers a dark scalar $S$ mixing with the \acrshort{sm} Higgs boson. The dark scalar can be produced both through rare meson decays (relevant for intensity-frontier experiments) and at the \acrshort{lhc} through Higgs boson decays (with $\text{BR}(h \to SS) = 0.01$). In Figure~\ref{fig:bc5_gray}, we show sensitivity projections and existing limits on the parameter space of this model in the $\sin\theta$ vs. $m_S$ plane under the assumption that both $\lambda$ and $\mu$ are nonzero, see~\cite{Beacham:2019nyx} for details. 

\begin{figure}[t]
\centering
\includegraphics[width=0.75\textwidth]{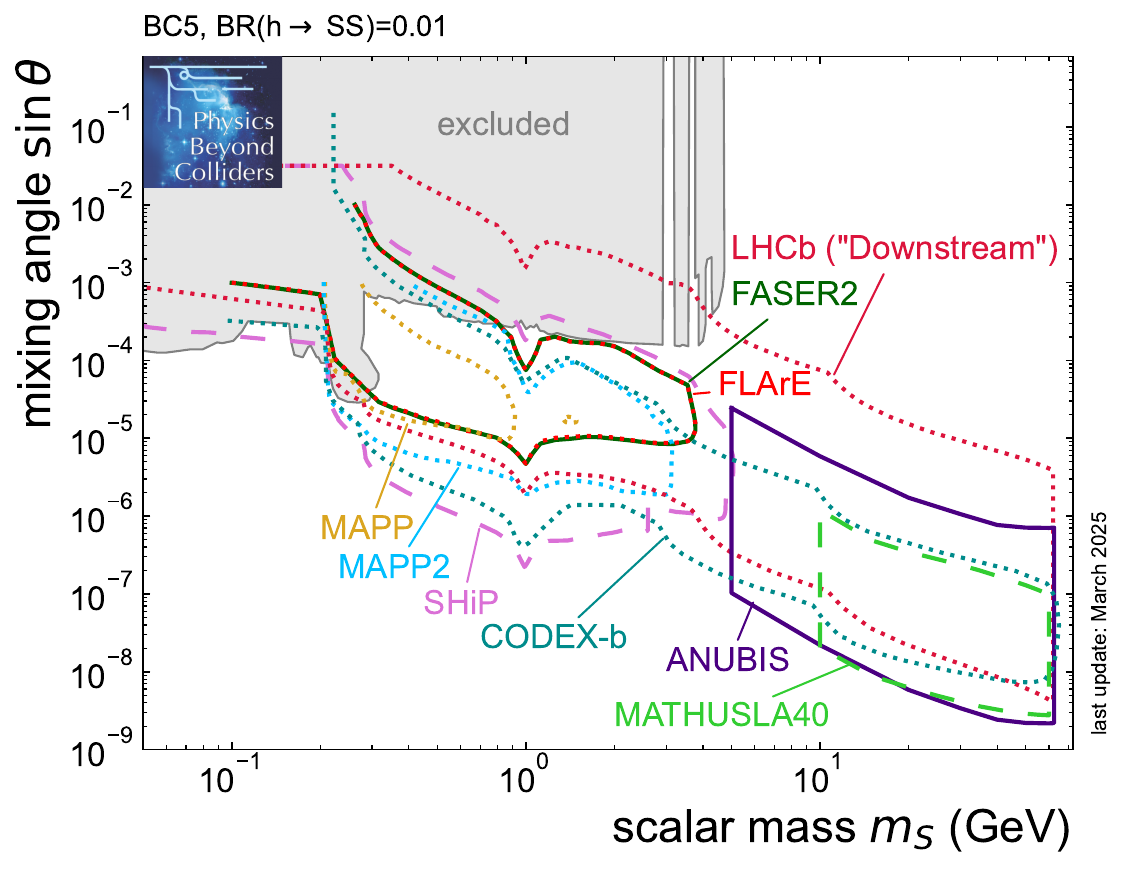}%
\caption{\label{fig:bc5_gray} Sensitivity projections for the \acrshort{fpc} benchmark \acrshort{bc}5 for the dark scalar mixing with the Higgs in the plane of mixing angle $\sin \theta$ versus dark scalar mass $m_S$ under the hypothesis that both parameters $\lambda$ and $\mu$ are different from zero. 
For the various projections, the line style reflects the maturity of the background estimates: solid lines correspond to background estimates based on the extrapolation of existing data sets, dashed lines indicate background estimates based on full Monte Carlo simulations, and dotted lines represent projections based on toy Monte Carlo simulations or on the assumption that backgrounds are negligible.
The sensitivity projections have been obtained assuming $\text{BR}(h \to SS) = 10^{-2}$.
The sensitivity projection labeled `LHCb~("Downstream")' is taken from \cite{Gorkavenko:2023nbk}.
Upper limits on the mixing angle $\sin\theta$ as function of scalar mass $m_S$ from
E949\,\cite{BNL-E949:2009dza},
\acrshort{na62}\,\cite{NA62:2020pwi,NA62:2021zjw},
MicroBooNE\,\cite{MicroBooNE:2021sov,MicroBooNE:2022ctm,Ferber:2023iso},
KOTO\,\cite{KOTO:2020prk, Ferber:2023iso}, 
ICARUS\,\cite{ICARUS:2024oqb}, 
\acrshort{lhcb} \,\cite{LHCb:2015nkv, LHCb:2016awg, Winkler:2018qyg},
\acrshort{belleii}\,\cite{Belle-II:2023ueh}) as well as reinterpretations of PS191\,\cite{Gorbunov:2021ccu}, CHARM\,\cite{CHARM:1985anb, Winkler:2018qyg}, and LSND~\cite{Foroughi-Abari:2020gju}.
Note that these reinterpretations were not performed by the experimental collaborations and without access to raw data.
The sensitivity of \acrshort{mathusla40} for smaller masses is under study.
\acrshort{belleii} is expected to providing a distinctive handle on the production mode by reconstructing the complete $B$-meson decay~\cite{DallaValleGarcia:2025aeq}.
}
\end{figure}

\acrshort{bc}7 considers \acrshort{hnls} coupling exclusively to the second lepton generation. They can be produced both in rare meson decays (relevant for intensity‐frontier experiments) and in electroweak processes at the \acrshort{lhc}. In Figure~\ref{fig:pbc_bc7_gray_final}, we show sensitivity projections and existing limits on the parameter space of this model in the $\vert U_{\mu}\vert^2 = \theta_{\mu}^2$ versus $m_N$ plane under the hypothesis that $\vert U_e\vert^2 = \vert U_{\tau}\vert^2 = 0$, see~\cite{Beacham:2019nyx} for details. We note that models of \acrshort{hnls} linked to the generation of neutrino masses typically have a more complex flavour structure~\cite{Drewes:2022akb} but lead to a rather similar phenomenology~\cite{Feng:2024zfe}.

\begin{figure}[t]
\centering
\includegraphics[width=0.75\textwidth]{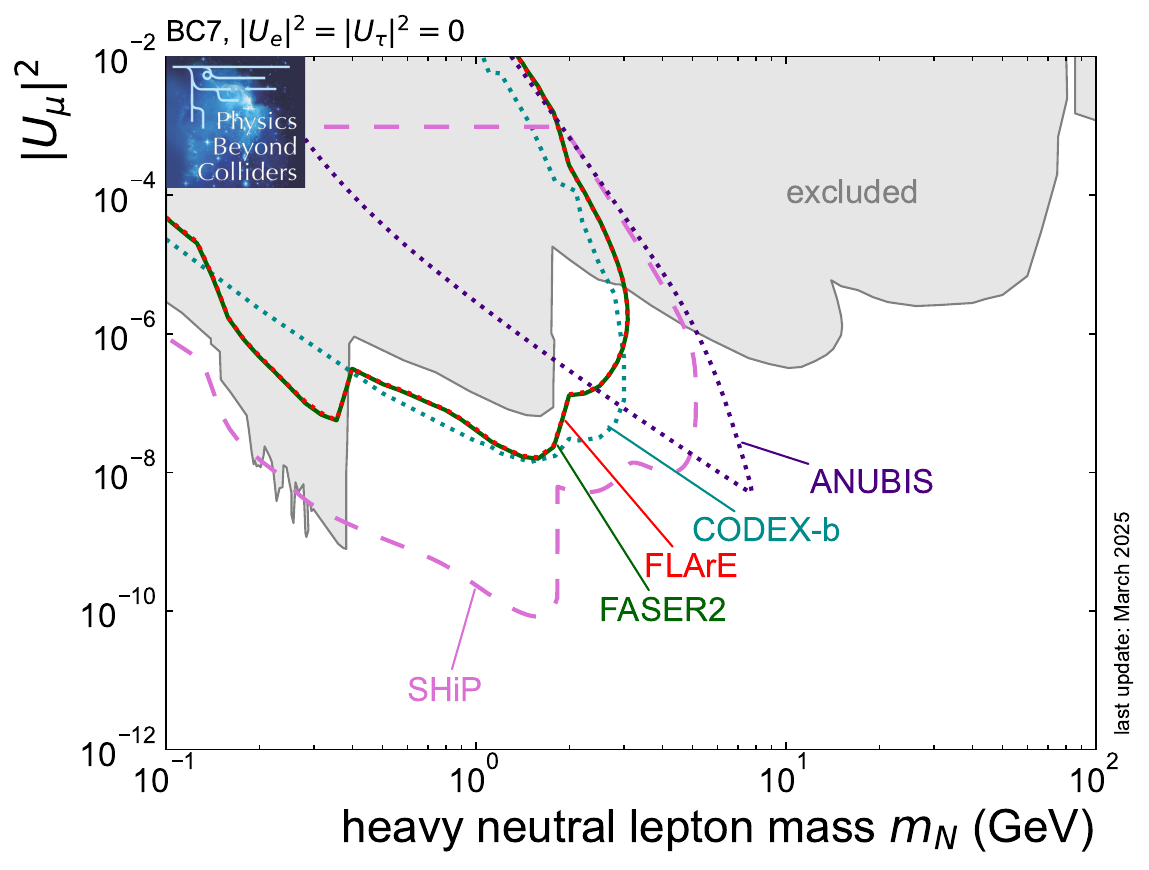}%
\caption{\label{fig:pbc_bc7_gray_final} Sensitivity projections for the \acrshort{fpc} benchmark \acrshort{bc}7 for \acrshort{hnls} with coupling to the second lepton generation only in the plane of the squared magnitude $\vert U_{\mu}\vert^2 = \theta_{\mu}^2$ of the mixing angle $\theta_{\mu}$ versus \acrshort{hnls} mass $m_N$ under the hypothesis that $\vert U_e\vert^2 = \vert U_{\tau}\vert^2 = 0$. 
For the various projections, the line style reflects the maturity of the background estimates: solid lines correspond to background estimates based on the extrapolation of existing data sets, dashed lines indicate background estimates based on full Monte Carlo simulations, and dotted lines represent projections based on toy Monte Carlo simulations or on the assumption that backgrounds are negligible.
Upper limits on the mixing angle $\vert U_{\mu}\vert^2$ as function of \acrshort{hnls} mass $m_N$ from 
NuTeV\,(\hbox{90\,\% CL}\,\cite{NuTeV:1999kej}), 
BEBC\,(\hbox{90\,\% CL}\,\cite{WA66:1985mfx}),
T2K\,(\hbox{90\,\% CL}\,\cite{T2K:2019jwa}),
\acrshort{na62}\,(\hbox{90\,\% CL}\,\cite{NA62:2021bji}),
E949\,(\hbox{90\,\% CL}\,\cite{E949:2014gsn}),
MicroBooNE\,(\hbox{90\,\% CL}\,\cite{MicroBooNE:2023eef}),
\acrshort{cms}\,(\hbox{95\,\% CL}\,\cite{CMS:2022fut, CMS:2024ake, CMS:2024xdq}),
\acrshort{atlas}\,(\hbox{95\,\% CL}\,\cite{ATLAS:2022atq}),
CHARM-II\,(\hbox{90\,\% CL}\,\cite{CHARMII:1994jjr}), and 
DELPHI\,(\hbox{90\,\% CL}\,\cite{DELPHI:1996qcc}).
Relevant sensitivity is expected also for \acrshort{mathusla40} as well as for long-lived particle searches at \acrshort{atlas} and \acrshort{cms} using \acrshort{hllhc} data, but no detailed estimates are available. Projected sensitivities for LHCb~("Downstream") using \acrshort{hllhc} data can be found in \cite{Gorkavenko:2023nbk}.}
\end{figure}

\subsubsection{Showcase models}

While the benchmark models established by the \acrshort{pbc} initiative~\cite{Beacham:2019nyx} have been a worldwide success and a valuable asset for the development and presentation of experimental proposals, the landscape of these models is not at all static. 
New or modified models are constantly being proposed and discussed by the \acrshort{fpc}, in order to refine the phenomenological description and determine benchmark cases of particular interest. 
In the following, we present several examples of such models, which have been selected to highlight exciting physics opportunities for specific \acrshort{pbc} experiments not covered by the usual benchmark models.
We emphasize that these models have not yet reached the same level of sophistication as the benchmark models discussed above, which means in particular that not all potentially relevant experiments have been able to perform the signal and background simulations needed for detailed sensitivity studies. 
Just because a specific experiment is not shown in these plots therefore does not mean that the experiment is insensitive. 
We comment on these additional opportunities in the respective figure captions.

\begin{figure}[t]
\centering
\includegraphics[width=0.75\textwidth]{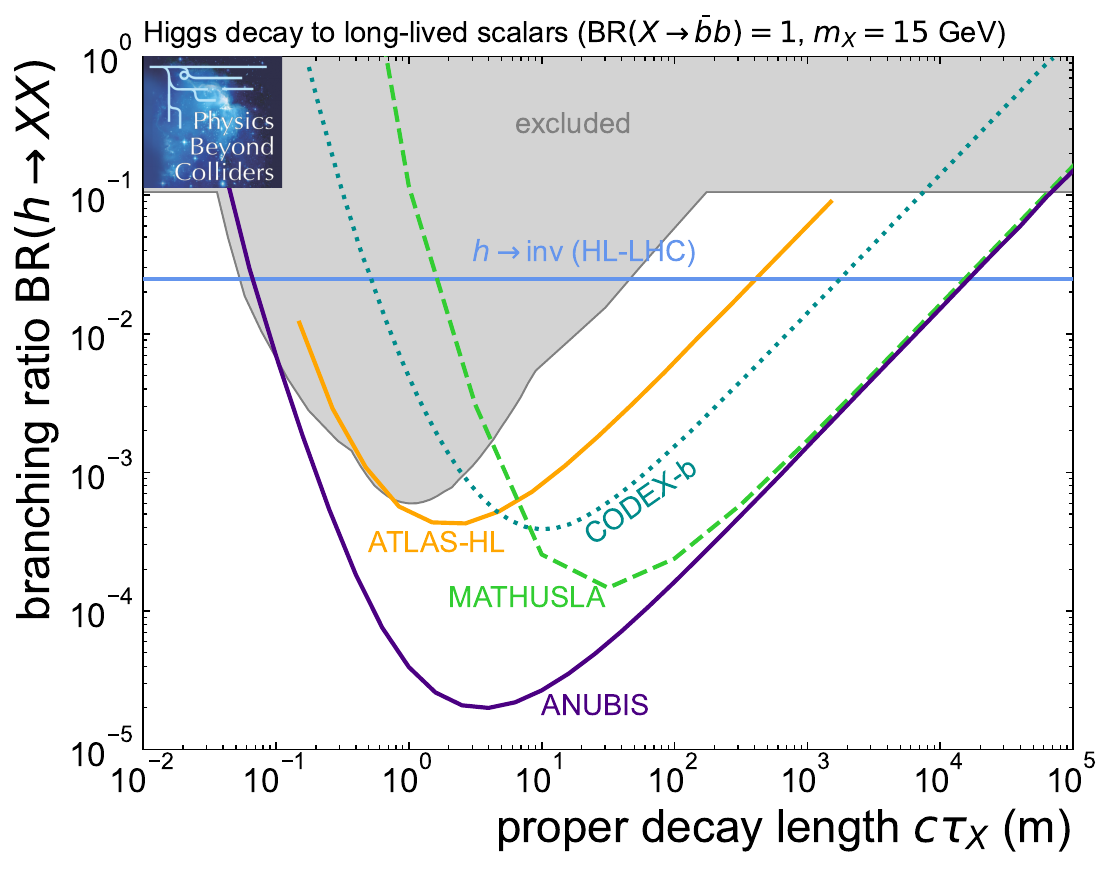}%
\caption{\label{fig:pbc_h_to_scalar} Sensitivity projections for Higgs boson decays into long-lived (pseudo)scalars with $m_X = 15\,\mathrm{GeV}$ in the plane of proper decay length $c \tau_X$ versus the exotic Higgs branching ratio $\text{BR}(h \to XX)$. For concreteness we assume that the (pseudo)scalars have 100\% branching ratio into $b$-quark pairs, but similar sensitivities are expected for other hadronic final states. 
For the various projections, the line style reflects the maturity of the background estimates: solid lines correspond to background estimates based on the extrapolation of existing data sets, dashed lines indicate background estimates based on full Monte Carlo simulations, and dotted lines represent projections based on toy Monte Carlo simulations or on the assumption that backgrounds are negligible. Existing exclusion limits come from \acrshort{atlas}~\cite{ATLAS:2022gbw} and \acrshort{cms}~\cite{CMS:2024bvl}, as well as from a combination of different bounds on the Higgs invisible branching ratio~\cite{ATLAS:2023tkt}. The corresponding projection for \acrshort{hllhc} is taken from ref.~\cite{Dainese:2019rgk}, while the projection for \acrshort{llp} searches at \acrshort{atlas} is taken from Ref.~\cite{Coccaro:2016lnz}. Relevant sensitivity is expected also for \acrshort{llp} searches at \acrshort{atlas} and \acrshort{cms} using \acrshort{hllhc} data, but no detailed estimates are available. In addition, LHCb ("Downstream") may also have sensitivity to this model, utilizing the same search strategy as for the \acrshort{bc}5 model~\cite{Gorkavenko:2023nbk}.}
\end{figure}

\paragraph*{Exotic Higgs decays to long-lived scalars}

Extended Higgs sectors often feature additional scalar or pseudoscalar states ($X$) lighter than the \acrshort{sm} Higgs boson, which couple to the latter through mixed quartic terms in the scalar potential~\cite{Craig:2015pha}. For $m_X < m_h / 2$, this opens up the possibility that the \acrshort{sm} Higgs boson decays into pairs of such (pseudo)scalars~\cite{Curtin:2013fra}. Based on the hypothesis of minimal flavour violation, these particles are expected to couple to fermions proportional to their mass. Couplings to down-type quarks may be further enhanced for example in Two-Higgs doublet models~\cite{Branco:2011iw}. In such a setup, the (pseudo)scalar is expected to decay with nearly 100\% probability into $b$-quark pairs. The decay width, and hence the proper decay length $c \tau_X$, is however independent from the production mode, such that detectors placed at different distance from the interaction points at the \acrshort{lhc} are needed for optimal coverage of the model parameter space. We show current constraints and sensitivity projections in Figure~\ref{fig:pbc_h_to_scalar}.

\begin{figure}[t]
\centering
\includegraphics[width=0.75\textwidth]{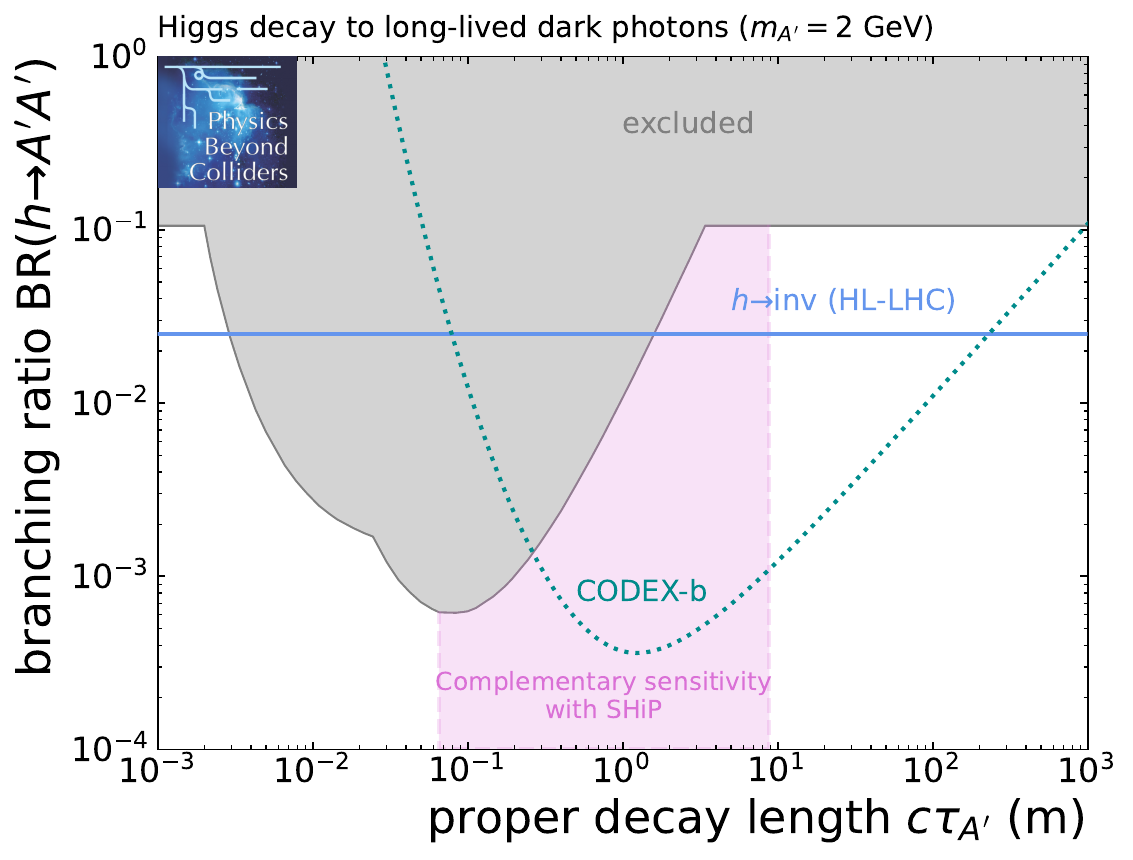}%
\caption{\label{fig:pbc_h_to_dark_photon} Sensitivity projections for Higgs boson decays into long-lived dark photons with mass $m_{A'} = 2 \, \mathrm{GeV}$ in the plane of proper decay length $c \tau_{A'}$ versus the exotic Higgs branching ratio $\text{BR}(h \to A' A')$. 
For the various projections, the line style reflects the maturity of the background estimates: solid lines correspond to background estimates based on the extrapolation of existing data sets, dashed lines indicate background estimates based on full Monte Carlo simulations, and dotted lines represent projections based on toy Monte Carlo simulations or on the assumption that backgrounds are negligible. Existing exclusion limits come from an \acrshort{atlas} search~\cite{ATLAS:2022izj} and from a combination of different bounds on the Higgs invisible branching ratio~\cite{ATLAS:2023tkt}. The corresponding projection for \acrshort{hllhc} is taken from Ref.~\cite{Dainese:2019rgk}. In the red shaded region, dark photons can be directly produced via kinetic mixing and detected at \acrshort{ship}, offering complementary detection prospects. Relevant sensitivity is expected also for \acrshort{mathusla40} and \acrshort{anubis}, as well as for \acrshort{atlas} and \acrshort{cms} using \acrshort{hllhc} data, but no detailed estimates are available.}
\end{figure}

\paragraph*{Exotic Higgs decays to long-lived dark photons}

Models of dark photons ($A'$) often do not specify the mechanism that generates the gauge boson mass. If the mass is generated via a new scalar field analogous to the Higgs mechanism of the \acrshort{sm}, the two Higgs fields can mix with each other~\cite{Schabinger:2005ei,Patt:2006fw}. As a result, the \acrshort{sm} Higgs boson obtains a new tree-level decay mode $h \to A^\prime A^\prime$ analogous to the \acrshort{sm} decay modes into $Z$ and $W$ bosons~\cite{Curtin:2014cca}. The branching ratio depends primarily on the Higgs mixing and can be varied independently from the dark photon mass $m_{A'}$ and the other couplings of the dark photon such as the kinetic mixing, which determine its proper decay length $c \tau_{A'}$~\cite{Fabbrichesi:2020wbt}. These so-called Hidden Abelian Higgs Models can be probed by searches for \acrshort{llps} produced in Higgs boson decays~\cite{Ferber:2023iso}. Dark photons with mass below a few GeV will decay from a displaced vertex into a low multiplicity of charged leptons or hadrons. This signature presents a challenge for the \acrshort{lhc} main detectors but can be targeted with dedicated detectors for \acrshort{llps} such as \acrshort{codexb}. Complementary constraints come from SHiP, which is not sensitive to exotic Higgs boson decays but can search for dark photons produced directly via kinetic mixing, which gives an independent parameter that does not directly affect the branching ratio. 
Other experiments that are sensitive to long-lived dark photons produced via kinetic mixing either probe shorter lifetimes (e.g., \acrshort{belleii}~\cite{Ferber:2022ewf}) or are not sensitive to such heavy dark photons. 
We show current constraints and sensitivity projections in Figure~\ref{fig:pbc_h_to_dark_photon}.

\begin{figure}[t]
\centering
\includegraphics[width=0.75\textwidth]{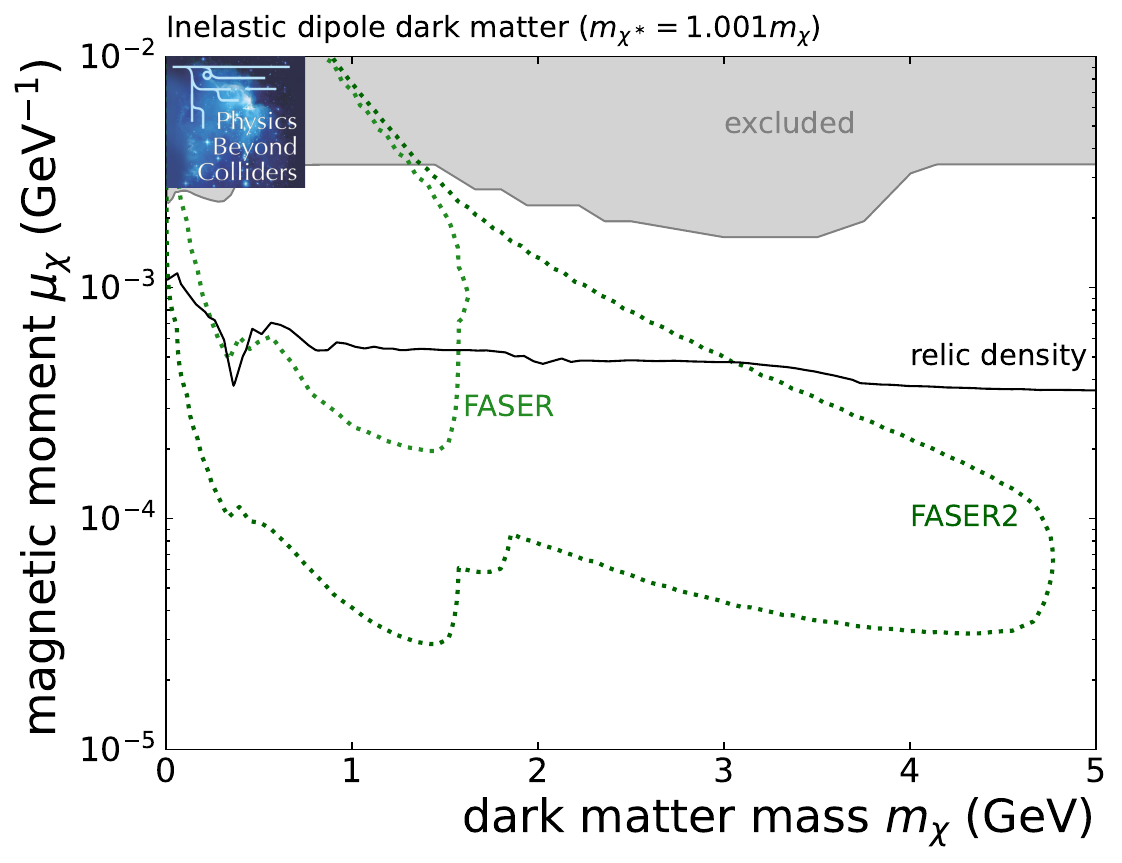}%
\caption{\label{fig:pbc_iDM} Sensitivity projections for inelastic-dipole \acrshort{dm} in the plane of \acrshort{dm} mass $m_\chi$ versus magnetic dipole moment $\mu_\chi$ for a very small mass splitting of $m_{\chi^\ast} = 1.001 m_\chi$. 
For the various projections, the line style reflects the maturity of the background estimates: solid lines correspond to background estimates based on the extrapolation of existing data sets, dashed lines indicate background estimates based on full Monte Carlo simulations, and dotted lines represent projections based on toy Monte Carlo simulations or on the assumption that backgrounds are negligible.
Existing exclusion limits come from reinterpretations~\cite{Izaguirre:2015zva} of  \acrshort{lep}~\cite{OPAL:1993ezs} and BaBar~\cite{BaBar:2017tiz} data.
Note that these reinterpretations were not performed by the experimental collaborations and without access to raw data. For larger mass splittings than considered here, relevant sensitivity is expected also for \acrshort{ship} (using both \acrshort{snd} and LAr detectors) and for \acrshort{llp} searches at \acrshort{belleii}.}
\end{figure}

\paragraph*{Inelastic-Dipole Dark Matter}

Models of inelastic \acrshort{dm}, which involve transitions between two states $\chi$ and $\chi^\ast$ of slightly different mass, provide attractive targets for accelerator experiments, since the strong constraints from direct and indirect detection experiments can be evaded and the observed \acrshort{dm} relic density can be reproduced~\cite{Izaguirre:2015zva}. A particularly simple realisation of this idea is inelastic dipole \acrshort{dm}, in which \acrshort{dm} particles possess a transition dipole moment that couples directly to the electromagnetic field strength tensor, such that the entire model is characterised by the two masses and the dipole moment~\cite{Dienes:2023uve}. For sufficiently large mass splitting ($m_{\chi^\ast} \gtrsim 1.1 m_\chi$) the model can be constrained with a variety of collider and fixed-target experiments, such as \acrshort{belleii}~\cite{Duerr:2019dmv} and long-lived particle detectors such as \acrshort{ship}~\cite{Berlin:2018jbm} or NA64~\cite{Mongillo:2023hbs}. However, for much smaller mass splittings, the energy released in the de-excitation process $\chi^\ast \to \chi + \gamma$ is so small that it can only be detected if $\chi^\ast$ is highly boosted. In this case, forward experiments like \acrshort{faser2} have particularly promising prospects~\cite{Dienes:2023uve}. We show current constraints and sensitivity projections in Figure~\ref{fig:pbc_iDM}.\\

\begin{figure}[t]
\centering
\includegraphics[width=0.75\textwidth]{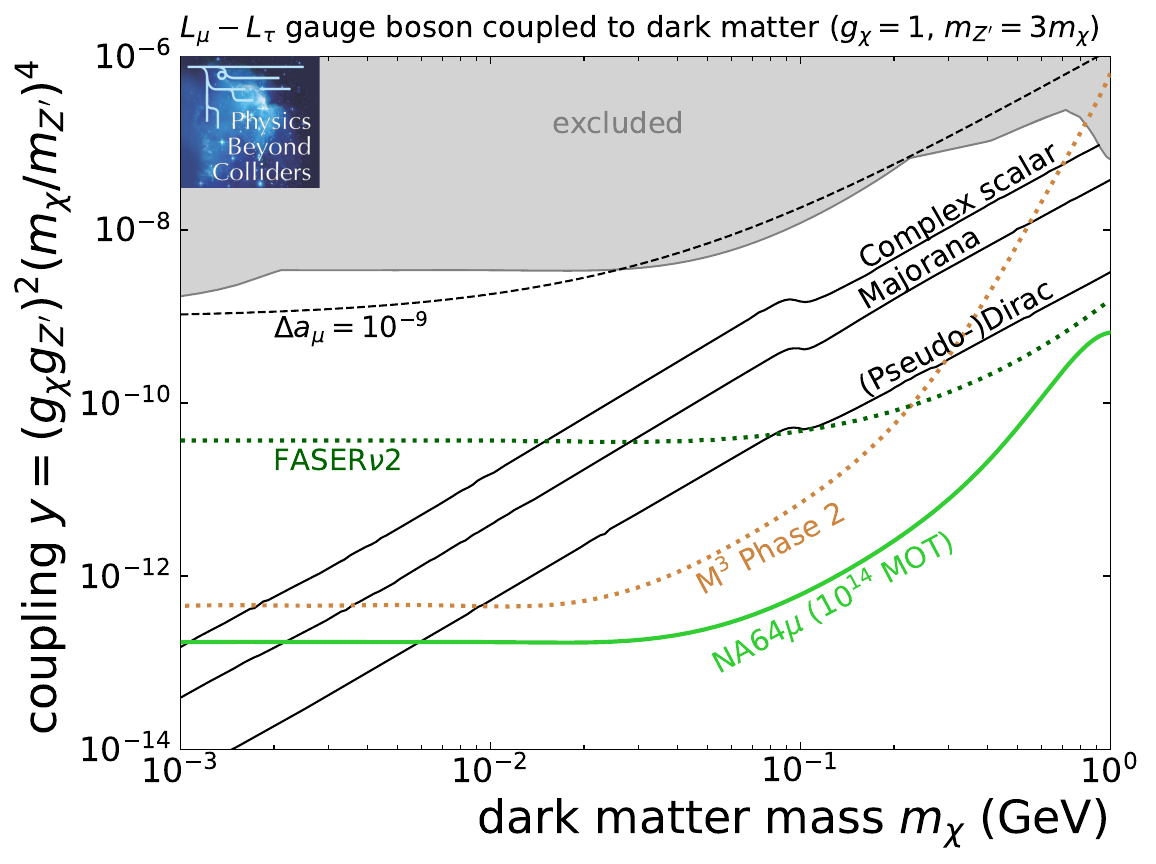}%
\caption{\label{fig:pbc_Lmu-Ltau} Sensitivity projections for a $L_\mu - L_\tau$ gauge boson coupled to \acrshort{dm} in the plane of \acrshort{dm} mass $m_\chi$ versus the effective \acrshort{dm}-\acrshort{sm} coupling $y = (g_\chi g_{Z'})^2 (m_\chi / m_{Z'})^4$.  Here $g_\chi = 1$ denotes the coupling of the \acrshort{dm} particle to the $Z'$ gauge boson and the mass ratio has been fixed to $m_{Z'} = 3 m_\chi$, while the gauge coupling $g_{Z'}$ is a free parameter. 
For the various projections, the line style reflects the maturity of the background estimates: solid lines correspond to background estimates based on the extrapolation of existing data sets, dashed lines indicate background estimates based on full Monte Carlo simulations, and dotted lines represent projections based on toy Monte Carlo simulations or on the assumption that backgrounds are negligible.
Existing exclusion limits come from \acrshort{na64}$e$~\cite{Andreev:2024lps}, \acrshort{na64}$\mu$~\cite{NA64:2024klw}, \acrshort{belleii}~\cite{Belle-II:2022yaw} and a reinterpretation~\cite{Altmannshofer:2014pba} of data from CCFR~\cite{CCFR:1991lpl}. Note that this reinterpretation was not performed by the experimental collaboration and without access to raw data. Relevant sensitivity is expected also for future \acrshort{belleii} searches with more data, but no detailed estimates are available.}
\end{figure}

\paragraph*{$L_{\mu}-L_{\tau}$ gauge boson}

Among the simplest gauge extensions of the \acrshort{sm} are models involving gauged lepton number $L$ and/or baryon number $B$~\cite{Ilten:2018crw}. These models can be made anomaly-free if one includes three right-handed neutrinos that are singlets under the \acrshort{sm} gauge group. While gauged $B-L$ is tightly constrained by electron-positron colliders and electron beam-dump experiments, these constraints are evaded in models of gauged $L_\mu - L_\tau$, where couplings to electrons are absent~\cite{Bauer:2018onh}. As a result, these models can give a sizable contribution to the muon anomalous magnetic moment $\Delta a_\mu$~\cite{Chen:2017awl}. If the corresponding gauge boson $Z'$ acts as mediator between the \acrshort{sm} and a \acrshort{dm} particle $\chi$, it is possible to reproduce the observed \acrshort{dm} relic abundance for complex scalar \acrshort{dm}, Majorana \acrshort{dm} and (Pseudo-)Dirac \acrshort{dm} \cite{Berlin:2018bsc}. The most promising experiments to probe the interesting regions of parameter space are those involving muon beams, such as \acrshort{na64}$\mu$~\cite{NA64:2024klw} or the proposed $M^3$ experiment~\cite{Kahn:2018cqs}, as well as experiments sensitive to particle scattering in forward detectors, such as \acrshort{fasernu2}. We show current constraints and sensitivity projections in Figure~\ref{fig:pbc_Lmu-Ltau}.\\[1cm]

\subsubsection{Experiments table}
Several experiments within \acrshort{pbc} have been proposed to investigate signatures of \acrshort{bsm} physics. 
These experiments primarily target \acrshort{llps}, \acrshort{dm} candidates, neutrinos, and milli-charged particles. 
Table~\ref{tab:bsm_comparison} provides a summary of the experimental parameters for these proposed experiments.
The comparison shows accelerator facilities, interaction points, detector acceptance in pseudorapidity and azimuthal coverage, decay volume specifications, and detector instrumentation.

\newcommand{\plainfootnote}[1]{%
  \begingroup
  \renewcommand{\thefootnote}{}
  \footnotetext{#1}%
  \addtocounter{footnote}{-1}%
  \endgroup
}

\begin{table}[ht!]
\scriptsize
\centering
\begin{tabular}{lccccccc}
\hline\hline
 & \acrshort{mathusla40} & \acrshort{codexb} & \acrshort{anubis} & \acrshort{faser2} & \acrshort{fasernu2} & \acrshort{formosa} & \acrshort{flare} \\
\midrule
Accelerator         & \acrshort{hllhc} & \acrshort{hllhc} & \acrshort{hllhc} & \acrshort{hllhc}   & \acrshort{hllhc}    & \acrshort{hllhc}  & \acrshort{hllhc} \\
IP/Facility         & \acrshort{cms}    & \acrshort{lhcb}   & \acrshort{atlas}  & \acrshort{fpf}      & \acrshort{fpf}        & \acrshort{fpf}      & \acrshort{fpf}  \\
Physics focus       & \acrshort{llp}    & \acrshort{llp}    & \acrshort{llp}    & $\nu$, \acrshort{llp}  & $\nu$, \acrshort{dm} & \acrshort{mqp}     & $\nu$, \acrshort{dm}, \acrshort{llp}, \acrshort{mqp} \\
$\eta$ coverage      & 0.69 -- 1.1 & 0.13 -- 0.54 & -0.965 -- 0.965$^{*}$ & $>6.7$  & $>8$    & $>7$ & $>5.6$ \\
$\phi$ coverage [rad]& 0.5  & 0.36   & 1.53$^{*}$  & $2\pi$  & $2\pi$   & $2\pi$  & $2\pi$ \\
Distance [m]        & 120   & 25     & 23$^{*}$    & 620     & 620      & 620     & 620 \\
Overburden [mwe$^{\diamond}$] & 0 & 250  & 150$^{**}$  & 250     & 250      & 250     & 250 \\
Decay volume length [m]    & 14    & 10     & 15    & 10      & 6$^{\dagger}$ & 5    & 7 \\
Magnetic field      & no    & no     & yes$^{\mathparagraph}$ & yes     & no       & no      & yes$^{\ddagger}$ \\
Timing [ps]         & 1000   & 100    & 250   & 100     & no       & 100     & 300 \\
Calorimeter         & no    & no     & no    & yes     & yes      & no      & yes \\
Tracking            & yes   & yes    & yes   & yes     & yes      & approximate & yes \\
Charge threshold    & 1e    & 1e     & 1e    & 1e      & 1e       & <0.001e & 0.01e \\
Reference           & \cite{Aitken:2025cjq}    & \cite{Aielli:2019ivi} & \cite{Bauer:2019vqk,Satterthwaite:2839063} & \cite{Adhikary:2024nlv,Feng:2022inv,Salin:2927003} & \cite{FASER:2024hoe} & \cite{Foroughi-Abari:2020qar,Feng:2022inv} & \cite{FLArESimNote,FLArETechNote} \\
\hline\hline
\end{tabular}

\vspace{1em} 

\begin{tabular}{lcccccc}
\hline\hline 
 & \acrshort{na64} & \acrshort{snd} & \acrshort{ship}  & \acrshort{mappoutrigger} & \acrshort{mapp2} & \acrshort{sndhllhc} \\
\midrule
Accelerator         & \acrshort{sps}  & \acrshort{sps}  & \acrshort{sps}    & \acrshort{lhc}, \acrshort{hllhc}   & \acrshort{lhc}, \acrshort{hllhc}   & \acrshort{hllhc} \\
IP/Facility         & -  & \acrshort{bdf} & \acrshort{bdf}   & \acrshort{lhcb}          & \acrshort{lhcb}          & \acrshort{atlas} \\
Physics focus       & \acrshort{llp}, \acrshort{dm} & $\nu$, \acrshort{dm}& \acrshort{llp}    & \acrshort{mqp}, \acrshort{llp}    & \acrshort{llp}          & $\nu$,\acrshort{dm}\\
$\eta$ coverage      & fixed target & $>2.2$ & $>1.7$ & -4.4 -- -3.3 & 1.4 -- 3.0 & 6.9 -- 7.6 \\
$\phi$ coverage [rad]& $2\pi$ & $2\pi$ & $2\pi$ & 0.07--0.2  & 0.2--0.9  & 0.79 \\
Distance [m]        & active dump & 28  & 32   & 120           & 25--55      & 480 \\
Overburden [mwe$^{\diamond}$] & 0 & 0   & 0    & 270           & 270         & 250 \\
Decay volume length [m]    & 5    & ~1.2$^{\dagger}$ & 50   & up to 60$^{\S}$ & 6--15       & 0.9$^{\dagger}$ \\
Magnetic field      & yes  & yes  & yes   & no            & no          & yes \\
Timing [ps]         & 1000 & 500  & 100   & 500          & 1000        & 50 \\
Calorimeter         & yes  & yes  & yes   & no            & no          & yes \\
Tracking            & yes  & yes  & yes   & approximate   & yes         & yes \\
Charge threshold    & 1e   & 1e   & 1e    & 0.01e  & 1e          & 1e \\
Reference           & \cite{NA64:2017vtt} & \cite{Golutvin:2917226} & \cite{Albanese:2878604,Golutvin:2917226} & \cite{Pinfold:2918254} & \cite{MoEDAL-MAPP:2022kyr} & \cite{Abbaneo:2926288} \\
\hline\hline
\end{tabular}
\caption{Summary of experimental parameters for proposed PBC BSM experiments designed to probe \acrshort{llps}, \acrshort{dm} candidates, neutrinos, and \acrshort{mqp}s. }
\label{tab:bsm_comparison}
\end{table}

\begin{minipage}{\textwidth}
\footnotesize
\plainfootnote{* This is the \acrshort{anubis} "ceiling" configuration described in \cite{Satterthwaite:2839063}.}
\plainfootnote{** This value does not account for the two access shafts to the \acrshort{atlas} cavern that are air-filled. Depending on the angle of the incoming cosmic ray and its position, the effective overburden is significantly lower.} 
\plainfootnote{$\dagger$ Length of instrumented target instead of an empty decay volume.} 
\plainfootnote{$\S$ The distance between \acrshort{lhcb} and the \acrshort{mappoutrigger} is filled with concrete and rock. \acrshort{llp} decay products must have sufficient momentum to avoid being absorbed. No decay vertex reconstruction is possible.} 
\plainfootnote{$\mathparagraph$ Stray magnetic field of the \acrshort{atlas} muon spectrometer.}
\plainfootnote{$\ddagger$ Magnet only in HCAL section, not in LAr.} 
\plainfootnote{$\diamond$ The approximate shielding against cosmic particles is given in meter water equivalent (mwe), which represents the depth of water that would provide the same level of attenuation, assuming a rock density of 2.51 g/cm$^3$.}
\end{minipage}

\subsection{Electric dipole moment}
\label{subsec:EDM}

Particles with spin, not identical to their antiparticles, possess an \acrshort{mdm}, which characterizes the strength of the spin coupling to external magnetic field. In addition to the \acrshort{mdm}, the spin can also contain a coupling to the external electric field, with the strength regulated by the \acrshort{edm}. 
These have long been identified as powerful probes of new \acrshort{bsm} physics (cf., e.g., \cite{Abel:2001vy,Pospelov:2005pr,Abel:2005er,Chupp:2014gka} for some overviews).
The \acrshort{edm}s are forbidden if $P$ and $T$ are exact symmetries, and the search for \acrshort{edm}s is equivalent to searches of \acrshort{cp} violation. Unlike the decays of $K,\,B,\,D$ mesons, where \acrshort{cp} violation has been observed, \acrshort{edm}s represent the test of \acrshort{cp} violation in the flavour-diagonal channel. Such violation is extremely well motivated by theories of baryogenesis, i.e., the dynamical generation of the baryon asymmetry of the Universe from the initially particle-antiparticle symmetric state. Multi-decade efforts to detect \acrshort{edm}s resulted in the upper limits on the \acrshort{edm}s of neutrons, nuclei, electrons, and muons. Only indirect limits exist for the \acrshort{edm}s of heavier flavours. Also, the proton \acrshort{edm} $d_p$  limit can be extracted from the atomic \acrshort{edm} experiments, but right now lag behind sensitivity to neutron \acrshort{edm}, $d_n$. 

The \acrshort{sm} is believed to have two sources of \acrshort{cp} violation. One firmly established source, the Kobayashi-Maskawa phase of the quark weak charged currents, predicts tiny \acrshort{edm}s not observable in the near future. The second source of \acrshort{cp} violation, the \acrshort{qcd} theta term, if large, would predict very large \acrshort{edm}s for neutrons, protons and nuclei, and therefore is severely limited. The absence of the \acrshort{edm}s induced by theta motivates the dynamical mechanism of its removal, connected to a new particle, the \acrshort{qcd} axion. Thus, searches of \acrshort{edm}s represent the searches of new \acrshort{cp}-violating physics, that manifest itself over the expected \acrshort{edm}s from the Kobayashi-Maskawa phase.

\begin{figure}[tb]
\centering
\includegraphics[width=0.75\textwidth]{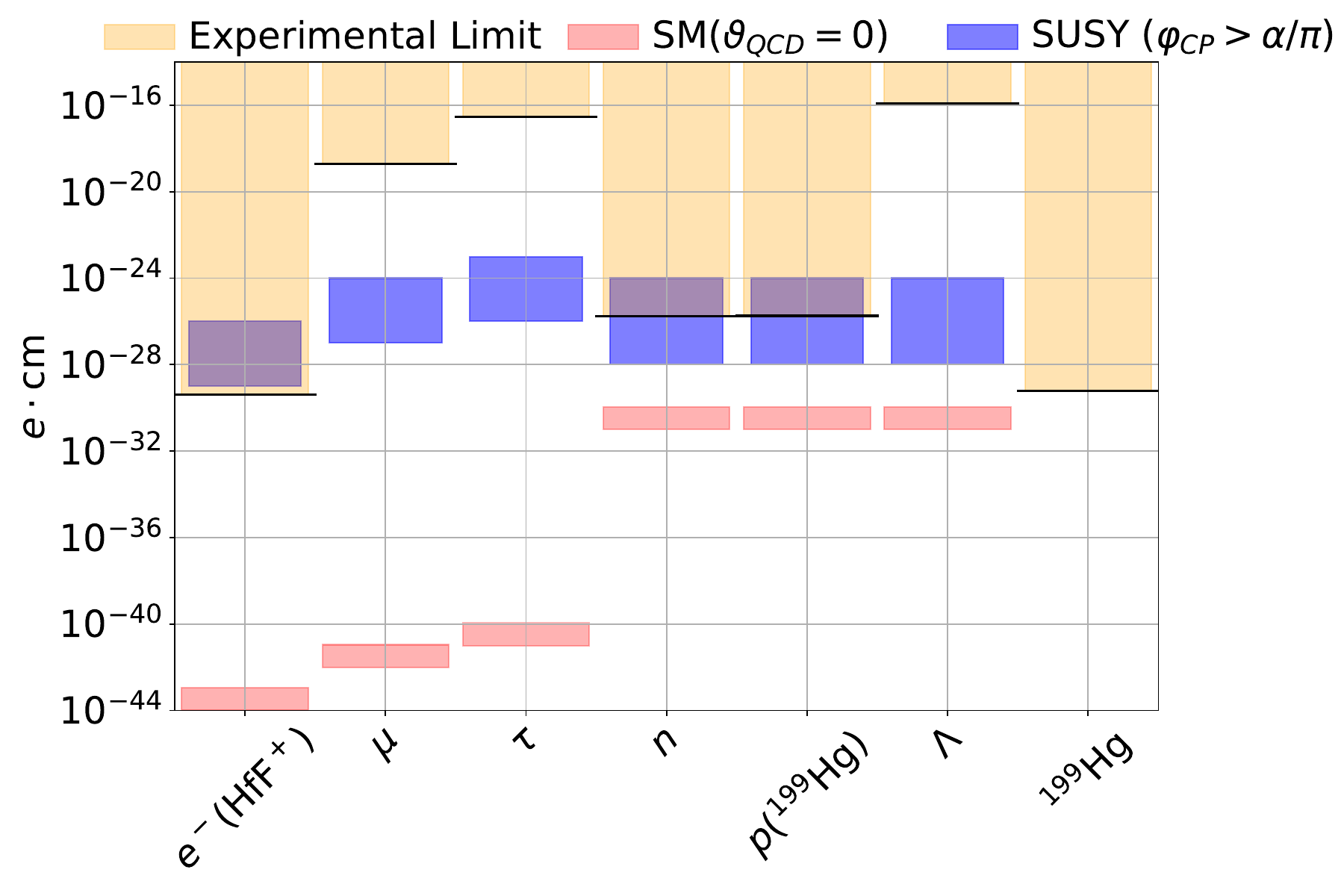}
\caption{Upper Limits on \acrshort{edm}s for various leptons, hadrons and the Hg atom~\cite{Roussy:2022cmp,Muong-2:2008ebm,Belle:2021ybo,Abel:2020pzs,ACME:2018yjb,Graner:2016ses,Pondrom:1981gu}, and corresponding expectations from the \acrshort{sm} and its super-symmetric extension. (Figure adapted from Ref.~\cite{Pretz:2024lbv}; cf., e.g., Refs.~\cite{Abel:2001vy,Pospelov:2005pr,Abel:2005er} for some SUSY estimates.)} 
\label{fig:EDMs}
\end{figure}

Current \acrshort{edm} sensitivities (with the limit on electron \acrshort{edm} being at $4\times 10^{-30}e\cdot$cm~\cite{Roussy:2022cmp}), depending on the model of New Physics, may be sensitive to energy scales vastly exceeding the electroweak scale, and indeed even the reach of the \acrshort{lhc}, being truly a probe ``beyond the collider reach''. Therefore, the continuation of the \acrshort{edm} program worldwide is one of the priorities of fundamental physics searches. Current experimental limits and predictions for selected cases is presented in Figure~\ref{fig:EDMs}.

The main goal of the \acrshort{cpedm} project is to advance sensitivity to the \acrshort{edm}s of charged particles, and in particular, proton \acrshort{edm} $d_p$. The main idea is to circulate a well controlled and spin-polarized proton beam in a storage ring in the configuration that minimizes spin evolution due to the proton \acrshort{mdm}. The stated goal (cf., e.g., \cite{CPEDM:2018tfu}) of the experiment is to reach the level of $10^{-29}e\cdot$cm, which would make proton probe even more valuable than the current $d_n$ sensitivity and on par with $d_e$, but naturally sensitive to a wider variety of new physics models that involve strongly-interacting particles. 
\acrshort{aladdin}  on the other hand aims at measuring the \acrshort{mdm} and \acrshort{edm} of charmed baryons, such as $\Lambda_c^+$ and $\Xi_c^+$, produced in a fixed-target setup in which protons from the \acrshort{lhc} halo are deflected on a tungsten target and produced baryons channeled through a bent crystal inducing spin precession. Simulation studies suggest that \acrshort{aladdin} can achieve a 10~\% precision on the \acrshort{mdm}s and can set first direct limits on the \acrshort{edm}s at the $10^{-16} e\cdot$cm level within some years of data-taking. It also offers the opportunity to explore the \acrshort{edm} of the $\tau$ lepton.

\subsection{Gravitational Wave Detection}
\label{subsec:GW}

Measurements of \acrfull{gws} have become essential tools for exploring astrophysics and cosmology, opening a new way
to study the Universe beyond traditional electromagnetic observations. Figure~\ref{fig:GW_Spectrum} illustrates some astrophysical sources of \acrshort{gws} and the sensitivities of selected detectors. Experiments based on \acrfull{ai}, such as that suggested for the PX46 access shaft at \acrshort{cern}~\cite{Arduini:2023wce}, offer
 sensitivity in a frequency range that is inaccessible to current and planned laser interferometers such as
\acrshort{ligo}~\cite{Aasi2015}, \acrshort{virgo}~\cite{VIRGO:2014yos}, \acrshort{kagra}~\cite{Aso:2013eba}, \acrshort{et}~\cite{ET:2019dnz} and
\acrshort{lisa}~\cite{LISA:2017pwj}. Among the targets in this frequency range are \acrshort{gws} from the
mergers of black holes with masses intermediate between the stellar-mass black holes discovered by \acrshort{ligo},
\acrshort{virgo} and  \acrshort{kagra} and the \acrfull{smbhs} found in the cores of galaxies~\cite{Badurina:2021rgt}. Measurements of the mergers of such intermediate mass black holes may
cast light on the formation of \acrshort{smbhs}. \acrshort{flash} targets \acrshort{gws} at very high frequencies beyond 100~MHz: for a survey of possible sources of high-frequency \acrshort{gws} from physics beyond the Standard Model, see~\cite{Aggarwal:2025noe}.

\begin{figure}[htb]
\centering
\includegraphics[width=0.8\textwidth]{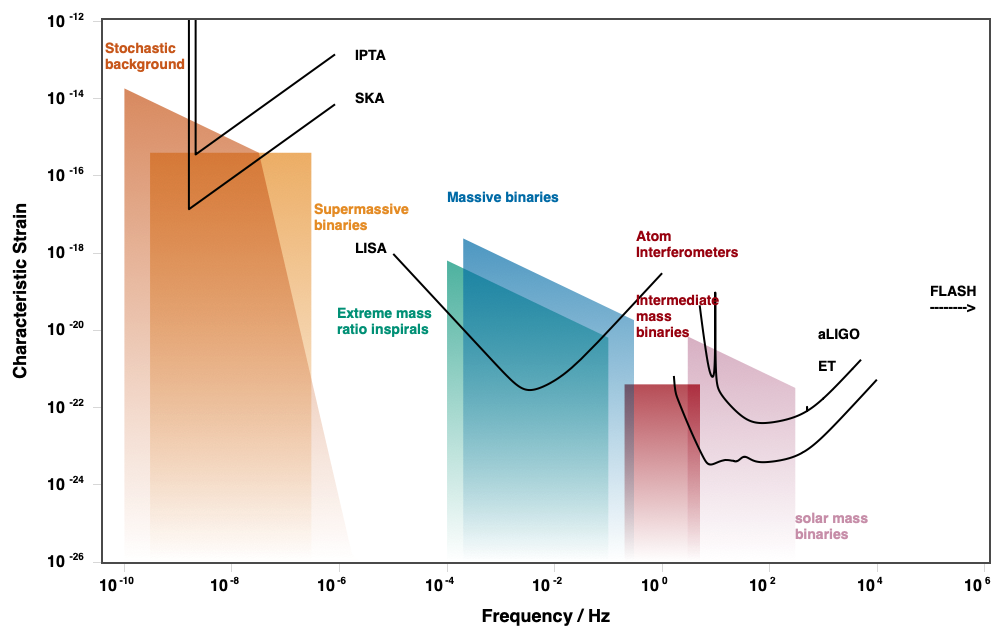}
\caption{Spectrum of possible sources of \acrshort{gws} and the sensitivities of existing and planned detectors. \acrshort{ai} experiments such as that suggested for \acrshort{cern}~\cite{Arduini:2023wce} target the range of frequencies intermediate between \acrshort{ligo}~\cite{Aasi2015}/\acrshort{et}~\cite{ET:2019dnz} and \acrshort{lisa}~\cite{LISA:2017pwj}, while \acrshort{flash}~\cite{Alesini:2023qed} targets \acrshort{hfgw} with frequencies above 100~MHz. Plot prepared using {\tt GWplotter}~\cite{Moore:2014lga}.} 
\label{fig:GW_Spectrum}
\end{figure}

\newpage

\section*{Acknowledgments}

The work of F.Kl.~was supported by the Deutsche Forschungsgemeinschaft under Germany's Excellence Strategy -- EXC 2121 Quantum Universe -- 390833306. 
G.S. acknowledges support by the State Agency for Research of the Spanish Ministry of Science and Innovation through the grant PID2022-136510NB-C33 as well as  via the QuantERA project T-NiSQ grant PCI2022-132984 funded by MCIN/AEI/10.13039/501100011033 and by the European Union “NextGenerationEU”/PRTR.
The work of I.S. was supported by the Swiss National Science Foundation under grant no. 214492.
The work of G.V. was supported by the Leverhulme Trust, LIP-2021-01.
M.M.P. and A.R. acknowledge the support by the Polish Minister of Education and Science (contract No. 2021/WK/10); the work of MMP was also supported by WUT IDUB.

{\small
\bibliographystyle{utphys}
\bibliography{PBCreport}
}

\newpage
\appendix


\newacronym{abs}{ABS}{Atomic Beam Source}
\newacronym{ad}{AD}{Antiproton Decelerator}
\newacronym{advsnd}{AdvSND}{Advanced Scattering and Neutrino Detector}
\newacronym{ai}{AI}{Atom Interferometry}
\newacronym{aion}{AION}{Atom Interferometer Observatory and Network}
\newacronym{akwisp}{aKWISP}{advanced Kinetic Weakly interacting Sub-eV Particle experiment}
\newacronym{aladdin}{ALADDIN}{An Lhc Apparatus for Direct Dipole moments INvestigation}
\newacronym{alice}{ALICE}{A Large Ion Collider Experiment}
\newacronym{alice3}{ALICE3}{A Large Ion Collider Experiment Upgrade~3}
\newacronym{alps}{ALPs}{Axion-Like Particles}
\newacronym{alpsii}{ALPS-II}{Any Light Particle Search-II}
\newacronym{amber}{NA66/AMBER}{Apparatus for Meson and Baryon Experimental Research}
\newacronym{anubis}{ANUBIS}{AN Underground Belayed In-Shaft search experiment}
\newacronym{astae}{ASTAE}{Advancing Science and Technology using Agile Experiments}
\newacronym{atlas}{ATLAS}{A Toroidal LHC ApparatuS Experiment}
\newacronym{auger}{Auger}{Pierre Auger Observatory}
\newacronym{awake}{AWAKE}{Advanced WAKEfield Experiment}
\newacronym{babyiaxo}{BabyIAXO}{Baby International AXion Observatory}
\newacronym{base}{BASE}{Baryon Antibaryon Symmetry Experiment}
\newacronym{bc}{BC}{Benchmark Case}
\newacronym{bcs}{BCs}{Benchmark Cases}
\newacronym{bd}{BD}{Beam Dump}
\newacronym{bdf}{BDF}{Beam Dump Facility}
\newacronym{bdx}{BDX}{Beam Dump Experiment}
\newacronym{belleii}{Belle~II}{Belle~II}
\newacronym{besiii}{BES~III}{Beijing Spectrometer III}
\newacronym{bms}{BMS}{Beam Momentum Spectrometer}
\newacronym{bnl}{BNL}{Brookhaven National Laboratory}
\newacronym{bsm}{BSM}{Beyond the Standard Model}
\newacronym{cbm}{CBM}{Compressed Baryonic Matter experiment}
\newacronym{cbwg}{CBWG}{Conventional Beams Working Group}
\newacronym{cc}{CC}{Charged-Current}
\newacronym{cdr}{CDR}{Conceptual Design Report}
\newacronym{cds}{CDS}{Comprehensive Design Study}
\newacronym{cern}{CERN}{European Organization for Nuclear Research}
\newacronym{cernrb}{CERN RB}{CERN Research Board}
\newacronym{charm}{CHARM}{CERN High Energy Accelerator Mixed-field}
\newacronym{charme}{CHARM Experiment}{CERN HAmburg Rome Moscow Experiment}
\newacronym{ckm}{CKM}{Cabibbo Kobayashi Maskawa}
\newacronym{clic}{CLIC}{Compact LInear Collider}
\newacronym{cloud}{CLOUD}{Cosmics Leaving Outdoor Droplets experiment}
\newacronym{cms}{CMS}{Compact Muon Solenoid}
\newacronym{cngs}{CNGS}{CERN Neutrino beam to Gran Sasso}
\newacronym{codexb}{CODEX-b}{COmpact Detector for EXotics at LHCb}
\newacronym{codexbeta}{CODEX-$\beta$}{Prototype of COmpact Detector for EXotics at LHCb}
\newacronym{com}{\textit{c.m.}}{centre-of-mass}
\newacronym{compass}{COMPASS/NA58}{Common Muon and Proton Apparatus for Structure and Spectroscopy}
\newacronym{cosy}{COSY}{COoler SYnchrotron}
\newacronym{cp}{CP}{Charge Parity}
\newacronym{cpedm}{cpEDM}{charged particle Electric Dipole Moment}
\newacronym{cpv}{CPV}{Charge Parity violation}
\newacronym{ctf2}{CTF2}{CLIC Test Facility 2}
\newacronym{critp}{{\textbf{CP}}}{Critical Point}
\newacronym{desy}{DESY}{Deutsche Elektronen-Synchrotron}
\newacronym{dis}{DIS}{Deep Inelastic Scattering}
\newacronym{dm}{DM}{Dark Matter}
\newacronym{drd}{DRD}{Detector Research and Development programme}
\newacronym{ds}{DS}{Dark Sector}
\newacronym{ds20}{DarkSide-20k}{DarkSide-20k}
\newacronym{dune}{DUNE}{Deep Underground Neutrino Experiment}
\newacronym{dy}{DY}{Drell-Yan}
\newacronym{e144}{E144}{E144 Experiment}
\newacronym{ea}{EA}{East Area}
\newacronym{eacons}{EA-CONS}{East Area Consolidation Project}
\newacronym{ebye}{EbyE}{Event-by-Event}
\newacronym{ecfa}{ECFA}{European Committee for Future Accelerators}
\newacronym{ecn3}{ECN3}{Experimental Cavern North 3}
\newacronym{ecn4}{ECN4}{Experimental Cavern North 4}
\newacronym{edm}{EDM}{Electric Dipole Moment}
\newacronym{ehn1}{EHN1}{Experimental Hall North~1}
\newacronym{ehn2}{EHN2}{Experimental Hall North~2}
\newacronym{eic}{EIC}{Electron Ion Collider}
\newacronym{elena}{ELENA}{Extra Low Energy Antiproton ring}
\newacronym{em}{EM}{Electro-Magnetic}
\newacronym{enubet}{ENUBET}{Enhanced NeUtrino BEams from kaon Tagging}
\newacronym{eoi}{EoI}{Expression of Interest}
\newacronym{eot}{EoT}{Electrons on Target}
\newacronym{eppsu}{EPPSU}{European Particle Physics Strategy Update}
\newacronym{erc}{ERC}{European Research Council}
\newacronym{es}{ES}{Electro-Static}
\newacronym{espp}{ESPP}{European Strategy for Particle Physics}
\newacronym{et}{ET}{Einstein Telescope}
\newacronym{ew}{EW}{Electro-Weak}
\newacronym{hk}{Hyper-K}{Hyper-Kamiokande}
\newacronym{fair}{FAIR}{Facility for Antiproton and Ion Research}
\newacronym{faser}{FASER}{Forward Search Experiment}
\newacronym{faser2}{FASER2}{Forward Search Experiment 2}
\newacronym{fasernu}{FASER$\nu$}{FASER Neutrino detector}
\newacronym{fasernu2}{FASER$\nu$2}{FASER Neutrino Detector 2}
\newacronym{fcc}{FCC}{Future Circular Collider}
\newacronym{fccee}{FCC-ee}{Future electron-positron Circular Collider}
\newacronym{fcchh}{FCC-hh}{Future hadron-hadron Circular Collider}
\newacronym{fnal}{Fermilab}{Fermilab}
\newacronym{finuda}{FINUDA}{FIsica NUcleare a DA$\phi$NE}
\newacronym{fips}{FIPs}{Feebly Interacting Particles}
\newacronym{fpc}{FPC}{FIP Physics Centre}
\newacronym{flare}{FLArE}{Forward Liquid Argon Experiment}
\newacronym{flash}{FLASH}{FINUDA magnet for Light Axion SearcH}
\newacronym{formosa}{FORMOSA}{FORward MicrOcharge SeArch detector}
\newacronym{fpf}{FPF}{Forward Physics Facility}
\newacronym{ft}{FT}{Fixed-Target}
\newacronym{gf}{GF}{Gamma Factory}
\newacronym{gif++}{GIF\texttt{++}}{Gamma Irradiation Facility\texttt{++}}
\newacronym{gfspspop}{GF SPS PoP}{Gamma Factory SPS Proof of Principle}
\newacronym{ghg}{GHG}{Greenhouse Gas}
\newacronym{gpd}{GPD}{Generalized Parton Distribution}
\newacronym{gps}{GPS}{General Purpose Separator}
\newacronym{grahal}{GrAHal}{Grenoble Axion Haloscope}
\newacronym{gtk}{GTK}{Giga-TracKer}
\newacronym{gws}{GWs}{Gravitational Waves}
\newacronym{hades}{HADES}{High Acceptance Di-Electron Spectrometer}
\newacronym{hsds}{HSDS}{Hidden Sector Decay Spectrometer}
\newacronym{hearts}{HEARTS}{High Energy Accelerators for Radiation Testing and Shielding}
\newacronym{hera}{HERA}{Hadron-Elektron-Ringanlage}
\newacronym{hfgw}{HFGW}{High Frequency Gravitational Waves}
\newacronym{hiecn3}{HI-ECN3}{High Intensity ECN3}
\newacronym{hieisolde}{HIE-ISOLDE}{High Intensity and Energy ISOLDE}
\newacronym{hike}{HIKE}{High Intensity Kaon Experiment}
\newacronym{hiradmat}{HiRadMat}{High Radiation to Material}
\newacronym{hllhc}{HL-LHC}{High Luminosity LHC}
\newacronym{hnl}{HNL}{Heavy Neutral Lepton}
\newacronym{hnls}{HNLs}{Heavy Neutral Leptons}
\newacronym{hs}{HS}{Hidden Sector}
\newacronym{hts}{HTS}{High-Temperature Superconductors}
\newacronym{hvp}{HVP}{Hadronic Vacuum Polarisation}
\newacronym{iaxo}{IAXO}{International AXion Observatory}
\newacronym{icecube}{IceCube}{IceCube Neutrino Observatory}
\newacronym{iefc}{IEFC}{Injectors and Experimental Facilities Committee} 
\newacronym{ijclab}{IJCLab}{Laboratory of the Physics of the two Infinities Irène Joliot-Curie}
\newacronym{ilc}{ILC}{International Linear Collider}
\newacronym{infn}{INFN}{Istituto Nazionale di Fisica Nucleare}
\newacronym{ip}{IP}{Interaction Point}
\newacronym{ir}{IR}{Insertion Region}
\newacronym{isolde}{ISOLDE}{Isotope Separator On Line DEvice}
\newacronym{its3}{ITS3}{Inner Tracking System~3}
\newacronym{jparc}{J-PARC}{Japan Proton Accelerator Research Complex}
\newacronym{jlab}{JLab}{Jefferson Lab}
\newacronym{kagra}{KAGRA}{KAmioka GRAvitational wave detector}
\newacronym{km3net}{KM3NeT}{Cubic Kilometre Neutrino Telescope}
\newacronym{l1}{L1}{Level 1}
\newacronym{last}{LAST}{Large Acceptance Silicon Tracker}
\newacronym{lc}{LC}{Linear Collider}
\newacronym{ldm}{LDM}{Light Dark Matter}
\newacronym{ldmx}{LDMX}{Light Dark Matter eXperiment}
\newacronym{leir}{LEIR}{Low Energy Ion Ring}
\newacronym{lep}{LEP}{Large Electron–Positron Collider}
\newacronym{lfu}{LFU}{Lepton Flavour Universality}
\newacronym{lfv}{LFV}{Lepton Flavour Violation}
\newacronym{lhc}{LHC}{Large Hadron Collider}
\newacronym{lhcb}{LHCb}{Large Hadron Collider beauty Experiment}
\newacronym{lhcc}{LHCC}{LHC Experiments Committee}
\newacronym{lhcft}{LHC-FT}{PBC LHC Fixed-Target}
\newacronym{lhcspin}{LHCspin}{LHCSpin}
\newacronym{ligo}{LIGO}{Laser Interferometer Gravitational-Wave Observatory}
\newacronym{linac}{Linac}{Linear Accelerator}
\newacronym{linac4}{Linac 4}{Linear Accelerator 4}
\newacronym{lisa}{LISA}{Laser Interferometer Space Antenna}
\newacronym{liu}{LIU}{LHC Injectors Upgrade}
\newacronym{llp}{LLP}{Long-Lived Particle}
\newacronym{llps}{LLPs}{Long-Lived Particles}
\newacronym{llrf}{LLRF}{Low Level Radio Frequency}
\newacronym{lnf}{LNF}{Laboratori Nazionali di Frascati}
\newacronym{loi}{LoI}{Letter of Intent}
\newacronym{ls2}{LS2}{Long Shutdown 2}
\newacronym{ls3}{LS3}{Long Shutdown 3}
\newacronym{ls4}{LS4}{Long Shutdown 4}
\newacronym{luxe}{LUXE}{Laser Und XFEL Experiment}
\newacronym{magis}{MAGIS}{Matter-wave Atomic Gradiometer Interferometric Sensor}
\newacronym{maps}{MAPS}{Monolithic Active Pixel Sensors}
\newacronym{mapp}{MAPP}{MoEDAL Apparatus for Penetrating Particles}
\newacronym{mappoutrigger}{MAPP-Outrigger}{Outrigger Detector for the MoEDAL Apparatus for Penetrating Particles}
\newacronym{mapp2}{MAPP-2}{MoEDAL Apparatus for Penetrating Particles, Phase-2}
\newacronym{mathusla}{MATHUSLA}{MAssive Timing Hodoscope for
Ultra-Stable neutraL pArticles)}
\newacronym{mathusla40}{MATHUSLA40}{MAssive Timing Hodoscope for
Ultra-Stable neutraL pArticles, $40\times 40$~m$^2$ area)}
\newacronym{md}{MD}{Machine Development}
\newacronym{mdm}{MDM}{Magnetic Dipole Moment}
\newacronym{milliqan}{MilliQan}{MilliQan Experiment}
\newacronym{mte}{MTE}{Multi Turn Extraction}
\newacronym{muone}{MUonE}{Muon On Electron Elastic Scattering}
\newacronym{mwpc}{MWPC}{Multi-Wire Proportional Chamber}
\newacronym{mqp}{mQP}{Milli-charged Particle}
\newacronym{mqps}{mQPs}{Milli-charged Particles}
\newacronym{na}{NA}{North Area}
\newacronym{na60+}{DICE/NA60+}{Dilepton and Charm Experiment/NA60+ Collaboration}
\newacronym{na62}{NA62}{NA62 Experiment}
\newacronym{na61/shine}{NA61/SHINE}{SPS Heavy Ion and Neutrino Experiment}
\newacronym{na64}{NA64}{NA64 Experiment}
\newacronym{nacons}{NA-CONS}{North Area Consolidation Project}
\newacronym{nboa}{NBOA}{Narrow-Band Off-Axis}
\newacronym{nc}{NC}{Neutral Current}
\newacronym{np06}{NP06}{Neutrino Platform 06 Experiment}
\newacronym{nsi}{NSI}{Non-Standard Interactions}
\newacronym{ntof}{nToF}{neutron Time of Flight facility}
\newacronym{nutag}{NuTag}{Neutrino Tagging}
\newacronym{p5}{P5}{Particle Physics Project Prioritization Panel}
\newacronym{pbc}{PBC}{Physics Beyond Colliders}
\newacronym{pdf}{PDF}{Parton Distribution Function}
\newacronym{pdg}{PDG}{Particle Data Group}
\newacronym{pid}{PID}{Particle Identification}
\newacronym{poc}{PoC}{Proof of Concept}
\newacronym{pot}{PoT}{Protons on Target}
\newacronym{ppb}{ppb}{particles per bunch}
\newacronym{ppp}{ppp}{particles per pulse}
\newacronym{proanubis}{proANUBIS}{ANUBIS prototype}
\newacronym{ps}{PS}{Proton Synchrotron}
\newacronym{psb}{PSB}{Proton Synchrotron Booster}
\newacronym{psi}{PSI}{Partially Stripped Ions}
\newacronym{pu}{PU}{average event pile-up ~$\left<\mu\right>$}
\newacronym{puma}{PUMA}{antiProton Unstable Matter Annihilation}
\newacronym{qed}{QED}{Quantum Electro-Dynamics}
\newacronym{qcd}{QCD}{Quantum Chromo-Dynamics}
\newacronym{qgp}{QGP}{Quark-Gluon Plasma}
\newacronym{qrades}{QRADES}{Quantum Relic Axion DEtection Sensors}
\newacronym{qti}{CERN QTI}{CERN Quantum Technology Initiative}
\newacronym{quaxlnf}{QUAX@LNF}{QUaerere AXions experiment @ LNF}
\newacronym{rades}{RADES}{Relic Axion Detector Exploratory Setup}
\newacronym{rd}{R\&D}{Research and Development}
\newacronym{rhic}{RHIC}{Relativistic Heavy-Ion Collider}
\newacronym{rich}{RICH}{Ring-Imaging Cherenkov}
\newacronym{rp}{RP}{Radiation Protection}
\newacronym{rpc}{RPC}{Resistive Plate Chamber}
\newacronym{rf}{RF}{Radio Frequency}
\newacronym{sbn}{SBN}{Short Baseline Neutrino}
\newacronym{sbnd}{SBND}{Short-Baseline Near Detector}
\newacronym{sc}{SC}{Superconducting}
\newacronym{scifi}{SciFi}{Scintillating Fibre}
\newacronym{sfqed}{SFQED}{strong-field QED}
\newacronym{shadows}{SHADOWS}{Search for Hidden And Dark Objects With the SPS}
\newacronym{ship}{SHiP/NA67}{Search For Hidden Particles experiment}
\newacronym{sipm}{SiPM}{Silicon Photo Multipliers}
\newacronym{slac}{SLAC}{Stanford Linear Accelerator Center}
\newacronym{sm}{SM}{Standard Model}
\newacronym{smbhs}{SMBHs}{supermassive black holes}
\newacronym{smog}{SMOG}{System to Measure the beam Overlap with Gas}
\newacronym{smog2}{SMOG2}{System to Measure the beam Overlap with Gas 2}
\newacronym{snd}{SND}{Scattering and Neutrino Detector}
\newacronym{sndlhc}{SND@LHC}{Scattering and Neutrino Detector at the LHC}
\newacronym{sndhllhc}{SND@HL-LHC}{Scattering and Neutrino Detector at the HL-LHC}
\newacronym{sps}{SPS}{Super Proton Synchrotron}
\newacronym{spsc}{SPSC}{SPS and PS Experiments Committee}
\newacronym{sqms}{SQMS}{Superconducting Quantum Materials and Systems Center}
\newacronym{squid}{SQUID}{Superconducting QUantum Interference Device}
\newacronym{star}{STAR}{Solenoidal Tracker at RHIC}
\newacronym{stax}{STAX}{Sub-THz Axion eXperiment}
\newacronym{step}{STEP}{Symmetry Tests in Experiments with Portable antiprotons}
\newacronym{tcc2}{TCC2}{Tunnel Target Cavern~2}
\newacronym{tcc4}{TCC4}{Tunnel Target Cavern~4}
\newacronym{tcc6}{TCC6}{Tunnel Target Cavern~6}
\newacronym{tcc8}{TCC8}{Tunnel Target Cavern~8}
\newacronym{tdr}{TDR}{Technical Design Report}
\newacronym{ti}{TI}{Tunnel Injection}
\newacronym{tmci}{TMCI}{Transverse Mode Coupling Instability}
\newacronym{tmd}{TMD}{Transverse Momentum Distribution}
\newacronym{tnc}{TNC}{Tunnel Neutrino Cavern}
\newacronym{tpc}{TPC}{Time Projection Chamber}
\newacronym{tt61}{TT61}{Transfer Tunnel~61}
\newacronym{tvlbai}{TVLBAI}{Terrestrial Very Long Baseline Atom Interferometry}
\newacronym{twocryst}{TWOCRYST}{Double Crystal set-up Proof of Principle Experiment}
\newacronym{uldm}{ULDM}{Ultra-Light Dark Matter}
\newacronym{us}{US}{United States}
\newacronym{ux}{UX}{Underground Experimental Cavern}
\newacronym{velo}{VELO}{Vertex Locator}
\newacronym{vhf}{VHF}{Very High Frequency}
\newacronym{virgo}{Virgo}{Virgo interferometric gravitational-wave observatory}
\newacronym{vmbcern}{VMB@CERN}{Vacuum Magnetic Birefringence experiment at CERN}
\newacronym{wa}{WA}{West Area}
\newacronym{wg}{WG}{Working Group}
\newacronym{wls}{WLS}{Wavelength Shifting}
\newacronym{yets}{YETS}{Year-End Technical Stop}

\printglossary[type=\acronymtype,title=List of Acronyms, toctitle=List of Acronyms]

\end{document}